\documentstyle[epsf,12pt,world_sci]{article}   

\begin{document}

\newcommand{\be}{\begin{equation}}
\newcommand{\ee}{\end{equation}}
\newcommand{\bea}{\begin{eqnarray}}
\newcommand{\eea}{\end{eqnarray}}
\newcommand{\dk}{\frac{d^{4}k}{(2\pi)^4i}}
\newcommand{\di}[1]{\frac{d^{4}{#1}}{(2\pi)^4i}}
\newcommand{\eq}[1]{eq.(\ref{#1})}
\newcommand{\f}[2]{\tilde {F}_{#1}(#2)}
\newcommand{\ns}{\hspace{-0.5ex}}
\newcommand{\p}{p \ns \cdot \ns \gamma }
\newcommand{\hf}{{\textstyle \frac{1}{2}}}
\newcommand{\Tr}{\,\mbox{Tr}\,}
\newcommand{\trc}{\,{\mbox{Tr}}_C\,}
\newcommand{\tr}{\,\mbox{tr}\,}
\newcommand{\Ln}{\,\mbox{Ln}\,}
\renewcommand{\ln}{\,\mbox{ln}\,}
\newcommand\half{{1\over 2}}
\renewcommand\Im{{\rm Im}}
\newcommand\mn{{\mu\nu}}
\newcommand\parm{\par\medskip}
\renewcommand\Re{{\rm Re}}
\renewcommand{\theequation}{\thesection.\arabic{equation}}
\newcommand{\cl}{\centerline}
\renewcommand{\thefootnote}{\fnsymbol{footnote}}

\def\pmb#1{\setbox0=\hbox{$#1$}%
\kern-.025em\copy0\kern-\wd0
\kern.05em\copy0\kern-\wd0
\kern-.025em\raise.0433em\box0}

\title{\bf \large SPIN, TWIST AND HADRON STRUCTURE\\ IN DEEP INELASTIC
PROCESSES\footnotemark[1]}

\author{\vspace{.2in}R. L. Jaffe\\[5ex]
{\small\em Center for Theoretical Physics} \\
{\small\em  Massachusetts Institute of Technology } \\
{\small\em                        Cambridge, MA ~02139~~U.S.A.}\\[1ex]
{\small\em and}\\[1ex]
{\small\em Department of Physics} \\
{\small\em  Harvard University} \\
{\small\em                        Cambridge, MA ~02138~~U.S.A.}}

\footnotetext[1]
{\baselineskip=16pt
This work is supported in part
by funds provided by 
National Science Foundation (N.S.F.) grant \break \#PHY 92-18167 and by
the U.S.~Department of Energy (D.O.E.)
under contracts \#DF-FC02-94ER40818 and \#DF-FG02-92ER40702.
\hfil\break
MIT-CTP-2506 and HUTP-96/A003\hfil
January 1996\break
}

\maketitle

\vskip10ex

\centerline{Notes for Lectures Presented at the}
\centerline{International School of Nucleon Structure}
\centerline{The Spin Structure of the Nucleon}
\centerline{Erice, 3 -- 10 August 1995}

\vskip10ex

\begin{abstract}
These notes provide an introduction to polarization effects in deep
inelastic processes in QCD.  We emphasize recent work on transverse
asymmetries, subdominant effects, and the role of polarization in 
fragmentation and in purely hadronic processes.  After a review of kinematics and
some basic tools of short distance analysis, we study the twist, helicity,
chirality and transversity dependence of a variety of high energy processes
sensitive to the quark and gluon substructure of hadrons.
\end{abstract}

\thispagestyle{empty}
\setcounter{page}{0}

\newpage
\bibliographystyle{unsrt}
\tableofcontents
\section*{Introduction}
\addcontentsline{toc}{section}{Introduction}

In recent years hadron spin physics has emerged as one of the most 
dynamic areas of particle physics.  During the same period the field
has got considerably more complicated.  In times past only
longitudinal asymmetries, that have simple parton model
interpretations, attracted much attention; only dominant effects,
that scale in the Bjorken limit, were experimentally accessible; and
only relatively crude experimental data were available.  Now
interest has spread to transverse polarization asymmetries,
subdominant effects, polarization effects in fragmentation and in
purely hadronic processes.  The aim of these lectures is to present an
introduction to spin dependent effects at dominant and subdominant
order in deep inelastic processes including deep inelastic scattering
of leptons, $e^+e^-$ annihilation, and Drell-Yan processes.  The
methods can be extended relatively straightforwardly to other spin
dependent effects in hard processes. 

In a short set of lectures some detail and background must be
sacrificed.  As for background, I will assume that readers are
familiar with the elementary parton model treatment of highly
inelastic  processes in the ``infinite momentum frame''.  Anyone who
is not familiar with basic parton model ideas should consult standard
textbook presentations. \cite{Ait89,Clo79,Nac90} Although I will have a
lot to say about the parton model, it may look poorly motivated to
someone who has not seen the ideas presented in their simplest form
first.  As for detail, I will mostly ignore the complications of QCD
radiative corrections, normally included via the renormalization
group.  There are many excellent treatments including books by
Collins\cite{Col84}, Muta \cite{Mut87} and most recently in a context
particularly well suited to these lectures, by Roberts. \cite{Rob90} 
Of course radiative corrections and the momentum scale dependence they
generate are central to the understanding of QCD.  Some important
aspects are covered in Al Mueller's lectures in this volume. Here we
will be interested in the {\it classification\/} of scattering
amplitudes in terms of helicity, chirality, twist, {\it etc.\/} -- a
classification which is largely (but not entirely) independent of
radiative corrections.  In many cases the soft, $\ln Q^2$ dependence
they generate can be regarded as decorations of our primary results. 
Where this is not the case, I will try to warn the reader and refer to
the appropriate literature.

The main question to be addressed here is: How can one classify and
interpret the wide variety of spin dependent phenomena expected in hard
processes?  Which phenomena are displayed in which experiments?  What
are the selection rules enforced by the symmetries of QCD?  Which
phenomena dominate at large-$Q^2$, which are suppressed, and what
is the physical origin of the suppression?  In short, the object is to
provide the background for both experimental and theoretical analysis
of spin effects in hard processes.  In contrast, I will resist almost
entirely the temptation to speculate about the origins of spin effects
based on models of hadron structure.  These notes are not intended to
be an introduction to the so-called ``spin crisis'' which grew out of
the observation that only a small fraction of the nucleon's spin is
carried by the spin of quarks.  Theorists will not find their
own or my own favorite explanation of the spin crisis in these
lectures.  That is a subject for another school.

Certain predictions of perturbative QCD are admired for being very 
general and independent of the difficult details of hadron structure. 
Examples include the cross section for $e^+e^-\rightarrow$ hadrons,
event shapes in $e^+e^-$ annihilation, the $\ln Q^2$ dependence of
deep inelastic structure functions, and the Gross-Llewellyn Smith and
Bjorken Sum Rules.  Studies of these processes provide essential tests
of QCD.  These will not be major topics here.  I will assume that
perturbative QCD is correct and use it as a sophisticated probe of the
poorly understood dynamics of confinement.  As we shall see,
perturbative QCD is by now so well understood that it is possible to
``tune'' the probe to measure the nucleon expectation values of a
variety of quark and gluon distributions and correlations within
hadrons.  Probes can be selected for spin, twist and flavor quantum
numbers, and can be used either to analyze the structure of hadronic
targets or reaction fragments.  No other approach yields such well
defined information about hadronic bound states.  This information may
help guide us to a better understanding of confinement from first
principles.

Many aspects of these lectures are based on work performed in
collaboration with Xiangdong Ji.  The reader who wishes to explore
subjects in greater depth should look at
refs.~\cite{JafJi91,JafJi1952,JafJi2005,JafJi2158,JafJi2365,JafJi2402}, as well as other references provided in the text.  I
would like to thank Xiangdong for the pleasure of this long
collaboration.  Thanks are also due to Matthias Burkardt, Gary
Goldstein and Aneesh Manohar who collaborated on other projects
related to this work.  In addition I have benefited greatly from
discussions with Guido Altarelli, Xavier Artru, Ian Balitsky, Vladimir
Braun, Gerry Bunce, John Collins, Vernon Hughes, Gerd Mallot, Al
Mueller, Richard Milner, John Ralston, Phil Ratcliffe, 
Klaus Rith, Jacques Soffer, and
Linda Stuart.

These lectures grew out of talks at schools and conferences in the
early 1990's.  A version presented at the 1992 Graduiertenkolleg of
the Universities of Erlangen and Regensberg at Kloster Banz, Germany,
was recorded by  Drs. H.~Meyer and G.~Piller.  The present version is
based on a manuscript prepared by Drs.~Meyer and Piller from their
notes.  I would like to thank them for the substantial work they
undertook at that time.  Subsequently I have edited, reformulated and
expanded the notes, most recently for the Erice School on the Internal
Spin Structure of the Nucleon.

\newpage

\section{Kinematics and other Generalities}
The organizers of the school asked if I would briefly introduce the
kinematic and dynamical variables common in the study of deep inelastic
processes.  So before getting down to the business of dynamics here is a
short summary --- the cogniscenti will certainly want to skip this
section.  Others, who may be familiar with less streamlined notation might
wish at least to look at
eqs.~(\ref{eq:WS}), (\ref{eq:WA}), (\ref{eq:xsection}),
and (\ref{eq:spindepxsection}).  I hope students with less background in
perturbative QCD will find this section useful. 
\subsection{Deep Inelastic Scattering}
\subsubsection{Basic Variables}
Deep inelastic scattering (DIS) is the archetype for hard processes in QCD: 
a lepton --- in practice an electron, muon or neutrino --- with high energy
scatters off a target hadron --- in practice a nucleon or nucleus, or
perhaps a photon --- transferring large quantities of both energy and
invariant squared-four-momentum. For charged leptons the dominant reaction
mechanism is electromagnetism and one photon exchange is a good
approximation.  For neutrinos either $W^\pm$ (charged current) or $Z^0$
(neutral current) exchange may occur.  The weak interactions of electrons
may also be studied either by means of small parity violating asymmetries
originating in $\gamma - Z^0$ interference, or by means of the charged
current reaction $e^-\rightarrow\nu_e$.  

\begin{figure}[b]
	\centerline{\epsffile{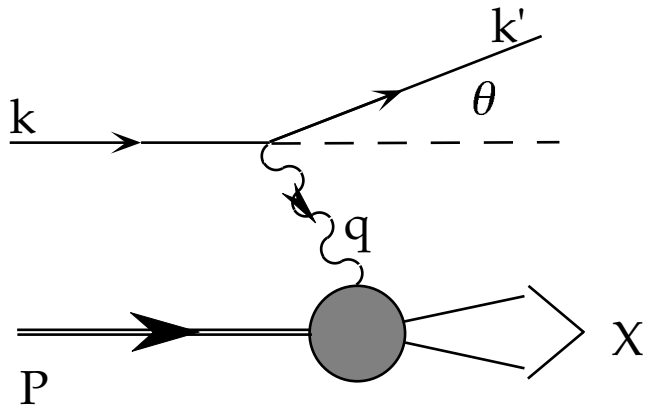}}
	\caption{{\sf Kinematics of lepton-hadron scattering in the target rest
	frame }}
	\label{fig:kin}
\end{figure}

We are primarily interested in experiments performed
with polarized targets.  Neutrino scattering experiments require far too
massive targets for polarization to be a practical option, so we will
ignore them, although $W$-exchange has been observed in
$e^-p\rightarrow\nu_e+X$, at HERA,\cite{ccatHera} and could be extended to 
a polarized target, at least in principle.  
Thus we are mainly limited to charged lepton scattering by one photon exchange. 
The kinematics is shown in fig.~(\ref{fig:kin}). The initial lepton with
momentum $k$ and energy $E$ exchanges a photon of momentum
$q$ with a the target with momentum $P$. Only the outgoing electron with
momentum $k^\prime$ and energy $E^\prime$ is detected.
One can define the two invariants
\bea
	& &q^2 \equiv (k-k^\prime )^2=q_0^2-\vec{q}\hspace{1mm}^2
	 =-4EE^\prime \sin^2(\theta /2)=-Q^2<0 \\
	& &\nu \equiv P\cdot q =M(E - E^\prime),
\eea
where the lepton mass has been neglected (and will be neglected henceforth).
The meaning of the scattering angle
$\theta$ is clear from fig.~(\ref{fig:kin}).  Unless otherwise noted, $E$,
$E^\prime$, $\theta$ and $q^0\equiv E - E^\prime$ refer to the target rest
frame.  The deep inelastic, or {\it Bjorken\/} limit is where $Q^2$ and
$\nu$ both go to infinity with the ratio, $x\equiv Q^2/2\nu$ fixed.  $x$ is
known as the Bjorken (scaling) variable.

Since the invariant mass of the hadronic final state is larger than
or equal to the mass of the target, $(P+q)^2\ge M^2$, one has
$0<x\leq 1$.  It is convenient also to measure the energy loss using
a dimensionless variable,
\be
	0\leq y \equiv {\nu \over ME} \leq 1.
\ee
We will find  $E$, $Q^2$, $x$, and $y$ to be a useful set of variables. 
Note that it is overcomplete since $xy=Q^2/2ME$, and note also that what
we define as $\nu$ differs from common usage by a factor of $M$.  The
behavior of cross sections at large $Q^2$ is much more transparent using
these variables than using the set ($E,E^\prime,
\theta$) favored by experimenters for the reason that $\theta\rightarrow 0$
as $Q^2\rightarrow\infty$ at fixed $x$ and $y$.
\subsubsection{Cross Section and Structure Functions}
\begin{figure}
\centerline{\epsffile{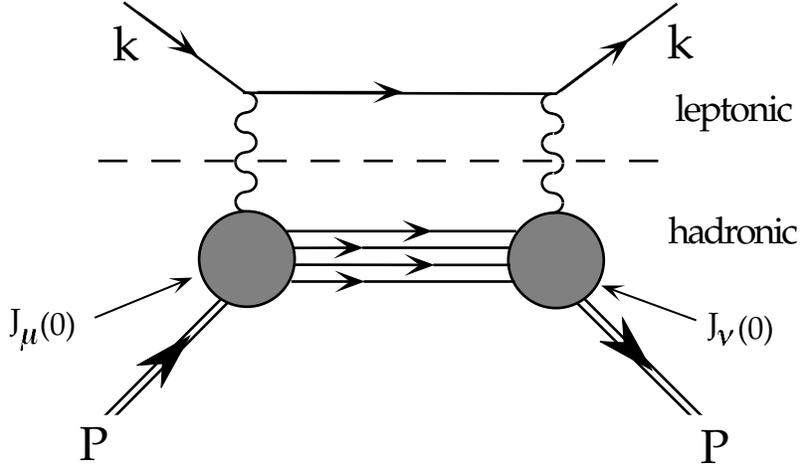}}
	\caption{{\sf The squared amplitude ${\cal A}$ for electron-hadron
	scattering can be separated into a leptonic tensor
	$l^{\mu \nu}$ and a hadronic tensor $W^{\mu \nu}$. }}
	\label{fig:amplsep}
\end{figure}

The differential cross-section for inclusive scattering 
($eP\rightarrow e^\prime X$) is given by:
\begin{equation}
	d\sigma = {1 \over J} {d^3 k^\prime \over 2E^\prime (2\pi )^3}
	\sum _X\prod _{i=1}^{n_X} \int {d^3p_i \over (2\pi )^3 2p_{i0}}
	\vert {\cal A} \vert ^2 (2\pi )^4 \delta ^4 (P+q -	\sum _{i=1}^{n_X} p_i).
	\label{cc}
\end{equation}
The flux factor for the incoming nucleon and electron is denoted by 
$J=4 P\cdot k$, which is equal to $J=4ME$ in the rest frame of the nucleon. 
The sum runs over all hadronic final states $X$ which are not observed. Each
hadronic final state consists of $n_X$ particles  with momenta $p_i$
($\sum_{i=1}^{n_X} p_i\equiv p_X$). The squared-amplitude
$\vert{\cal A}\vert^2$ can be separated into a leptonic
($l^{\mu \nu}$) and a hadronic ($W^{\mu \nu}$) tensor (see
fig.~(\ref{fig:amplsep})):
\begin{equation}
	\left| {{\cal A}\over 4\pi} \right|^2
	={\alpha^2 \over Q^4}l^{\mu \nu}W_{\mu \nu},
\end{equation}
where $\alpha \sim 1/137$ is the electromagnetic fine structure constant.
The leptonic tensor $l^{\mu \nu}$ is given by the square of
the elementary spin $1/2$ current (summed over final spins):
\begin{eqnarray}
	l^{\mu \nu}
	&=&\sum _{s^\prime }
	\bar{u} (k,s) \gamma ^\mu u(k^\prime,s^\prime)
	\bar{u} (k^\prime,s^\prime)\gamma ^\nu u (k,s)\nonumber\\
	&=&2(k^{\prime \mu} k^\nu + k^{\prime \nu} k^\mu )
	- 2 g^{\mu \nu} k\cdot k^\prime + 2i\epsilon ^{\mu \nu \lambda \sigma}
	q_\lambda s_\sigma,\nonumber\\
	\label{lepten}
\end{eqnarray}
and consists of parts symmetric and antisymmetric in
$\mu$ and $\nu$. The antisymmetric part is linear in the spin vector 
$s$, which is normalized to $s^2 = -m^2$. While the leptonic tensor is known
completely, $W^{\mu \nu}$, which describes the internal structure of the
nucleon,  depends on non-perturbative strong interaction dynamics.  It is
expressed in terms of the current $J^\mu$ as:
\begin{eqnarray}
	4\pi W^{\mu \nu} & = &
	\sum_{X} \langle PS\vert J^{ \mu}\vert X\rangle
	\langle X \vert J^{\nu} \vert PS \rangle (2\pi )^4 \delta
	(P+q-p_X) \label{eq:Wmunu} \label{eq:csec} \\
	& = & \int d^4\xi e^{iq\cdot \xi}
	\langle PS\vert [ J^\mu(\xi) ,
	J^{\nu}(0) ] \vert PS \rangle _c. 
 	\label{eq:com} 
\end{eqnarray}
The steps leading from eq.~(\ref{eq:csec}) to eq.~(\ref{eq:com}) include writing
the $\delta$ function as an exponential,
\be
	(2\pi)^4\delta^4(K)=\int d^4\xi e^{i\xi\cdot K},
\ee
translating the current, $e^{i\xi\cdot(P-p_X)}\langle PS\vert J^{
\mu}(0)\vert X\rangle = \langle PS\vert J^{ \mu}(\xi)\vert X\rangle$, and
using completeness, 
$\sum_X\vert X\rangle\langle X \vert=1$.  Note that another term has been
subtracted to convert the current product into a commutator.  It is easy to
check that the new term vanishes for $q^0> 0$ which is the case for physical
lepton scattering from a stable target. The subscript $_c$ means that the
graphs associated with the matrix element must be connected. Finally, note
that the states are covariantly normalized to:
\be
	\langle P\vert P^\prime \rangle = 2E (2\pi )^3 \delta^3 (P-P^\prime).
\ee

\begin{figure}
\centerline{\epsffile{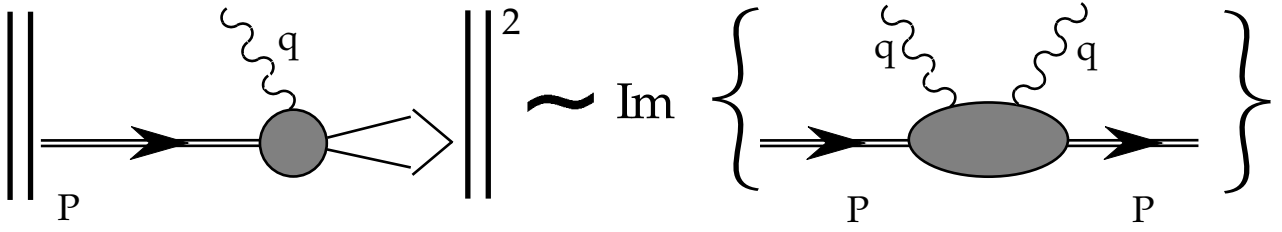}}
	\caption{{\sf The optical theorem relates the hadronic
	structure tensor, $W^{\mu\nu}$ to the imaginary part of
	forward ($P=P'$, $q=q'$), virtual ($Q^2<0$) Compton
	scattering}} 
        \label{fig:opti}
\end{figure}

The optical theorem:
\be
	2\pi W^{\mu \nu} = 
	{\rm Im} \hspace{2mm} T^{\mu \nu}     
\ee
relates the hadronic tensor to the imaginary part of the forward virtual
Compton scattering amplitude, $T$:
\be
	T _{\mu \nu} =i \int d^4 \xi e^{i q\cdot \xi}
	\langle PS\vert T(J_\mu (\xi ) J_\nu (0) )\vert PS\rangle
\ee
as shown graphically in fig.~(\ref{fig:opti}).

\subsubsection{Structure Functions}
Using Lorentz covariance, gauge invariance, parity conservation in
electromagnetism and standard discrete symmetries of the strong interactions,
$W^{\mu \nu}$ can be parametrized in terms of four
scalar dimensionless structure functions $F_1(x,Q^{2})$, $F_2(x,Q^{2})$,
$g_1(x,Q^{2})$ and $g_2(x,Q^{2})$. They depend only on the two invariants
$Q^{2}$ and $\nu$, or  alternatively on $Q^2$ and the dimensionless Bjorken
variable $x$. Splitting $W^{\mu\nu}$ into symmetric and anti-symmetric parts
we have,
\begin{equation}
	W^{\mu\nu}=W^{\{\mu\nu\}}+W^{[\mu\nu]},
\end{equation}
with
\begin{eqnarray}
	W^{\{\mu \nu\}} & = 
	& \left( -g^{\mu \nu}+{q^{\mu}q^{\nu}\over q^2}\right) 
   	F_1 +\left[ \left( P^{\mu}-{\nu \over q^2}q^{\mu} \right) 
  	\left( P^{\nu}-{\nu \over q^2}q^{\nu} \right) \right] {F_2 \over \nu}, 
	\label{eq:WS}\\
 	W^{[\mu \nu]} & = & -i\epsilon ^{\mu \nu \lambda \sigma} q_{\lambda}
  	\left( {S_{\sigma}\over \nu} \left( g_1 + g_2 \right)
	  -{q\cdot S P_{\sigma}\over \nu^2}g_2 \right), \label{eq:WA}
\end{eqnarray}
where $S^\sigma$ is the polarization vector of the nucleon $(S^2=-M^2)$,
$P\cdot S=0$.  $S^\sigma$ is a pseudovector.   Since $W^{[\mu\nu]}$ is a
normal tensor, parity demands that the $S^\mu$ appear with another 
pseudotensor, and the only one available is the
$\epsilon^{\mu\nu\sigma\lambda}$.  Students often ask why
$W^{\mu\nu}$ depends only linearly on $S^\mu$ -- what is wrong with $S^\mu
S^\nu$, for example?  Lorentz invariance demands  that $W^{\mu\nu}$, defined
in eq.~(\ref{eq:com}) be linear in the initial and final nucleon spinors,
$U(P,S)$ and $\bar U(P,S)$.  Tensors constructed from these are either spin
independent ($\bar U(P,S)\gamma^\mu U(P,S) = 2P^\mu$) or linear in $S^\mu$
(($\bar U(P,S)\gamma^\mu\gamma_5 U(P,S) = 2S^\mu$), but that is the end of
it.

Note also that  $W^{\mu\nu}$ is dimensionless (we shall have more to say
about operator dimensions shortly).  Factors of $\nu$ have been judiciously
introduced into eqs.~(\ref{eq:WA}) and (\ref{eq:WS}) so that the four structure
functions, $F_1$, $F_2$, $g_1$, and $g_2$ are dimensionless.  These structure
functions are related to others in common use by:
\be
	W_1 = F_1, \quad
	W_2 = {M^2 \over \nu} F_2, \quad
	G_1 = {M^2 g_1 \over \nu}, \quad
	G_2 = {M^4 g_2\over \nu ^2}.
\ee

\subsubsection{Scaling and Kinematic Domains}

\begin{figure}
\centerline{\epsffile{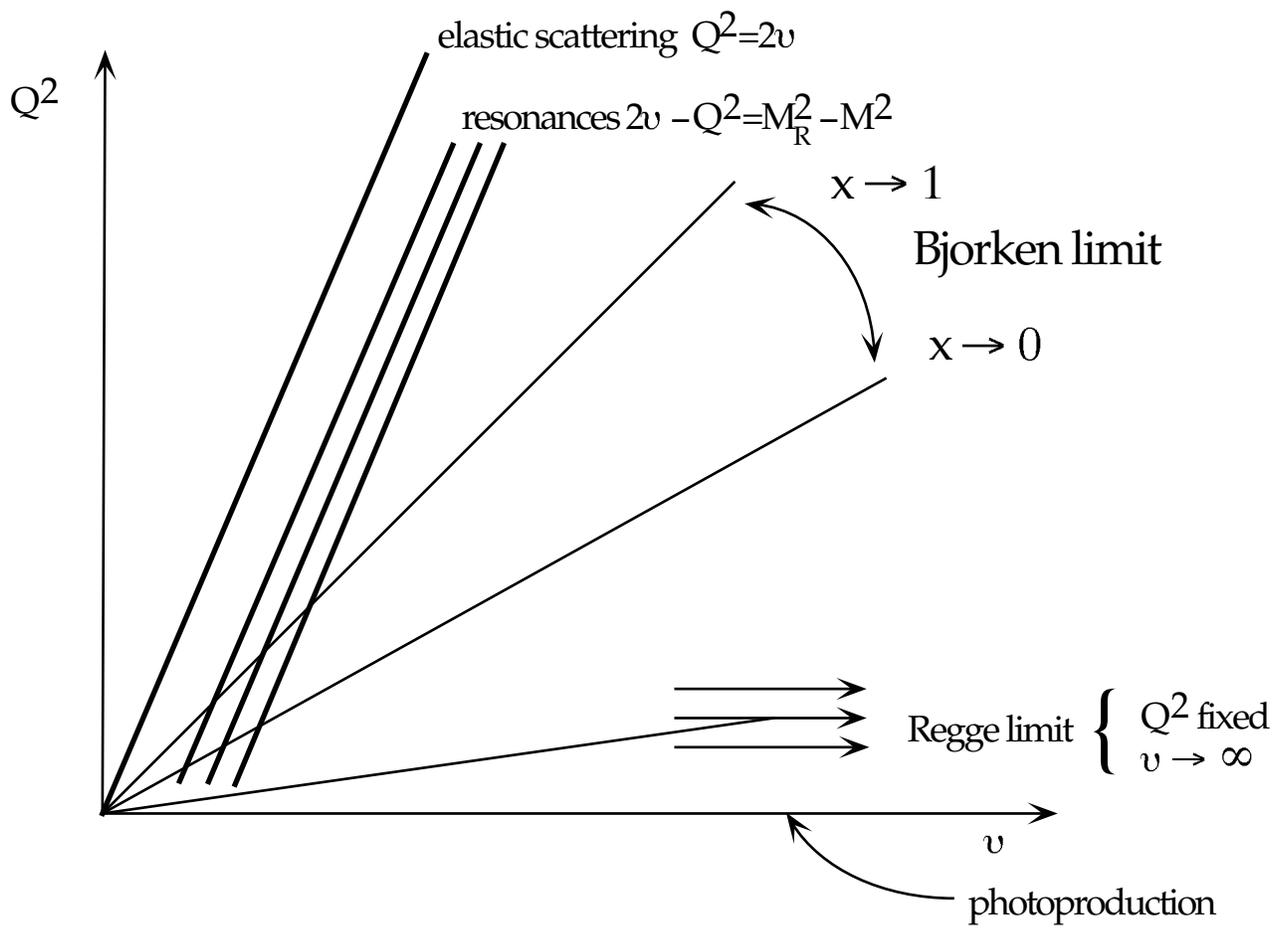}}
	\caption{{\sf Kinematic domains in electron-nucleon scattering.}}
	\label{fig:kinmap}
\end{figure}

Our choice of invariant structure functions makes the determination of
scaling behavior at large $Q^2$ almost trivial.  In the Bjorken limit where
$Q^2 \rightarrow \infty$ and 
$\nu \rightarrow \infty$,
$x=Q^2/2\nu$ fixed, QCD becomes scale invariant up to logarithms of
$Q^2$ generated by radiative corrections.  Under a scale transformation,
$P\rightarrow \lambda P$, $q\rightarrow\lambda q$, and $M\rightarrow\lambda
M\ne M$, so a theory with a discrete spectrum of massive particles cannot be
scale invariant except in a limit in which all masses are negligible.  Thus
no masses can appear in $W^{\mu\nu}$ in the Bjorken limit; it must be a
dimensionless function of $P^\mu$, $q^\mu$, $S^\mu$, and the invariants
$Q^2$ and $\nu$.  In particular, it cannot depend explicitly on the target
mass, $M$.  If, for example, a term like ${{P^\mu P^\nu}\over M^2}W_2$
appeared in $W^{\{\mu\nu\}}$, it would violate scale invariance unless $W_2$
vanished like ${M^2\over\nu}$ at large $Q^2$.  Clearly, the way to avoid such
pathological choices of structure functions is to write the dimensionless
tensor $W^{\mu\nu}$ in terms of dimensionless invariant functions using
$\nu$ (or $Q^2$) to supply dimensional factors as needed.  The immediate
conclusion is that the functions $F_1$, $F_2$, $g_1$, and $g_2$ defined in
eqs.~(\ref{eq:WS}) and (\ref{eq:WA}), become functions only of the dimensionless
ratio $x = Q^2/2\nu$, modulo logarithms, in the Bjorken limit,
\bea
	F_1(Q^2,\nu ) &\rightarrow F_1(x, \ln Q^2),\quad F_2(Q^2,\nu )
	 &\rightarrow F_2(x, \ln Q^2)\nonumber\\
	g_1(Q^2,\nu ) &\rightarrow g_1(x, \ln Q^2),\quad	g_2(Q^2,\nu )
	&\rightarrow g_2(x, \ln Q^2)\nonumber\\
\eea
as $Q^2$ and $\nu$ become large at fixed $x$. In practice it is observed that
for $Q^2 > 1 GeV^2$, the structure functions depend only very weakly on
$Q^2$. Furthermore one observes an approximate relationship between $F_1$
and $F_2$, known as the Callan-Gross relation,\cite{CalGross}
\be
	F_1-{1\over 2x}F_2 \sim {1\over \ln Q^2},
\ee
which indicates that the particles that carry electric charge (the
quarks) have spin ${1\over 2}$.  The different kinematic domains of interest
in inelastic electron scattering are shown in fig.~(\ref{fig:kinmap}).

 \subsubsection{Flavor Generalizations}
Only up, down and strange quarks appear to be important constituents of
light hadrons.  The processes of interest to us, therefore are mediated by
currents lying in the $SU(3)_R\times SU(3)_L$ space of $u,d,s$ vector and
axial currents,
\bea
	J_{\mu}^a & \equiv {1\over 2} \bar{\psi} \gamma_\mu \lambda ^a
	\psi,\nonumber\\
	 J_{\mu 5}^a & 
	\equiv {1\over 2} \bar{\psi} \gamma_\mu \gamma_5\lambda ^a 
	 \psi,\nonumber\\
\eea
where $ \lambda ^a$ for $a=1,2 \ldots 8$ are the flavor $SU(3)$ matrices,
which are normalized to ${\rm Tr} \lambda^a \lambda^b =2\delta ^{ab}$.  Note, in
particular, that $\lambda_3 = {\rm diag}(1,-1,0)$ and $\lambda_8 = {\rm
diag}{1\over\sqrt{3}}(1,1,-2)$. In addition one has the flavor singlet
current $J_\mu^0=\sqrt{2/3} \bar{\psi} \gamma_\mu \psi$, acting like
$\sqrt{2\over 3}{\rm diag}(1,1,1)$ in flavor space. 
      
\subsubsection{Cross Section for Electron-Hadron Scattering}
The differential cross section for unpolarized electron-hadron scattering
can now be expanded in the Lorentz scalar structure functions by contracting
the symmetric tensor, eq.~(\ref{eq:WS}), with the leptonic tensor,
eq.~(\ref{lepten}).  Likewise the cross section for polarized scattering is
obtained by contracting the antisymmetric tensor, eq.~(\ref{eq:WA}), with the
same lepton tensor.   The result is often quoted in terms of the experimenter's
variables, $Q^2$, $\nu$, $\theta$, $E$ and $E'$, {\it e.g.\/} for the spin
average case,
\begin{equation}
	{d^3 \bar\sigma \over dE ^\prime d\Omega }
	=  {4 \alpha ^2 \over M Q^4} E^{\prime 2}
	\left\{ 2 W_1(q^2,\nu ) \sin ^2 {\theta \over 2}+W_2(q^2, \nu) 
 	\cos ^2 {\theta \over 2} \right\}.
\end{equation}
The relative importance of the two terms is difficult to judge. 
Superficially  it looks as though $W_1$ and $W_2$ are equally important.  On
second thought, $W_1$ is multiplied by $\sin^2{\theta\over 2}$ which gets
small in the Bjorken limit.   On third thought, $W_2$ vanishes like
${M^2\over\nu}$.  To disentangle all this, we rewrite $d\bar\sigma$ in terms
of $F_1$, $F_2$, $x$, $y$, and $Q^2$, where scaling behavior should be
manifest,
\begin{equation}
 	{d^3\bar{\sigma}\over dx\,dy\,d\phi} = {e^4\over 4\pi^2 Q^2} 
 	\left\{ {y\over 2}
	 F_1(x,Q^2) + {1\over 2xy}\left( 1 - y - {y^2\over 4}(\kappa-1)\right) 
	 F_2(x,Q^2)\right\}
	 \label{eq:xsection}
\end{equation}
with $\kappa \equiv 1-{4x^2M^2\over Q^2}$. No scaling approximations have
been made in eq.~(\ref{eq:xsection}).  Under typical experimental conditions
$y$ and $x$ are of order unity, though experiments are now being carried out
at very low-$x$.  Since $F_1\sim{1\over x}$ and $F_2\sim$ const. for small
$x$, the two terms are comparable. There is no significant dependence on the
azimuthal angle $\phi$, which cannot even be uniquely defined for inclusive
scattering with an unpolarized target.

\begin{figure}
\centerline{\epsffile{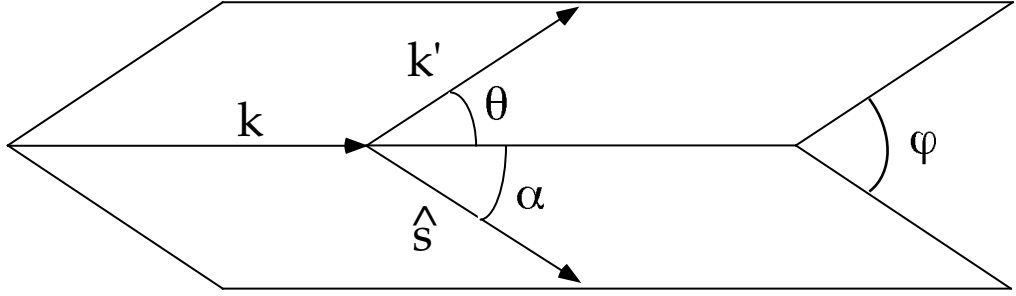}}
	\caption{{\sf Kinematic variables in polarized lepton scattering from a
        polarized target.}}
	\label{fig:spinkin} 
\end{figure}

It is clear from the tensor structure of $\ell_{\mu\nu}$ and $W_{\mu\nu}$
that no target spin dependent effects survive if the beam is unpolarized. 
Therefore we lose no generality by defining the spin dependent cross
section, $\Delta\sigma$ as half the difference between right-
and left-handed incident electron cross sections,\cite{Jafg2}
\begin{eqnarray}
 	{d\Delta\sigma(\alpha)\over dx\,dy\,d\phi}&=&{e^4\over 4\pi^2
	 Q^2}\Biggl\{\cos\alpha\left\{ \left[ 1 - {y\over 2} - 
 	{y^2\over 4} (\kappa-1) \right] g_1
 	(x,Q^2) - {y\over 2} (\kappa-1) g_2 (x,Q^2) \right\}\nonumber \\ 
 	&-&\sin\alpha\cos\phi\sqrt{ (\kappa-1) \left( 1 - y - {y^2\over 4}
 	(\kappa-1)\right)}\,\left( {y\over 2} g_1(x,Q^2) + g_2
 	(x,Q^2)\right)\Biggr\}\nonumber\\
 	\label{eq:spindepxsection}
\end{eqnarray}
Now the azimuthal angle $\phi$ and the angle, $\alpha$, between the target
spin $\hat S$ and the incident electron momentum, $\hat k$, make non-trivial
appearances.  These and other kinematic variables are defined in
fig.~(\ref{fig:spinkin}).  Note the following:
\begin{itemize}
 \item  $\alpha$ is the angle between the spin vector of the target $(\hat
   S)$ and the incident electron beam $(\hat k)$, {\it not\/} the virtual
   photon direction $(\hat q)$.
 \item $\phi$ is the azimuthal angle between the plane defined by ${\vec
   k}$ and ${\vec k}'$ and the plane defined by ${\vec k}$ and $\hat S$.
 \item  Eqs.~(\ref{eq:xsection}) and (\ref{eq:spindepxsection}) are 
   exact (except that lepton masses have been ignored): no scaling limit
   has been taken.  $\kappa-1 \equiv M^2Q^2/\nu^2 = 4M^2 x^2/Q^2$ is a
   measure of the approach to the scaling limit, $Q^2\to\infty$.
 \item To eliminate {\it spin-independent\/} effects one may either (i)
   subtract cross sections for different values of $\alpha$; (ii) subtract
   cross sections for right- and left-handed leptons; or (iii) measure
   $\phi$-dependence.
\end{itemize}
Notice that effects associated with $g_2(x,Q^2)$ are suppressed by a factor
$\sqrt{\kappa-1} = {2Mx\over \sqrt{Q^2}}$ with respect to the dominant
structure function $g_1(x,Q^2)$.  In technical terms, this means that
effects associated with $g_2$ are ``higher twist'' --- suppressed by  a
power of $Q$ relative to the leading phenomena in the Bjorken limit.
However, at $90^\circ$ the coefficient of the dominant term vanishes
identically and allows the combination ${y\over 2} g_1 + g_2$ to be
extracted cleanly at large $Q^2$.  This is a unique feature of the
spin-dependent scattering.  Only very rarely, to my knowledge, can a higher
twist effect be selected by an adroit kinematic arrangement, thereby
avoiding the difficult process of fitting and subtracting away a leading
twist effect to expose the higher twist correction underneath.
 
\subsection{Other Basic Deep Inelastic Processes}
\subsubsection{Inclusive $e^+$ $e^-$ Annihilation}
\begin{figure}
\centerline{\epsffile{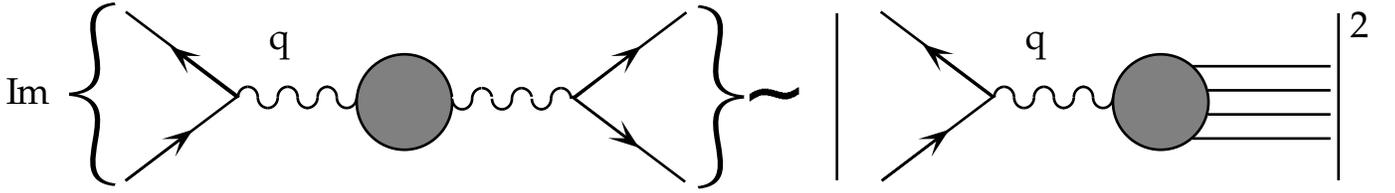}}
	\caption{{\sf The optical theorem relates the total
	cross-section for $e^+e^-$ annihilation to the imaginary
	part of the photon propagator.}}
	\label{fig:phot}
\end{figure}

In this process an electron with momentum $k$ and a positron with momentum
$k^\prime$ annihilate to form a massive time-like photon with momentum
$q=k+k^\prime$ ($Q^2\equiv q^2 > 0$), which decays into an unobserved final
state. Through the optical theorem, the total cross section is proportional
to the imaginary part of the photon propagator (see fig.~(\ref{fig:phot})),
\be
	\sigma _{tot} ={16 \pi ^2 \alpha ^2 \over Q^2} \Pi (Q^2),
	\label{eq:schaap}
\ee
where $\Pi (Q^2)$ is the Lorentz scalar spectral function appearing in the
photon propagator:
\be
	\Pi _{\mu \nu} = (q_\mu q_\nu-q^2 g_{\mu \nu}) \Pi (Q^2),
\ee
and
\be
	\Pi (Q^2)=-{1\over 6Q^2} \int d^4 \xi e^{iq\cdot \xi}
	\langle 0 \vert \left[ J_\mu (\xi ),J^\mu (0) \right] \vert  0 \rangle.
\ee
Usually the data are expressed as a ratio to the pointlike
annihilation cross section to muons (to lowest order in $\alpha_{EM}$):
\be
	R(Q^2) \equiv {\sigma _{tot} (e^+e^- \rightarrow hadrons) \over
	\sigma (e^+e^- \rightarrow \mu^+\mu^-)}=
	12 \pi \Pi(Q^2).
	\label{eq:pruttel}
\ee
Since the hadronic process is initiated by the creation of a
$q\bar{q}$ pair, $R$ directly measures the number of colors.
At large $Q^2$ it is modified only by perturbative QCD corrections:
\bea
	R(Q^2)&=&\sum _q 3 e_q^2 \left\{ 1+{\alpha_{QCD}(Q^2) \over \pi}
	+1.409 {\alpha_{QCD} (Q^2) \over \pi^2} - 12.805 {\alpha _{QCD}^3 (Q^2) 
 	\over \pi^3}
 	+ \cdots \right\}\nonumber \\
	&+& \mbox{quark mass corrections}.
	\label{eq:rpert})
\eea
The coefficients in eq.~(\ref{eq:rpert}) are renormalization scheme dependent
beyond lowest order.  Those quoted in eq.~(\ref{eq:rpert}) were calculated in
$\overline{MS}$ scheme with five flavors.\cite{MSBar}  The formula for $R$
does not depend on any details of hadronic structure, so it provides an
important test of QCD (and measurement of $\alpha_{QCD}$).  Similar remarks
apply to processes in which jets are observed in the final state of $e^+e^-$
annihilation.  Two jet events have the angular distribution that one expects
for two spin $1/2$ quarks; a third jet is associated with gluonic
bremsstrahlung. These processes however are not sensitive to the structure
of hadrons and we will not discuss them further here.
\subsubsection{Inclusive $e^+e^-$ Annihilation with One Observed
Hadron}
This process looks very much like a timelike version of deep inelastic
scattering.  Indeed it shares many important characteristics, but it also
differs in essential ways.  From the point of view of a theorist interested
in hadron structure, the opportunity to study unstable hadrons makes this
process very attractive.  Deep inelastic scattering from $\Lambda$-hyperons
or $\pi$ or $\rho$-mesons will never be more than a {\it gedanken\/}
experiment.  However, these and other unstable hadrons have already been
studied in $e^+e^-$-annihilation.  The physical basis of ``fragmentation''
--- the process by which a quark created by the current from the vacuum
fragments into the observed hadron --- is not as well understood as DIS,
making this an area of considerable interest at the present time.

\begin{figure}[b]
\centerline{\epsffile{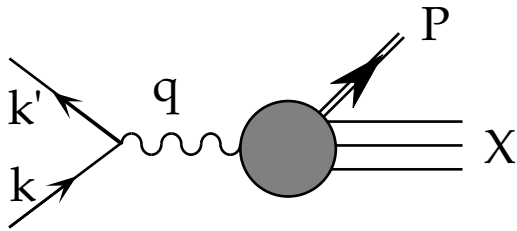}}
	\caption{{\sf Kinematics for single particle 
   	inclusive annihilation --- $e^+e^-\rightarrow PX$.}}
	\label{fig:eekin}
\end{figure}

The kinematics for $e^+e^-\rightarrow P + X$ is illustrated in
fig.~(\ref{fig:eekin}).  Once again two kinematic invariants, $Q^2$ and $\nu
= P\cdot q$, define the process.  The limit of interest is $Q^2, \nu
\rightarrow\infty$, with $z\equiv {2\nu\over Q^2}$ fixed.
The momentum of the virtual photon is 
time-like, and that makes a major difference as we shall see 
in \S 2.  The invariants are often expressed in terms of quantities measured
in the $e^+e^-$ center of mass:
\bea
	q^2 &\equiv& Q^2 = 4E_e^2 > 0 \nonumber\\
	P\cdot q &\equiv& \nu =E \sqrt{Q^2} \nonumber\\
	0<z &\equiv& {2P\cdot q \over q^2} = {E \over E_e} \leq 1,\nonumber\\
\eea
where E is the energy of the observed hadron.  We shall usually be
interested the polarization dependence, but here we illustrate the
kinematics for the simpler, spin-averaged case. The cross section can be
written as the product of a leptonic
$\hat l_{\mu \nu}$  and a hadronic tensor $\hat{W} ^{\mu \nu}$:
\be
	d\sigma \sim \hat l_{\mu \nu} \hat{W} ^{\mu \nu} 
	{d^3P \over (2\pi ^3) 2E}
\ee
The hadronic tensor is determined by the electromagnetic current
and depends on two invariant ``fragmentation functions'' due to 
current conservation and C, P and T invariance:
\bea
	\hat{W} ^{\mu \nu} &=&{1\over 4\pi} \sum _X (2\pi )^4 \delta ^4
	(P+P_X-q) \langle 0 \vert J_\mu \vert PX \rangle_{\rm out\,\,out}
	 \langle PX \vert J_\nu \vert 0 \rangle \nonumber\\
	&=&{1\over 4\pi} \int d^4 \xi e^{iq\cdot \xi} 
	\sum _X \langle 0 \vert J_\mu (\xi ) \vert PX \rangle_{\rm out\,\,out}
	\langle PX \vert J_\nu (0) \vert 0 \rangle \nonumber\\
	&=&-\left( g_{\mu \nu} -q_\mu q_\nu \over q^2 \right) 
	\hat{F}_1(z,q^2)+ {1\over \nu} \left( P_\mu -{\nu 
	\over q^2} q_\mu \right)
	\left( P_\nu -{\nu \over q^2} q_\nu \right) \hat{F}_2(z,q^2).\nonumber\\
\eea
In contrast to DIS, the sum over unobserved hadrons $X$ cannot be completed
because the state $\vert PX \rangle_{\rm out}$ depends non-trivially on the
observed  hadron.  Even if $P$ and $X$ did not interact, Bose or Fermi
statistics prevents the  states $X$ from being complete.  In practice
$P$ and $X$ interact dynamically, as indicated by the subscript ``out''. 
For simplicity we will generally suppress this subscript.   Thus,
$e^+e^-\rightarrow P + X$ is not controlled by the product of two operators
(electroweak currents), a feature which complicates the study of
$e^+e^-\rightarrow P + X$ significantly.

If $q^2 \rightarrow \infty$ at fixed $z$, the structure functions, 
$\hat F_1$ and $\hat F_2$ scale (up to logarithmic corrections) and obey a
``Callan-Gross'' relation, $\hat{F}_1 + (z/2) \hat{F}_2 \sim 1/\ln q^2$. In
this limit the cross section is:
\be
	{d\sigma \over dzd\Omega}={\alpha ^2 \over Q^2} z
	\left\{ \hat{F}_1(z,\ln q^2) + {1\over 4} z \sin ^2\theta
	\hat{F}_2(z,\ln q^2) \right\}.
\ee
In leading logarithmic order, using ``Callan-Gross'', the inclusive spectrum
reduces to
\be
	{1\over \sigma} {d\sigma \over dz} \sim {2\over R}z 
	\hat{F}_1(z,q^2) + \cdots,
\ee
where $R$ is defined by eq.~(\ref{eq:pruttel}).
\subsubsection{Lepton Pair Production}
\begin{figure}
\centerline{\epsffile{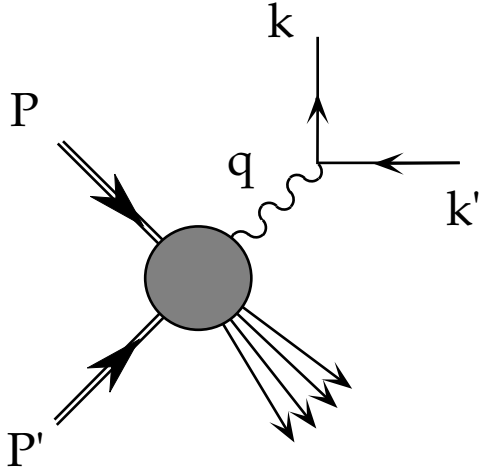}}
	\caption{{\sf Kinematics of the Drell-Yan process.}}
	\label{fig:DYkin}
\end{figure}

\renewcommand{\ln}{\,\mbox{ln}\,}
\renewcommand\Im{{\rm Im}}
\renewcommand\Re{{\rm Re}}
\renewcommand{\theequation}{\thesection.\arabic{equation}}
\renewcommand{\thefootnote}{\fnsymbol{footnote}}
The final process we will consider in detail is massive lepton-pair 
creation in hadron-hadron collisions -- the so-called ``Drell-Yan''
process.   The opportunities for study of novel aspects of hadron structure
by means of polarized Drell-Yan  experiments have motivated a major spin
physics program at RHIC.\cite{RHICSpin}  The kinematics of the lepton pair
production are illustrated in fig.~(\ref{fig:DYkin}). Two hadrons with
momenta $P$ and $P'$ collide at a 
center of mass energy $s=(P+P')^2=4E_{CM}^2$.  Two leptons with momenta
$k_1^{\mu}$ and $k_2^{\mu}$ respectively are produced. They result from the 
decay of a timelike photon, $W^\pm$, or $Z^0$ carrying a momentum $q$, with 
$q^2=Q^2>0$.  Two dimensionless scaling variables are defined by $x={2P\cdot
q\over q^2}$ and $y={2P'\cdot q\over q^2}$. It is easy to see that
$xy=Q^2/s$.  The differential cross section is 
\be
 	d\sigma \propto L_{\mu\nu} W^{\mu\nu} \frac{d^3k_1}{(2\pi)^32k_1^0}
 	\frac{d^3k_2}{(2\pi)^32k_2^0}.
\ee
The decay of the virtual gauge boson is described by the leptonic tensor
$L_{\mu\nu}$,  whereas all information about the hadronic process are
contained  in $W^{\mu\nu}$:
\bea
 	W^{\mu\nu} &=& \frac 12 s \sum\limits_X (2\pi)^4 \delta^4(P+P'-q-X) 
	      \phantom{\rangle}_{\rm in}\left<PP'|J^{\mu}(0)|X\right>
 	\left<X|J^{\nu}(0)|PP'\right>_{\rm in}\nonumber\\
	   &=& \frac 12 s \int d^4\xi e^{-iq\cdot\xi}          
	       \phantom{\rangle}_{\rm in}\left<
	PP'|J^{\mu}(\xi)J^{\nu}(0)|PP'\right>_{\rm in},\nonumber\\
\eea
where the in-state label on $\vert PP'\rangle$ will usually be suppressed. 
$W^{\mu\nu}$ contains many Lorentz invariant structure functions $W_k$.
Depending on the experimental circumstances different combinations of the
$W_k$ and differential cross sections are of interest.  As an example we
consider the inclusive cross section where the lepton momenta have been
integrated out, leaving $d\sigma/dq^4$, 
\be
 	\frac{d\sigma}{dq^4} = \frac {1}{6\pi^3}\frac{\alpha^2}{Q^2s^2} 
 	\left(-W_{\mu}^{\mu}\right). 
\ee
The scaling limit ($s, Q^2\rightarrow \infty$ but $\tau\equiv Q^2/s$ fixed)
once again yields a function of the dimensionless variables ($x$ and $y$)
modulo logarithms induced by QCD radiative corrections, and in this case
$W^\mu_\mu$ is of interest.
\be
	W_{\mu}^{\mu}\rightarrow W(x,y,\ln{Q^2}).
\ee

\section{Deep Inelastic Processes from a Coordinate Space Viewpoint}
\setcounter{equation}{0}

Traditional introductions to the parton model stay fixed in momentum space,
where they use the device of the ``infinite momentum frame'' to simplify
dynamical arguments.  More sophistication is necessary to handle the
complexities introduced by spin dependence and the subdominant effects
associated with transverse spin in DIS.  It is particularly useful to employ
coordinate space methods, mixing parton phenomenology with somewhat more
formal methods of the operator product expansion.\cite{Jaf85}  Certainly,
sophisticated momentum space methods\cite{EFP82} can achieve the same
results.  However, it is particularly easy to distinguish and catalogue
dominant and sub-dominant contributions using the operator product expansion
in coordinate space.  

In this section we will explore the coordinate space structure of the hard
processes introduced in \S 1.  Much of this material is to be found in
modern field theory texts,\cite{Itzub} however there is an  advantage to
providing a brief, self-contained introduction which stresses only those
elementary aspects of the formalism that are useful in characterizing
deep inelastic spin physics.

\subsection{$e^+ e^- \rightarrow$ hadrons --
 The Short-distance Expansion} 

Inclusive $e^+e^-$ annihilation into   
hadrons is the simplest process to analyze and illustrates the importance of
Wilson's short distance expansion. As shown in \S 1, this process is 
described  by the vacuum expectation value of a current commutator, 
\be 
 	\Pi (Q^2) \propto {1\over Q^2}\int d^4 \xi \, e^{i q \cdot \xi} 
 	\left< 0 | \left [ J_{\mu}(\xi), J^{\mu}(0) \right] | 0 \right>. 
	 \label{eq:burp}
\ee
In the center of mass system we have 
$q = (\sqrt{Q^2},\vec 0\,)$.  Since the commutator is causal,
\be
	\left[ J_\mu(\xi ), J_\nu (0) \right]=0\quad {\rm for}\quad \xi ^2 <0,
\ee
then $\vert \vec{\xi} \vert < \xi^0$ in the integral.
Using the symmetry of the commutator one obtains:
\be
	\Pi (Q^2) \propto\int _0^\infty d\xi^0 \sin{Q\xi^0}
	\int _{\vert \vec{\xi} \vert \leq \xi ^0} d^3\xi \langle 0\vert 
 	\left[ J_\mu (\xi),J^\mu (0)\right] \vert 0 \rangle
	\label{eq:knots}
\ee
In the high energy limit, $Q\rightarrow \infty$, $\sin{Q\xi^0}$ oscillates
rapidly, averaging out contributions except at the $\xi^0=0$ boundary of the
integration region.  This argument can be made more formal,  leading to the
conclusion that $\xi^0\sim{1\over q^0}$ gives the dominant contribution to
the integral.  Since $\xi^0>\vert\vec\xi\vert$
we can conclude that $e^+ e^-$ annihilation into hadrons at high $Q^2$  is
dominated by interactions at short  distances, $\xi^{\mu}\rightarrow 0$. 
This is, of course, a Lorentz invariant condition.

The leading contribution to the annihilation process can now be found 
via the operator product expansion (OPE).\cite{Itzub}  First postulated by
Wilson, the existence of the OPE has been demonstrated to all orders in
perturbation theory in renormalizable theories
and also in various toy models which can be solved
exactly.  According to the OPE, a product of local operators  
$\hat A(\xi)$ and $\hat B(0)$ at short distances (here $\xi_{\mu} 
\rightarrow 0$) can be expanded in a series of {\it non-singular\/} 
local operators multiplying c-number {\it singular\/} functions,
\be
 	\hat A(\xi) \hat B(0) \sim
 	\sum\limits_{[\alpha]} C_{[\alpha]}(\xi) \hat{\theta}_{[\alpha]}(0)\quad
 	{\rm as}\quad\xi^\mu\rightarrow 0.
 	\label{eq:ope}
\ee
In general the product $\hat A\hat B$ is singular as
$\xi\rightarrow 0$.  The substance of the expansion is that the
singularities can be isolated in the c-number ``Wilson
coefficients'', $C_{[\alpha]}$.  The operators in eq.~(\ref{eq:ope}) are cutoff
independent renormalized operators and the Wilson coefficients are likewise
cutoff independent.

The behavior of the Wilson coefficients at  $\xi_{\mu}\rightarrow 0$ follows
from dimensional analysis. In natural units, all quantities are measured in
dimensions of mass to the appropriate power.  For simplicity, if a quantity,
$\theta$ has units $m^{d_\theta}$, we write $d_\theta=[\theta]$.  This is a
simple concept, not to be confused with more subtle ones like anomalous
dimensions or scale dimensions.\cite{Cole}  The dimension of all operators
of  interest to us can be deduced from the fact that charge and action are
dimensionless.  Thus $[J^\mu] = 3$ because $\int d^3x J^0(x)=Q$.  For the
quark field $[\psi]= {3\over 2}$ because the free Dirac action is $\int d^4x
\bar\psi i\gamma\cdot\partial\psi+\ldots$, likewise for the gluon field
strength $[G_{\mu\nu}]=2$.  Since we normalize our states covariantly,
$\langle P\vert P^\prime\rangle =2E (2\pi )^3 \delta ^3
(\vec{P}-\vec{P}^\prime )$, $[\langle P\vert P'\rangle] = -2$.  For the
vectors $[P^\mu]=[S^\mu]=1$.  We see that
$W_{\mu\nu}$ is dimensionless, as reflected in the form of
eqs.~(\ref{eq:WS}) and (\ref{eq:WA}).

Dimensional consistency applied to the OPE requires,
\be
 	[\hat A] + [\hat B] = [C_{[\alpha]}] + [\hat\theta_{[\alpha]}].
\ee
What can account for the dimensions of the singular function $C_{[\alpha]}$?
If the operators $\hat A$ and $\hat B$ are finite in the
$m_{\rm quark}\rightarrow 0$ limit, then powers of
$m_{\rm quark}$ can only appear in the numerator of $C$.  The renormalization
scale, $\mu$, necessary to render the theory finite can only appear in
logarithms (of the form $\ln{(\mu\xi)}$) order by order in perturbation
theory.  This leaves the coordinate $\xi$ itself to absorb the dimensions.
\be 
 	C_{[\alpha]}(\xi)\sim 
 	\frac{1}{\xi^{[\hat A] + [\hat B] - [\hat{\theta}_{[\alpha]}]}}
 	\left(\ln^{\gamma_\theta}(\mu\xi)+\ldots\right)\quad 
 	{\rm as}\quad\xi\rightarrow 0.
 	\label{eq:hgnh}
\ee
The exponent $\gamma_\theta$ is the ``anomalous dimension'' of the operator
$\theta$ generated by radiative corrections.  Without minimizing the
importance of these logarithms, we will usually ignore them and focus on the
gross, power law, behavior required by dimensional analysis.  
For given operators $\hat A$ and $\hat B$ the leading contribution 
at short distances comes from that term in the OPE 
having the lowest operator dimension $[\hat{\theta}_{[\alpha]}]$. 

This can now easily be applied to $e^+ e^-$ annihilation.  The
dimension of the hadronic electromagnetic current  is $[J_{\mu}]=3$.  No fields
have negative dimensions, so the lowest dimension operator is the unit
operator, $\hat\theta_0\equiv {\bf 1}$, with $[{\bf 1}]=0$.  The
$C_0(\xi)\sim 1/\xi^6$ and the dominant contribution in the OPE is,  
\be
 	\left< 0 | \left [ J_{\mu}(\xi), J^{\mu}(0) \right] | 0 \right> 
 	\sim \frac{1}{\xi^6},\quad \mbox{modulo logarithms}. 
\ee
Consequently the current correlation function scales like  
\be 
 	\Pi (Q^2) \sim {1\over Q^2}\int d^4 \xi \, e^{i q \cdot \xi}
 	\frac{1}{\xi^6} \sim 1, 
\ee
again modulo logarithms, and the cross section eq.~(\ref{eq:schaap}) scales
like:
\be
 	\sigma\left(e^+ e^- \rightarrow {\rm hadrons}\right)\sim \frac{1}{Q^2}.
\ee
The logarithms can be gathered together into powers of ${\alpha_s\over\pi}$
as anticipated in eq.~(\ref{eq:rpert}).  Of course, having made no attempt to
derive the OPE or to study the effects of radiative corrections and
renormalization in detail, the example of the total  $e^+e^-$ annihilation
cross section becomes rather trivial.  Nevertheless it provides a useful
introduction to the more complicated cases which follow.

\subsection{$l p\rightarrow l X$ -- The Light-Cone Expansion}
\label{sub:kanarie}
Next we turn to deep inelastic scattering, which is characterized by two
large invariants -- $Q^2$ and $\nu$.  As we shall see, such processes are
dominated by physics close to the light-cone.
\subsubsection{Light-Cone Coordinates and Formulation of Deep Inelastic
Scattering}
The four-momenta $P^\mu$ and $q^\mu$ can be used to define a frame and a
spatial direction.  Without loss of generality we can choose our frame such
that $P^\mu$ and $q^\mu$ have components only in the time and $\hat e_3$
directions.  It is helpful to introduce the light-like vectors 
\bea
	p^{\mu} &=& \frac{p}{\sqrt{2}} \,(1,0,0,1), \nonumber\\
	n^{\mu} &=& \frac {1}{\sqrt{2}p} \,(1,0,0,-1)\nonumber\\
 \label{eq:peer}
\eea
with $n^2=p^2=0$ and $n\cdot p = 1$.  Up to the scale factor $p$, the vectors
$p^\mu$ and $n^\mu$ function as unit vectors along opposite tangents to the
light-cone.  They may be used to expand $P^\mu$ and $q^\mu$,
\bea
 	q^{\mu} &=& \frac{1}{M^2}\left(\nu-\sqrt{\nu^2+M^2Q^2}\right)p^{\mu} + 
	    \frac 12\left(\nu+\sqrt{\nu^2+M^2Q^2}\right)n^{\mu},\\ 
 	P^{\mu} &=& p^{\mu} + \frac{M^2}{2}n^{\mu},
\eea
In the Bjorken limit $q^{\mu}$  
simplifies to 
\be
 	\lim_{Bj} q^{\mu}\sim \left(\nu + \frac 12 M^2x\right)n^{\mu} - xp^{\mu}
 	+{\cal O}\left({1\over Q^2}\right).
\ee 

$p$ selects a specific frame. For example 
$p=M/\sqrt 2$ yields the target rest frame, while $p\rightarrow \infty$ 
selects the infinite momentum frame.  The decomposition along $p^{\mu}$ and
$n^{\mu}$ is equivalent  to the use of light-cone coordinates, which are
defined as follows.  An arbitrary four-vector $a^\mu=(a^0,a^1,a^2,a^3)$ can be
rewritten in terms of the four components $a^\pm={1\over\sqrt{2}}(a^0\pm
a^3)$, and $\vec a^\perp=(a^1,a^2)$.  In this basis, the metric $g_{\mu\nu}$
has non-zero components, $g_{+-}=g_{-+}=1$ and $g_{ij}=-\delta_{ij}$, so
$a\cdot b = a^+b^-+a^-b^+-\vec a^\perp\cdot\vec b^\perp$. 
The transformation to light-cone components can be recast as an expansion in
the basis vectors $p^\mu$ and $n^\mu$,
\be
	 a^{\mu} = \left(\frac{\sqrt 2 a^-}{p}\right) p^{\mu} + 
	  \left(\sqrt 2 a^+ p \right) n^{\mu} + a^{\perp \mu}.
\ee

With these preliminaries it is easy to find the space-time region which
dominates the DIS. Consider the hadronic tensor $W^{\mu \nu}$ defined in
eq.~(\ref{eq:com}): 
\be
 	W^{\mu\nu} = \frac 1{4\pi}\int d^4 \,\xi e^{iq\cdot \xi}
 	\left< P | \left [ J^{\mu}(\xi), J^{\nu}(0) \right] | P \right>, 
 	\label{eq:pier}
\ee
Take the Bjorken limit by keeping $P$ fixed and $q\rightarrow
\infty$. Define
\be 
 	\xi^{\mu} \equiv \eta p^{\mu} + \lambda n^{\mu} + \xi^{\perp \mu},  
\ee
we find in the Bjorken limit:
\be
 	\lim_{Bj} q\cdot \xi = \eta \nu - x\lambda.
\ee
Arguments similar to those used in the previous section show that the
integral in eq.~(\ref{eq:pier}) is dominated by  
$\eta\sim 1/\nu\sim 0$ and $\lambda\sim 1/x$, which is equivalent to 
$\xi^-\sim 0$  and $\xi^+ \sim 1/xp$ respectively.\cite{Ioffe,Jaf72}  As in
the previous case the commutator in  eq.~(\ref{eq:pier}) vanishes unless
$\xi^2=2\lambda \eta - \vec
\xi_{\perp}^2\ge 0$ because of causality.   Combining these results we find
that the Bjorken limit of DIS probes a current correlation function near the
light-cone $\xi^2=0$, extending out to distances ($\xi^3$ and $\xi^0$) of
order ${1\over{xp}}$.

\subsubsection{Deep Inelastic Scattering and the Short Distance Expansion}

QCD simplifies at short distances on account of asymptotic freedom.
The analysis of $e^+e^- \rightarrow$ hadrons simplifies greatly 
for this reason. Deep inelastic scattering is \underline{not} a short 
distance process; it is light-cone dominated. Nevertheless it can be 
related to the OPE and to short distances with considerable 
resulting simplification.

\begin{figure}
\centerline{\epsffile{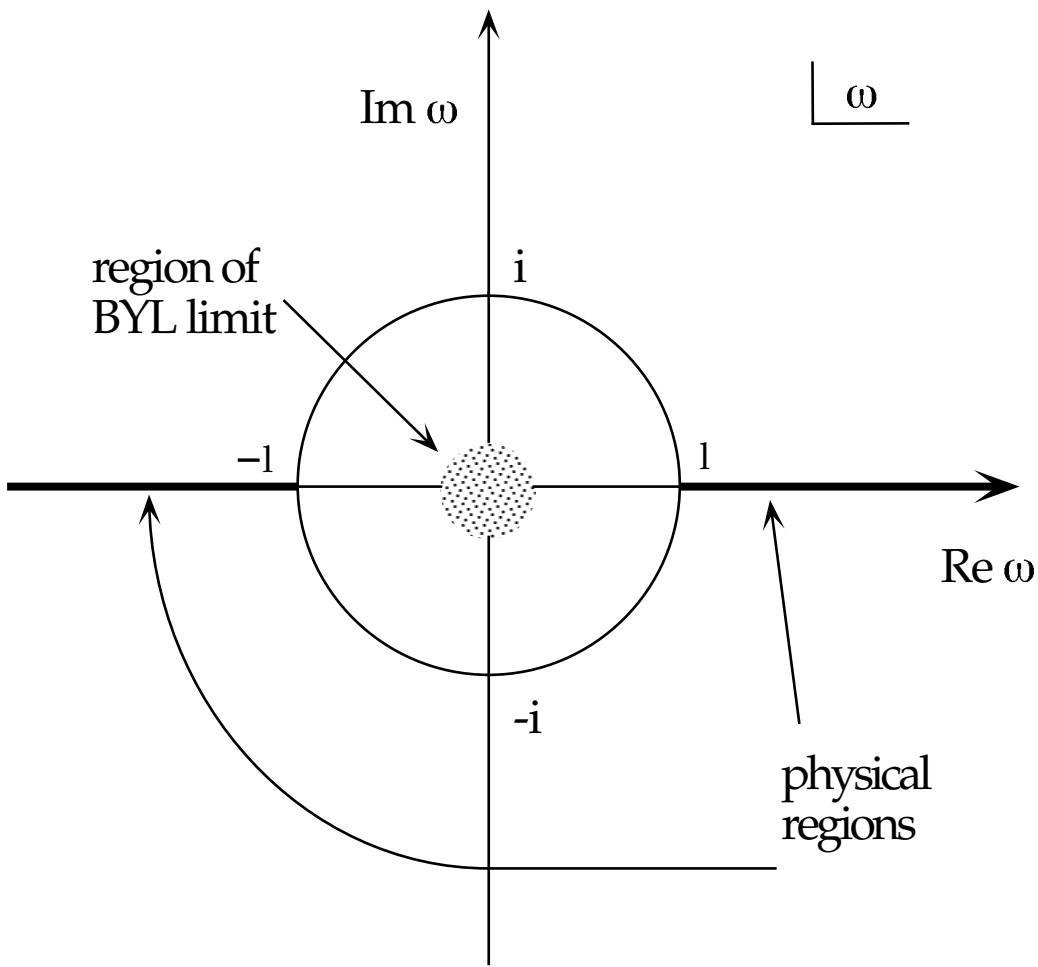}}
	\caption{{\sf Physical region of the forward 
	Compton amplitude and the BJL limit 
	in the complex $\omega$ plane.}}
	\label{fig:appel}
\end{figure}

To show this we consider the so-called the Bjorken-Johnson-Low limit 
($\lim_{BJL}$).\cite{Bjo66,Joh66}  This is a somewhat old fashioned method,
mostly supplanted by Wilson's operator product expansion.  It has the virtue
that the connection between measurable structure functions and local
operators is extremely clear (via dispersion relations).  Use of the BJL
limit prevents one making mistakes in subtle cases.\cite{EllJaf73,BGJ} In
the BJL limit one takes 
$\vec q=0$ and $q^0\rightarrow i\infty$,  which yields $q^2\rightarrow
-\infty$ and $x\rightarrow -i\infty$.  In the physical region $x$ is
restricted to be real and between $0$ and
$1$.  So the hadronic tensor $W_{\mu \nu}$ cannot be measured in the BJL
limit. It is useful because 1) it {\it is\/} dominated by short distances,
and 2) it can be related to $W_{\mu\nu}$ in the physical region through
dispersion relations. Remember that $W_{\mu\nu}$ is the imaginary part of
the forward, virtual Compton amplitude, $T_{\mu\nu}$, by the optical
theorem,
\be
 	T_{\mu\nu}(q^2,\nu) = i \int d^4\xi \,e^{iq\cdot \xi}
	 \left< P | T \left ( J_\mu(\xi) J_\nu(0) \right ) | P \right>,
 	\label{comptamp}
\ee
For simplicity we suppress Lorentz indices and spin degrees of freedom for a
while.  Standard dispersion theory arguments show that
$T(q^2,\nu)$ is an analytic function of $\nu$ at fixed spacelike $q^2$ with
branch points on the real--$\nu$ axis at $\nu=\pm{Q^2\over{2M}}$, the
threshold for the elastic process $\gamma^* p \rightarrow \gamma^* p$.  In
fig.~(\ref{fig:appel}) one can see the physical region of this process and
the area of the BJL limit in the complex $\omega = 1/x$ plane.  The physical
cuts lie on the real axis from
${1\over x}=\pm 1$ to $\pm\infty$. This means that $T(q^2,{1\over x})$ is
analytic within the unit circle about the origin.  The BJL limit takes
$1\over x$ to zero along the imaginary axis.  Thus $T(q^2,\nu)$ can be
expanded in a Taylor series in $({1\over x})$ about the origin in the
BJL-limit.
The coefficients in the Taylor expansion can be obtained from the dispersion
relation obeyed by $T$.  First remember that the optical theorem relates the
imaginary part of $T$ to the hadronic tensor $W(q^2,\nu)$ in the physical
region,
\be
	{\rm Im} T(q^2,\nu) = 4\pi W(q^2,\nu).
\ee  
Dispersion theory tells us that an analytic function can be represented in
terms of its singularities in the complex plane,\cite{BjD} in this case the
physical cut on the real axis,
\be
	T(q^2,\omega) = 4\int_1^{\infty} d\omega' \,\omega'\, 
	\frac{W(q^2,\omega')}{\omega'^2-\omega^2}.
	\label{eq:dispint}
\ee
Crossing, {\it i.e.\/} $T(\omega) = T(-\omega)$, has been used.   
Since $T(q^2, x)$ is analytic for $|{1\over x}|<1$ it may be expanded in 
a Taylor series in powers of ${1\over x}$: 
\be
	\lim_{\rm BJL} T(q^2,x) = 4 \sum\limits_{\rm n\,even} 
	M^n(q^2) \,{1\over x^n},
	\label{eq:verklchen}
\ee
with
\be
	M^n(q^2) = \int_0^1 dx x^{n-1} W(q^2,x).
\ee

Now consider where the BJL limit leads us in coordinate space.  With $\vec q
= 0$ and $q^0\rightarrow i\infty$, the factor $e^{iq\xi}$ in eq.~(\ref{comptamp})
reduces to $e^{-|q^0|\xi^0}$ and forces $\xi^0$ to zero.  Although the time
ordered product does not vanish outside the light-cone, it can be exchanged
for a ``retarded commutator'',\cite{Joh66,EllJaf73} which does.  Thus
$\xi^0\rightarrow 0$ forces $\xi^\mu\rightarrow 0$ and we conclude that the
BJL-limit takes us to short distances where Wilson's operator product
expansion may be used.  The OPE analysis of the product of currents yields
a power series in ${1\over x}$ multiplying the matrix elements of local
operators.  Identifying terms in this Taylor series with the terms in
eq.~\ref{eq:verklchen} we obtain the celebrated ``moment sum
rules'' relating integrals over deep inelastic structure functions to target
matrix elements of local operators.  We will not pursue this direction
further here --- it is treated in standard references.\cite{ChengLi,Itzub}
\subsection{$e^+ e^-\rightarrow h\,X$-- Once Again, the Light-Cone}

Like deep inelastic scattering, single particle inclusive production in
$e^+e^-$ annihilation is dominated by the light-cone.  However, the operator
product expansion does not apply and no short distance analysis exists. The
process is described by the tensor introduced in \S 1,
\be
	\bar W_{\mu\nu} = \frac 1{4\pi} \int d^4\xi \,e^{iq\cdot \xi}
	\sum_X\left< 0 | J_\mu(\xi) |PX\right>\left< PX | J(0)_\nu | 0 \right>
	\label{eq:plons}
\ee
Once again, the nucleon and photon momenta may be expanded
in terms of the light-like vectors 
introduced in \S \ref{sub:kanarie},
\bea
 	P^{\mu} &=& p^{\mu} + \frac{M^2}{2}n^{\mu},\nonumber\\
	q^{\mu} &=& \frac{1}{M^2}\left(\nu-\sqrt{\nu^2-M^2Q^2}\right)p^{\mu} + 
	    \frac 12\left(\nu+\sqrt{\nu^2-M^2Q^2}\right)n^{\mu},\nonumber\\
\eea
and in the Bjorken limit ($Q^2, \nu \rightarrow \infty$ with $z$ finite),
\be
	\lim_{Bj} q^{\mu} = \left(\nu - \frac 12 {M^2\over z}\right)n^{\mu} + 
	{1\over z}p^{\mu}.
\ee  
It is traditional to use the photon rest frame ($\vec{q}=0,
p\sim \sqrt{\nu}$) to analyze the process.  However, the label on the state,
($P^\mu$) changes as the limit $Q^2 \rightarrow \infty$ is taken in this
frame, making it difficult to sort out the important regions of the
$\xi$-integration. Things are simpler in a frame where
$P$ is fixed, {\it e.g.\/} the rest frame of the produced hadron, where
$p={M\over\sqrt{2}}$.  In such a frame, the analysis of the fourier integral
in eq.~(\ref{eq:plons}) proceeds exactly in the same way as for the
electroproduction process  of \S \ref{sub:kanarie}. With
\be 
	\xi^{\mu} = \eta p^{\mu} + \lambda n^{\mu} + \xi^{\mu \perp}  
\ee
we find in the Bjorken limit 
\be
	\lim_{Bj} q\cdot \xi = \eta \nu - {\lambda \over z}.
\ee
So, $\nu \rightarrow \infty$ implies $\eta \rightarrow 0$ and 
$\lambda \sim z$, since $z$ is finite. So light-like separations $\xi^\mu
\xi_\mu \sim 0$ dominate again unless unusual variations occur in the matrix
elements
\be
	 \sum\limits_X\left< 0 | J(\xi) |PX\right>\left< PX | J(0) | 0 \right>,
\ee
which will not happen in the frame where $P$ is independent of
$Q^2$ and $\nu$.  Also the frequencies associated with
the states in the sum $\sum\limits_X |X\left>\right<X|$ know nothing about
$Q^2$ and $\nu$, and will not spoil the argument.  For a contrasting
situation see the discussion of Drell-Yan in the following sub-section.

One can thus conclude that light-cone distances dominate fragmentation.
However, in contrast to DIS the OPE cannot be applied 
here since the observed hadron state, $\vert P\rangle$ interferes with the
attempt to complete the sum on $X$.  Nevertheless nearly all of the QCD
phenomenology developed for DIS can be carried over to this case, primarily
using momentum space methods we will not discuss here.\cite{QCDFrag}  
In \S 6 we will see that the limitations on the coordinate space analysis do
not prevent us from analyzing spin, twist and chirality in fragmentation.
\subsection{$ PP \rightarrow l^+l^- X$ -- The Drell-Yan Process}
Finally we consider the Drell-Yan process. Here the relevant 
hadronic tensor is (see \S 1):
\be
	W^{\mu\nu} = \frac {1}{2} s \int d^4\xi \,e^{iq\cdot \xi}
	\left< P P' | J^{\mu}(\xi) J^{\nu}(0) | P P' \right>. 
	\label{eq:braak}
\ee
It is simplest to consider the case where only the 
dilepton invariant mass distribution 
$d\sigma/dq^2$ is measured, though other observables behave similarly. 
${d\sigma \over{dq^2}}$ depends only on $W^\mu_\mu$.  We define a function,
$W(s,Q^2)$ by integrating $W^\mu_\mu$ over all $q^\mu$ with $q^2=Q^2$ and
$q^0>0$,
\bea
	W(s,Q^2) &=& \frac {1}{(2\pi)^4} \int_R d^4q\, 
	\delta(q^2-Q^2) (-g^{\mu\nu}
		W_{\mu\nu})\theta(q^0),\\
	    &=& -4\pi^2 s \int_R d^4q \,\delta(q^2-Q^2)  
		\sum\limits_X(2\pi^4) \delta^4(P+P'-q-X)\nonumber\\
	    &\times&  
		\left< P P'| J^{\mu}|X\right>\left< X | J_{\mu}| P P' \right>.
\eea
The virtual photon's momentum is integrated over all values consistent with
the constraint $q^2=Q^2$ and conservation of energy,
$\sqrt{Q^2} < q_0< (s+Q^2)/ 2\sqrt s$, which defines the region $R$. If we
introduce the function 
\be
 	\Delta_+^R(\xi,Q^2) = \int_R\frac{d^4q}{(2\pi)^3} \,e^{-iq\cdot\xi} 
 	\delta(q^2-Q^2)\theta(q_0),
\ee 
then $W(s,Q^2)$ can be written as 
\be 
 	W(s,Q^2) = -s\int d^4\xi \,\Delta_+^R(\xi,Q^2)  
		\left< P P'| J^{\mu}(\xi) J_{\mu}(0) | P P' \right>.
\ee 
In the scaling limit ($Q^2, s\rightarrow \infty, \tau = Q^2/s$ fixed)   
$\Delta_+^R(\xi,Q^2)$ approaches a well-studied function of quantum field
theory, the free field  singular function, $\Delta_+(\xi,Q^2),$\cite{Bog57},
\bea
	\Delta_+(\xi,Q^2) &=& \frac 1{4\pi} \epsilon(\xi_0) \delta(\xi^2) - 
		    \frac{mi}{8\pi\sqrt{\xi^2}}\theta(\xi^2) 
		    \left(N_1\left(Q\sqrt{\xi^2}\right) - i \epsilon (\xi_0) 
		    J_1\left(Q\sqrt{\xi^2}\right)\right) \nonumber\\
		  &+& 
		    \frac{mi}{4\pi^2\sqrt{-\xi^2}}\theta(-\xi^2) 
		    K_1\left(Q\sqrt{\xi^2}\right). 
\eea
$\Delta_+(\xi,Q^2)$ is singular on the light-cone and would select out
light-cone contributions were it not for high frequency variations in the
matrix elements. These can occur because the hadron momenta $P$ and $P'$
cannot be kept fixed as $Q^2\rightarrow \infty$ ($s\approx 2 P\cdot
P'>Q^2$)  in any frame.  Even in free field theory or the parton model the
matrix element behaves like 
\be
 	\left< P P'| J^{\mu}(\xi) J_{\mu}(0) | P P' \right>\sim \int d\alpha
 	\,d\beta\,  e^{i\alpha P\cdot \xi + 
	i \beta P'\cdot \xi} f(\alpha,\beta), 
\ee
where $f(\alpha,\beta)$ labels the momentum components of the partons that
contribute to the current.  To see that such variation can lead to
contributions off the light-cone, consider a frame defined through the two
vectors 
\bea
 	p^{\mu}  &=& p\,(1,0,0,1),\nonumber\\
 	p'^{\mu } &=& \frac {s}{4p} \,(1,0,0,-1).\nonumber\\ 
\eea
The hadron and photon momenta can be written as 
\bea
	P^{\mu}  &=& p^{\mu} + \frac {M^2}{s} p'{}^{\mu},\nonumber\\
	P'{}^{\mu} &=& p'{}^{\mu} + \frac {M^2}{s} p^{\mu},\nonumber\\
	q^{\mu}  &=& yp^{\mu}+xp'{}^{\mu}, \quad \mbox{for} 
	     \;|q_T|\ll\sqrt{Q^2}. \nonumber\\
\eea
With $\xi^{\mu} = \eta p^{\mu}+\lambda p'{}^{\mu} + \xi_\perp^\mu$ 
the hadronic tensor eq.~(\ref{eq:braak}) is then equal to 
\be
	W_{\mu}^{\mu} = \int d\lambda \,d\eta \,d^2\xi_T\,
		\int d\alpha \,d\beta \,f(\alpha,\beta) 
		e^{ i\frac{s}{2} [(\alpha-y)\lambda + (\beta-x)\eta]}.
\ee
Therefore the phases will cancel and the Drell-Yan process will escape 
from the light-cone if $\alpha\approx y=2q\cdot P'/s$ and  
$\beta\approx x=2q\cdot P/s$. In \S 5 we will return to this process
and see that such phases are generated in a natural way. 
\subsection{Dominant and Subdominant Diagrams}
Guided by our understanding of the regions of coordinate space important for
various deep inelastic processes, we can return to the more familiar world of
Feynman graphs and learn which diagrams are likely to give dominant and
subdominant contributions.  The quarks that couple to electroweak currents
propagate according to $S_F(\xi)$, the Feynman propagator.  In coordinate
space, $S_F(\xi)$ behaves like ${1\over \xi^3}$ at short distances (note
$S_F(\xi)\sim\int d^4p { e^{ip\xi}\over{\gamma\cdot p - m}}$),

\vspace{.1in}
\hspace{.2in}{\epsffile{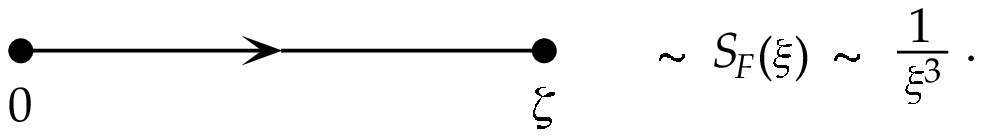}}

\noindent
Interactions will not increase the singularity. For example, coupling a gluon
to the propagating~quark~gives,

\vspace{.2in}
\hspace{.2in}{\epsffile{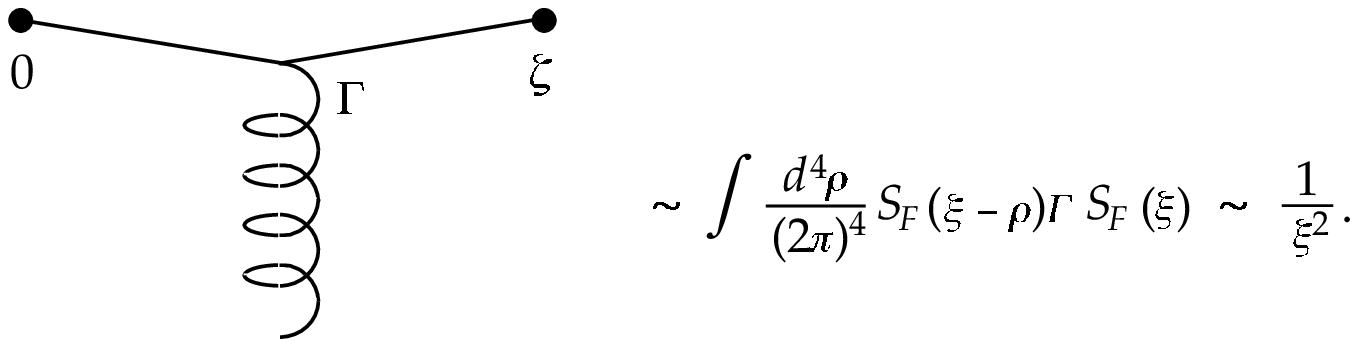}}

\vspace{-2in}
\noindent
Generally speaking in renormalizable field theories, interactions on
propagating lines do not increase the order of the short-distance or
light-cone singularity by more than logarithmic terms beyond free field
theory.  This can be used as a guideline to estimate the importance of
different perturbative diagrams for hard processes. 

As a first example, consider $e^+ e^-\rightarrow$ hadrons. 
The total cross section is proportional to the vacuum polarization of 
the photon propagator, whose leading contribution results from 

\vspace{.1in}
\hspace{1in}{\epsffile{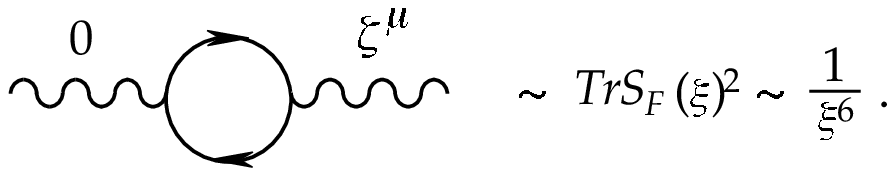}}
\vspace{.2in}

\noindent
the quark-antiquark loop fourier transformed, this ${\cal O}({1\over
\xi^6})$ behavior generates a cross section which scales like ${1\over
Q^2}$.  Radiative corrections introduce logarithmic dependence on
$\xi^2$, $\ln \xi^2 \mu^2$, where $\mu^2$ is the renormalization
point, but they do not change the power of the singularity in a
renormalizable theory.  The renormalization group may be used to sum
classes of diagrams giving modifications of the ${1\over \xi^6}$
behavior which go like powers of logarithms in an asymptotically free
theory like QCD.

\begin{figure}
\centerline{\epsffile{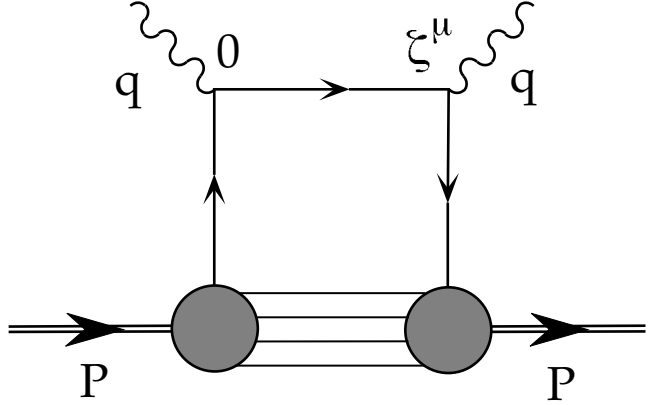}}
	\caption{{\sf Leading diagram in deep inelastic scattering.  The quark
propagator between the two currents carries the large momentum $q^\mu$ and
leads to a $1/\xi^3$ behavior at small distances.}}
	\label{fig:bloem}
\end{figure}

\begin{figure}
\centerline{\epsffile{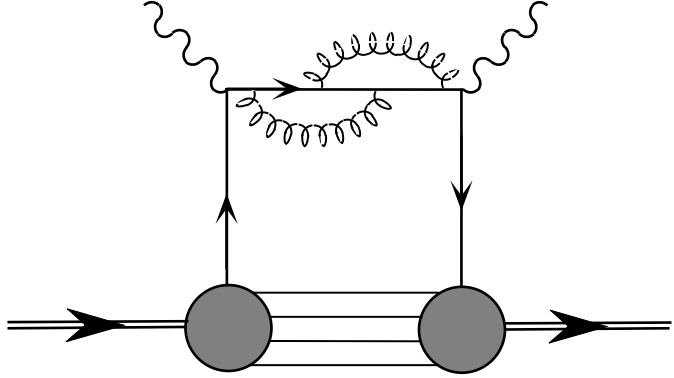}}
	\caption{{\sf Radiative corrections to the quark propagator
         lead to $O\left(\alpha_s(\xi^2) \right)$ corrections
	 to the coefficient of the leading $1/\xi^3$ term.}}
	\label{fig:boom}
\end{figure}

\begin{figure}
\centerline{\epsffile{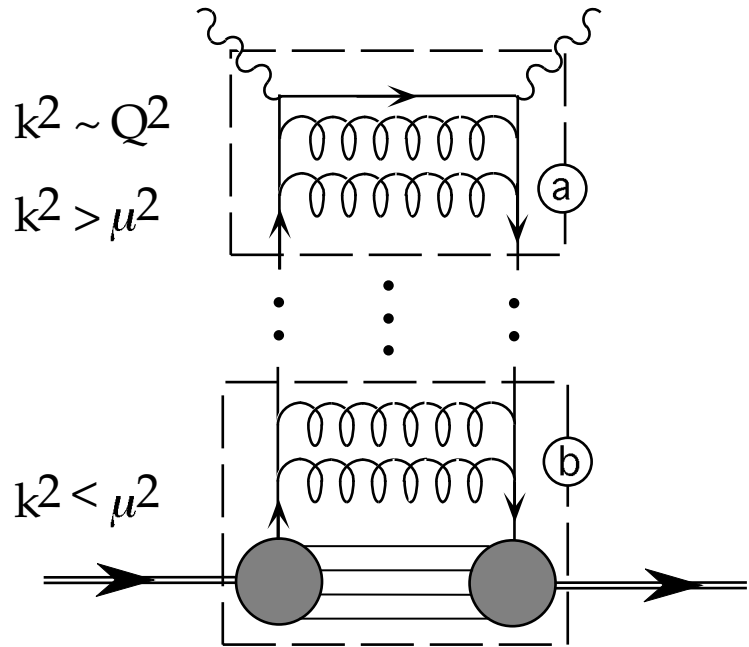}}
	\caption{{\sf 	The higher order contributions in the upper part 
	of the diagram $(a)$, where
	the quark virtuality is greater than $\mu^2$
	lead to more $O\left(\alpha_s(\xi^2)\,\ln\mu^2 \xi^2  \right)$ 
	corrections. The lower radiative
	corrections $(b)$ can be absorbed into the 
	quark-hadron amplitude which 
	will then depend on the renormalization scale 
	$\mu^2$. }}
	\label{fig:huis}
\end{figure}

\begin{figure}
\centerline{\epsffile{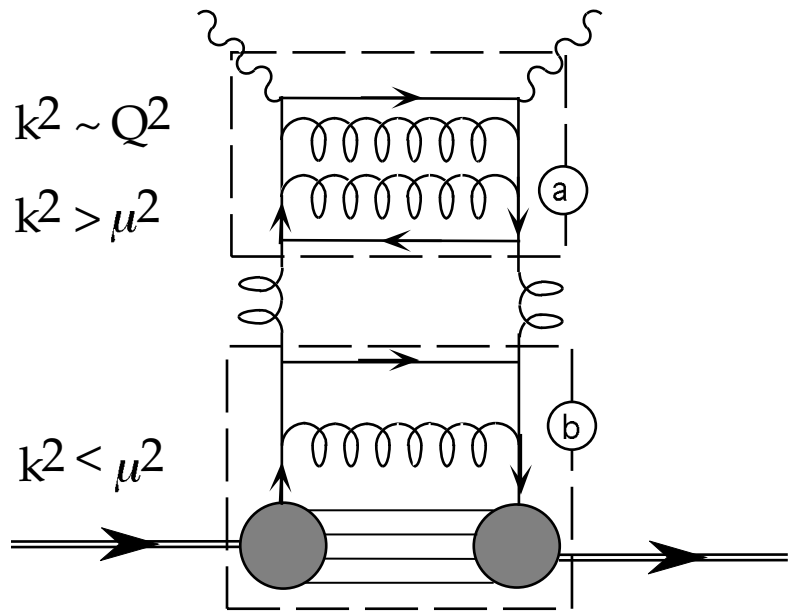}}
	\caption{{\sf 
	The upper part $(a)$ here generates 
	a $c$-number coefficient function for
	gluonic operators in the product of two currents, beginning at 
	$O\left(\alpha_s(\xi^2) \,\ln\mu^2\xi^2  \right)$,
	while the lower part 
	$(b)$ can be absorbed into a new gluon-hadron amplitude. }}
	\label{fig:beestje}
\end{figure}

\begin{figure}
\centerline{\epsffile{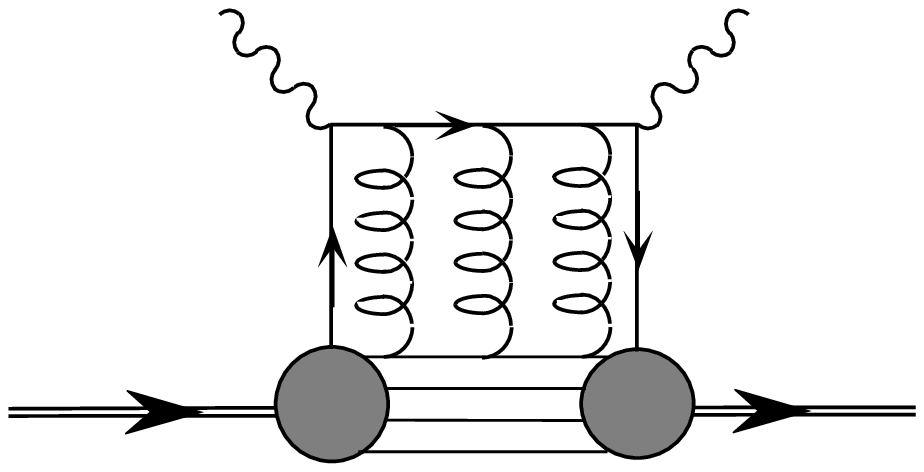}}
	\caption{{\sf 
	These corrections are either gauge artifacts or modifications of 
	lower order in $1/\xi$. }}
	\label{fig:auto}
\end{figure}

\begin{figure}
\centerline{\epsffile{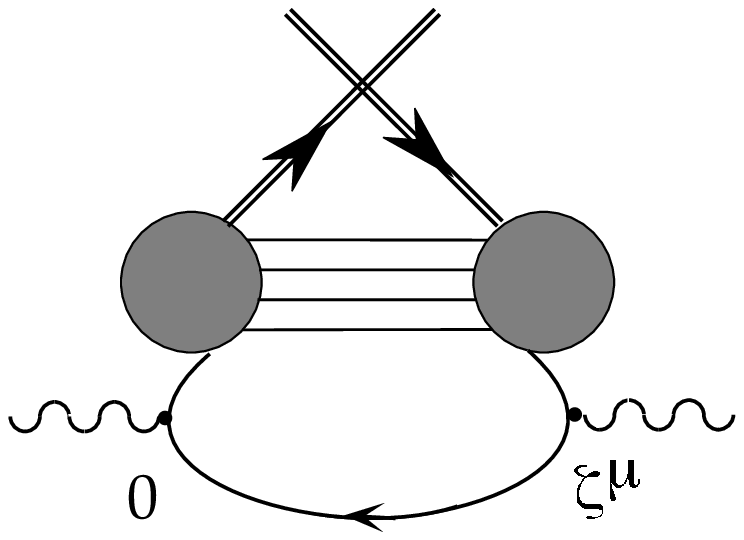}}
	\caption{{\sf 
	The dominant diagram for the single particle inclusive $e^+e^-$ 
	annihilation.}}
	\label{fig:lamp}
\end{figure}

\begin{figure}
\centerline{\epsffile{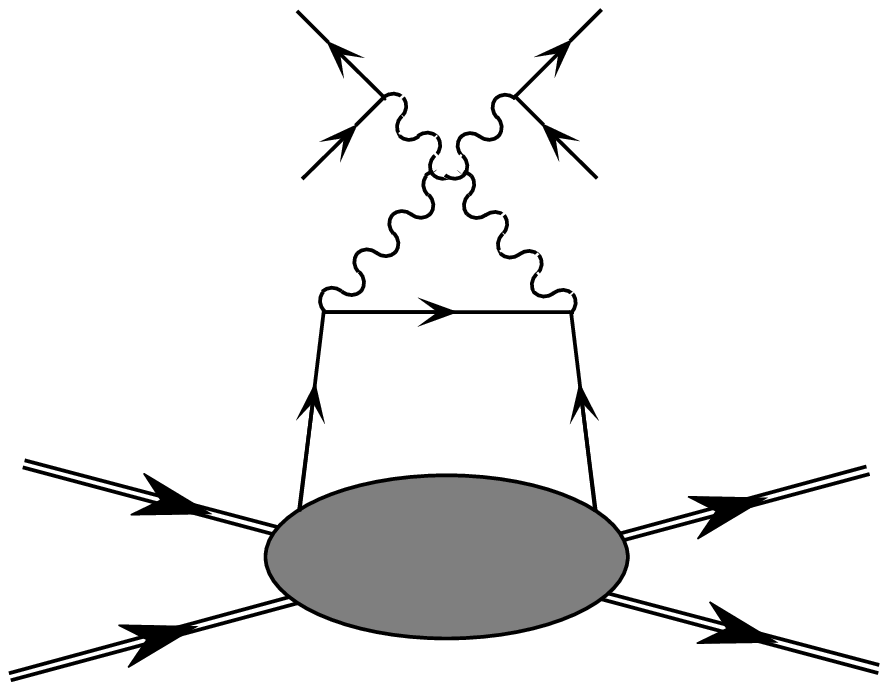}}
	\caption{{\sf 
	The diagram that has the leading light-cone singularity of 
	the Drell-Yan process, but does not dominate.}}
	\label{fig:gorilla}
\end{figure}

\begin{figure}
\centerline{\epsffile{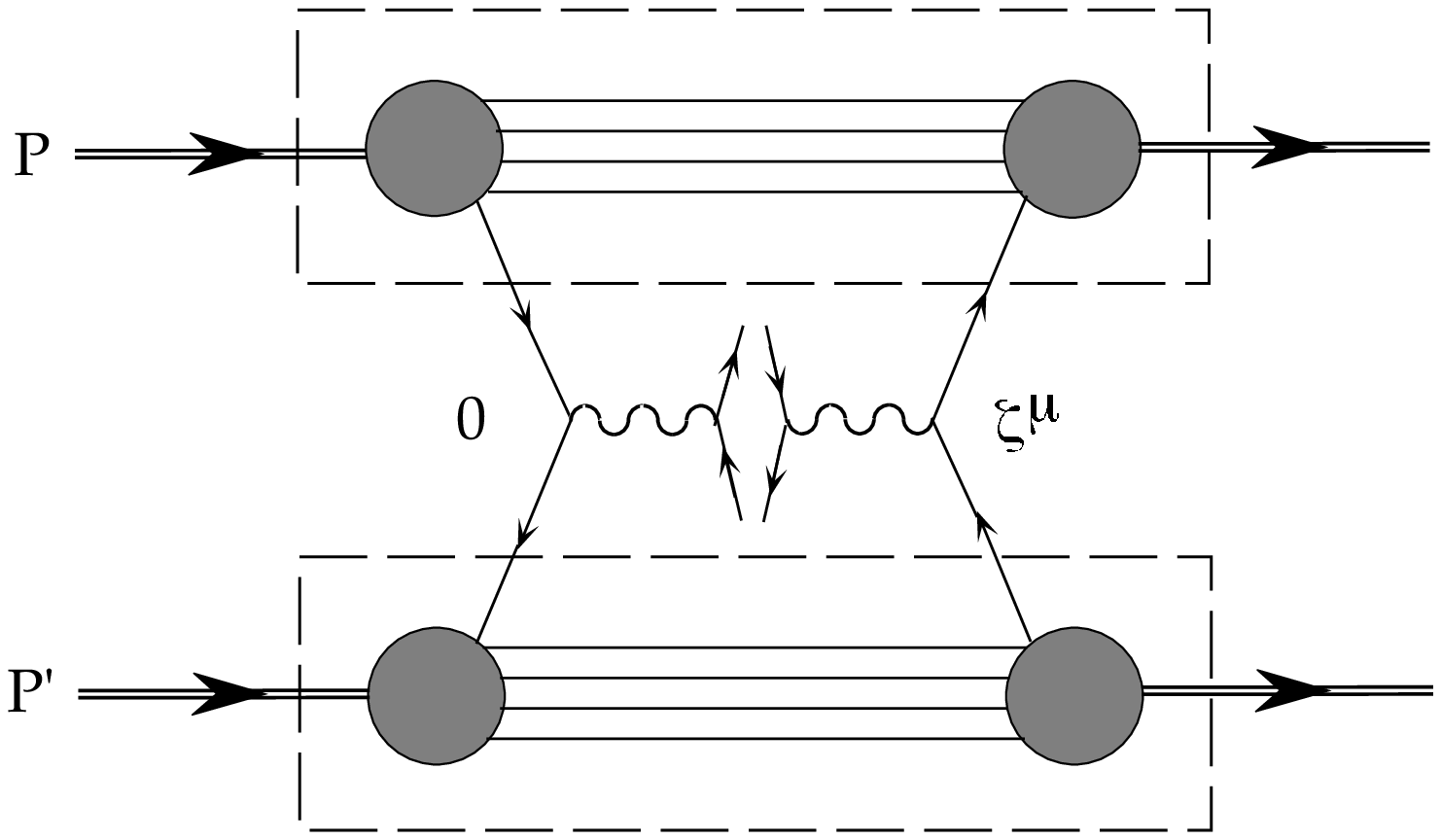}}
	\caption{{\sf 
	Dominant diagram of the Drell-Yan process.}}
	\label{fig:aap}
\end{figure}

In deep inelastic scattering the leading contribution to the 
cross section or the forward Compton amplitude is shown 
in fig.~(\ref{fig:bloem}). It dominates because the free quark propagator
has the greatest possible light-cone singularity. 
The modifications shown in figs.~(\ref{fig:boom}-\ref{fig:auto})
introduce only logarithmic modifications of the $\sim 1/\xi^3$
singularity. Renormalization group summation of leading $\ln \xi^2$
dependence leads to powers of $\ln \xi^2 \mu^2$ but no change in 
the fundamental power singularity. All radiative corrections can be
classified in the fashion outlined by figs. (\ref{fig:boom}-\ref{fig:auto}).
%
%
%
%
%
For single particle inclusive $e^+e^-$  annihilation one finds in analogy to
deep inelastic scattering the  leading diagram fig.~(\ref{fig:lamp}) which
has a singularity $1/\xi^3$ from a propagating quark.  Radiative corrections
can be treated as before.  
%
%
If light-cone dominance were the only consideration, the diagram in 
fig.~(\ref{fig:gorilla}) would dominate the Drell-Yan process.  However, if
one studies the  flow of hard momentum, this diagram turns out to be
suppressed.  The quark which brehmsstrahlungs the massive photon must be far
off-shell, which is unnatural in a hadron-hadron collision.  In coordinate
space this is reflected by the fact that no large phases are generated by
the matrix element.
The dominant contribution to the Drell-Yan process is shown in
fig.~(\ref{fig:aap}). 
The enclosed parts appear to be identical to the quark-hadron amplitude  
that occurs in the diagram that dominates in deep inelastic scattering.  This
means that at tree level, the same structure functions that appear in deep
inelastic scattering also contribute to the Drell-Yan process.   The diagram
should still be dressed with QCD radiative corrections.  The factorization
theorem of QCD\cite{QCDfactorize} says that this correspondence survives even
in the presence of radiative corrections.

 A subtlety of the Drell-Yan process is
that the term most singular on the light-cone does not dominate, nevertheless
the diagram gets its dominant contribution from
$\xi^2\approx 0$.  Returning to the definition of $W(Q^2,s)$, we see that
$\Delta_R(\xi,Q^2)$ forces $\xi^2$ to zero, but the phase factors generated by
the two separate quark-hadron amplitudes select tangent planes to the
light-cone that contribute to the $\xi^+$ and $\xi^-$ integrals respectively.

\begin{figure}
\centerline{\epsffile{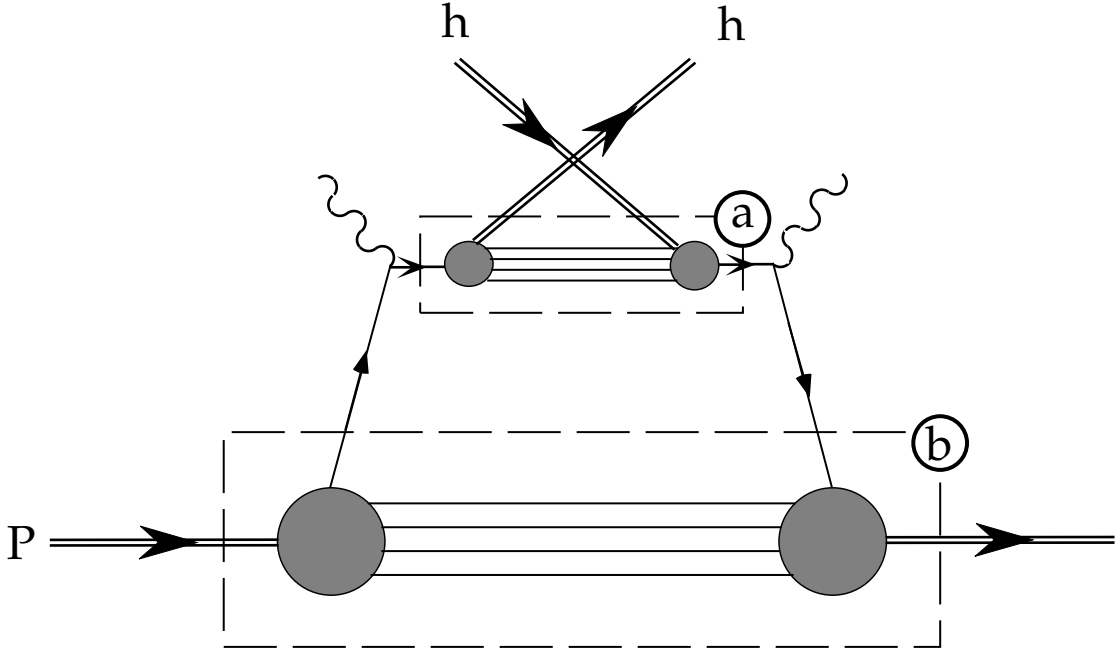}}
	\caption{{\sf
	Dominant diagram for the one particle 
	inclusive deep inelastic scattering.}}
	\label{fig:noot}
\end{figure}

One can now generalize these results to other processes, appealing to
factorization.\cite{QCDfactorize} 
For example fig.~(\ref{fig:noot})
shows the dominant contribution to one particle inclusive deep inelastic
scattering in the current fragmentation region.  
Building blocks of these calculations are $(a)$ and $(b)$, defined and
measured in $e^+e^-\rightarrow$ hadrons and DIS respectively.  Factorization
allows them to be carried from one process to another. 
They are the fundamental objects of study in hard inclusive 
QCD and command our attention. 

\section{Deep Inelastic Scattering and Generalized Distribution Functions I}
\setcounter{equation}{0}
In this first of two sections on deep inelastic scattering the focus will
be on developing the tools necessary to perform a complete classification of
effects at leading and next-to-leading order in ${1\over Q^2}$.  We begin
with some simple considerations of dimensional analysis, which we then apply
to introduce the operator product expansion (OPE) and introduce the concept
of ``twist'' which is useful to classify contributions to hard processes. 
To proceed further we must understand how to treat the Dirac structure of
quark fields on the light-cone.  This leads us briefly to explore
light-cone quantization and introduce helicity, chirality and transversity
as they apply to this problem.  We will then look in some detail at a
typical leading twist $({\cal O}(1))$ and next-to-leading twist $({\cal
O}({1\over Q}))$ distribution before attacking the complete problem in
\S 4.
 
\subsection{Twist}

In \S 2 we introduced the operator product expansion (OPE) as a tool
for analyzing $e^+e^-\rightarrow {\rm hadrons}$, where the operator with
lowest dimension dominates.  We also argued that light-like distances
($\xi^2\sim 0$) dominate deep inelastic scattering.  However, operators
 of high dimension can be important in this case.
Instead a new quantum number, ``twist'', related to both the dimension and
spin of an operator, orders the dominant effects.  

\subsubsection {Twist and the OPE}

As we learned in \S 2, the hadronic structure tensor of deep inelastic
scattering,
\be
	4\pi W_{\mu \nu}  = 
	 \int d^4\xi e^{iq\cdot \xi}
	\langle P,S\vert [ J_\mu(\xi) ,
	J_{\nu}(0) ] \vert P,S \rangle 
\ee
is dominated by $\xi ^2 \sim 0$ in the
$Q^2 \rightarrow \infty$ limit.  To make use of this, we expand the current
commutator in terms of decreasing singularity around $\xi ^2=0$,
\be
	\left[ J(\xi ),J(0) \right] \sim 
	\sum _{[\theta]} K_{[\theta ]}(\xi ^2) \xi ^{\mu _1} \ldots
	\xi^{\mu _{n_\theta}} \theta_{ \mu _1 \ldots \mu _{n_\theta}} (0),
	\label{eq:OPE}
\ee
where $\theta_{ \mu _1 \ldots \mu _{n_\theta}} (0)$ are local
operators, and
$K_{[\theta ]}(\xi ^2)$ are singular c-number functions that can be ordered
according to their degree of singularity at $\xi^2=0$.  Operators of the
same singularity at $\xi^2=0$ will be of the same importance as 
$\xi^\mu\xi_\mu\rightarrow 0$ even though
numerator factors of $\xi_\mu$ render some less singular than others as
$\xi_\mu\rightarrow 0$.
 For simplicity we have
suppressed all labels, including spin, on the currents $J(\xi)$. It often
convenient (and sometimes essential) to regroup the terms in eq.~(\ref{eq:OPE})
so that the operators $\theta _{ \mu _1 \ldots \mu _{n_\theta}}$ are
traceless ({\it i.e.\/} $g^{\mu_1\mu_2}\theta_ {\mu
_1\ldots\mu_{n_\theta}}=0$, {\it etc.\/}) and symmetric in their Lorentz
indices.  We will assume this has been done.

Substituting the OPE into the definition 
of the structure function gives:
\be
	4\pi W =
	\int d^4\xi e^{iq\cdot \xi}
	\sum _{[\theta ]} K_{[\theta ]}(\xi ^2) \xi ^{\mu _1} \ldots
	^{\mu _{n_\theta}} 
	\langle P\vert \theta _{ \mu _1 \ldots \mu _{n_\theta}} (0)\vert P 
	\rangle ,
\ee
where the matrix elements have the form:
\be
	\langle P\vert \theta _{ \mu _1 \ldots \mu _{n_\theta}} (0)
	  \vert P \rangle = P_{\mu _1} \ldots P_{\mu _{n_\theta}}
	  M^{d_\theta -n_\theta -2} f_\theta + \ldots.
\label{eq:matrix}
\ee
The $\ldots$ represent several types of terms which are less important in
the Bjorken limit.  We will return to them after looking at the dominant
term. 

Note that the power of a mass scale which appears in this expression is
determined by dimensional analysis alone.  We use the parameter $M$
generically for a typical hadronic mass scale
$M \sim \Lambda _{QCD} \sim R^{-1}_{Bag} \sim M_N /3$. The power with which
$M$ occurs defines the {\it twist\/} of the operator $\theta$,
\be
	t_\theta \equiv d_\theta - n_\theta 
\ee
The degree of the light-cone singularity of $K_{[\theta ]} \sim \xi
^{-6+t_\theta}$ is also determined by dimensional analysis and depends only
on the twist, $t_\theta$. 

To carry out the fourier transformation make the substitution,
\be
	\xi ^\mu \rightarrow -2iq^\mu {\partial \over \partial q^2}
\ee
which yields,
\be
	4\pi W \sim  \sum_{\{ \theta \}} 
	\left( {M\over\sqrt{ q^2}} \right)^{t_\theta -2}
	\left({1\over x}\right)^{n_\theta}
	f_\theta
\ee
So the importance of an operator as $q^2\rightarrow\infty$ is determined by its
twist.  As we shall see, it is typical for towers of operators with the same
twist (and other quantum numbers such as flavor) and increasing spin to appear in
the OPE.  Then it is convenient to sum over spin --- $\sum_{n_\theta} 
f_\theta{1\over x}^{n_\theta}\equiv \tilde f_\theta(x)$ -- where we now use
the label $\theta$ to refer to the entire tower of operators.

The effect of radiative corrections is to introduce logarithmic dependence
on $Q^2$ into the function $\tilde f_\theta$.  Note however that the power
law dependence on $Q^2$ is fixed by twist through dimensional analysis. 
Let us now return to the terms omitted in eq.~(\ref{eq:matrix}).  These include
terms like $P_{\mu_1}\ldots P_{\mu_{n-2}}g_{\mu_{n-1}\mu_n}M_N^2$ that make
the expression traceless.  It is easy to see that these contribute at most
corrections of order ${M_N^2\over Q^2}$ to the term we have kept.  To carry
through a complete analysis beyond order ${1\over Q}$ it is necessary to
keep careful track of these terms.  This, and the fact that interesting
spin effects appear at ${\cal O}({1\over Q})$, are the reasons we do not
consider ${\cal O}({1\over Q^2})$ here.

The lowest twist operator towers in QCD have
$t_\theta=2$ and scale -- modulo logarithms -- in the Bjorken
limit.  This reflects the underlying scale invariance of the classical
lagrangian.  The matrix elements of higher twist operators, or the higher
twist manifestations of twist-two operators are invariably signalled by the
appearance of positive powers of mass in expressions analogous to
eq.~(\ref{eq:matrix}).  Dimensional analysis then forces compensating factors of
large kinematic invariants in the denominator, suppressing the contribution. 
The simple conclusion is that {\it we can order the importance of effects in the
deep inelastic limit simply by keeping track of 
masses we are forced to introduce
into the numerators of parton-hadron amplitudes in order to maintain  the 
correct dimensions.}
\subsubsection{Examples and a Working Redefinition of Twist}
To make the preceding discussion clearer, here are some explicit examples
from free field theories.  These examples are not only pedagogical -- the
second one generates the leading twist effects in QCD up to logarithms.  The
light-cone singularities can be isolated easily. For the time ordered
product of two scalar currents built from scalar fields,
$J(\xi)=:\phi (\xi) \phi (\xi): $, one can use Wick's theorem to show that,
\be
	T(J(\xi)J(0))=-2 \Delta _F^2 (\xi) +4i\Delta _F(\xi) 
	:\phi (\xi) \phi (0):
	+ :\phi (\xi) \phi (\xi) \phi (0) \phi (0):\,,
\ee
where the normal ordering operation is sufficient to render the operator
products finite (in free field theory) as
$\xi\rightarrow 0$, and 
\be
	\Delta _F (\xi)= {i\over 4 \pi ^2}{1\over \xi^2 -i\epsilon }
\ee
for a massless scalar field.  To finally obtain the form of eq.~(\ref{eq:OPE}),
simply Taylor expand the bilocal operators -- 
\be
	 :\phi(\xi)\phi(0): =\sum_n\xi_{\mu_1}\ldots\xi_{\mu_n}\,
	 :\{\partial^{\mu_1}\ldots\partial^{\mu_n}\phi(0)\}\phi(0):
\ee

The current associated with a vector flavor symmetry of a fermion field
is
\be
	J_\mu ^a (\xi)=:\bar{\psi}(\xi) {\lambda ^a \over 2}
	\gamma _\mu \psi (\xi):.
\ee
Making use of the identity
\be
	[\bar\psi_1\psi_1,\bar\psi_2\psi_2]
	=\bar\psi_1\{\psi_1,\bar\psi_2\}\psi_2-
	\bar\psi_2\{\psi_2,\bar\psi_1\}\psi_1
\ee
(because $\{\psi_1,\psi_2\}=0$ in free field theory),
and, for a massless field,
\be
	\{\psi(\xi),\bar\psi(0)\}=
	{1\over 2 \pi} \partial \hspace{-2mm}/ \epsilon (\xi_0) \delta (\xi^2),
\ee
one can now express the commutator of two currents in terms of bilocal
operators:\cite{GelFri}
\bea
	\left[ J^{\mu a} (\xi), J^{\nu b} (0) \right]&=&
	-{1\over 4 \pi} \left( \partial _\rho \epsilon (\xi_0) 
	 \delta (\xi^2) \right)
	\Big[d^{abc} S^{\mu \rho \nu \alpha} A^c _\alpha (\xi,0) 
	-id^{abc} \epsilon ^{\mu \rho \nu \alpha} 
	 S^{5c} _\alpha (\xi,0)\nonumber\\
	 &+&i f^{abc} S^{\mu \rho \nu \alpha} S^{c} _\alpha (\xi,0) +
	f^{abc} \epsilon ^{\mu \rho \nu \alpha} A^{5c} _\alpha (\xi,0) \Big],
	\label{eq:puttel}
\eea
where the Lorentz structure is split into a symmetric and
an antisymmetric part according to:
\bea
	\gamma ^\mu \gamma ^\rho \gamma ^\nu &=&
	S^{\mu \rho \nu \alpha}\gamma_\alpha-i \epsilon^{\mu \rho \nu \alpha}
	 \gamma_\alpha\gamma ^5, \nonumber\\
	S_{\mu \rho \nu \alpha} &\equiv& {1\over 4}
	{\rm Tr} \gamma _\mu \gamma _\rho \gamma _\nu \gamma _\alpha 
	=g_{\mu \rho} g_{\nu \alpha} + g_{\mu \alpha} g_{\nu \rho} -
	g_{\mu \nu} g_{\alpha \rho},
\eea
and the flavor structure is split in a similar way:
\be
	\lambda ^a \lambda ^b =
	\left( d^{abc} \lambda ^c 
	+if^{abc} \lambda ^c \right).
\ee
The symmetric and anti-symmetric vector and axial {\it bilocal\/} currents
are defined by,
\be
	S/A^{[5]c}_\alpha \equiv
	\bar{\psi}(\xi) {\lambda ^c \over 2} 
	\gamma _\alpha [ \gamma ^5 ] \psi (0) \pm
	\bar{\psi}(0) {\lambda ^c \over 2} 
	\gamma _\alpha [ \gamma ^5 ] \psi (\xi) 
	\ee
Once again the form of eq.~(\ref{eq:OPE}) is obtained by Taylor expanding the
bilocal operators. 

We have presented these formulas in their full complexity because they
summarize the algebra of free quarks at short distances.  All of the
traditional results of the quark parton model applied to DIS (scaling
relations, the Adler, Bjorken, Gross-Llewellyn Smith and other sum rules,
the Callan Gross relation, {\it etc.\/}) can be obtained directly from these
relations.\cite{EllJaf73}

The steps of first expanding the bilocal operators, then resumming the
tower after fourier transformation are very inefficient.  Clearly it should
be possible to work directly with the bilocal operators.  The twist content
of a {\it bilocal operator\/} is somewhat more complicated than that of a
local operator.  Consider, for example, the bilocal current,
$\bar{\psi}(0) \gamma ^\mu\psi (\xi)$, which occurs in eq.~(\ref{eq:puttel}). 
The operator has dimension three and, were it a local operator, it would
have spin-one.  In fact it sums an infinite tower of operators of
increasing spin and dimension, with $t\ge 2$.  For example at short distance
one can write:
\bea
	\bar{\psi}(0) \gamma ^\mu \psi (\xi) &=&
	\bar{\psi}(0) \gamma ^\mu \psi (0) + 
	\xi_\nu \bar{\psi}(0) \gamma ^\mu \partial ^\nu \psi (0) + \ldots \\
	&\equiv& J^\mu (0)+ \xi_\nu \theta ^{\mu \nu}(0) + \ldots.
\eea
$J^\mu (0)$ is traceless, symmetric and local and has twist-two. The
operator $\theta ^{\mu \nu}$ can be decomposed into a traceless operator and
a ``trace'':
\be
	\theta ^{\mu \nu}=\left\{ \theta ^{\mu \nu}-{1\over 4} g^{\mu \nu}
	\theta ^\lambda _\lambda \right\} +
	{1\over 4} g^{\mu \nu}
	\theta ^\lambda _\lambda.
\ee
The first term is traceless, symmetric, with twist-two.  The second operator has
spin-$0$, hence its twist is four.  Further terms in the Taylor expansion
of the {\it bilocal operator\/} each yield a tower of local operators
beginning at twist-two and increasing in steps of two.

Up to now we have used {\it twist\/} only in the sense in which it was
originally introduced --- $t_{\theta}=d_{\theta}-n_{\theta}$.  In practice,
twist is used in a less formal way, to denote the order in ${1\over
Q^2}$ (modulo logarithms) at which a particular effect is seen in a
particular experiment.  If it behaves like  $(1/Q^2)^p$, then the object of
interest is said to have twist
$t=2+2p$.  A traceless symmetric operator of twist $t$ will generate
contributions that go like $(1/Q^2)^{(2-t)}$, $(1/Q^2)^{(4-t)}\ldots$ as we saw
explicitly for the operator $\theta_{\mu\nu}$. Although the
two meanings of twist do not coincide perfectly, both are in common use. 

We will make a definition of the {\it twist\/} of an
{\it invariant matrix element\/}
of a light-cone bilocal operators, that determines the scaling
behavior of the matrix element.  Matrix elements of operators like
{\it e.g.\/} $\bar\psi(0)\gamma_\mu\psi(\lambda n)$ are the
basic building blocks of the description of hard processes in QCD.
So we will call ``twist'' the order in ${M/Q}$ at which an
operator matrix element contributes to deep inelastic processes.   A few
virtues of our working definition are a) that it is easily read off by
inspection of matrix elements; b) that it directly corresponds to
suppression in hard processes; and c) that effects we label twist-$t$ never
enter hard processes with suppression less than
$(M/Q)^{t-2}$.  The twist we associate with the invariant matrix element of
a specific bilocal operator can be determined simply by considering the
powers of mass which must be introduced to perform a Lorentz-tensor
decomposition of the matrix element.  The powers of mass carry through the
entire calculation to the end where each power is compensated by a power of
$Q$ in the denominator.  Twist-two results in no suppression, therefore
$t-2$ is to be associated with the number of powers of mass introduced in
the tensor decomposition of a matrix element.  

The method is best explained by example.  Consider the spin
averaged matrix element of the bilocal current, $\bar{\psi}(\lambda
n)\gamma^\mu\psi (0)$ on the light-cone,
\be
	\langle P \vert \bar{\psi }(\lambda n) \gamma ^\mu \psi (0) 
	 \vert P \rangle
	=p^\mu f_1(\lambda ) + n ^\mu M^2 f_2 (\lambda ),
\ee
where the factor of $M^2$ must be introduced because $[n^\mu]=-1$.  The
twist of the first term is two but, due to the appearance of the factor
$M^2$, the twist of the second term is four.  In a physical application we
assert that the factor of $M^2$ will survive all manipulations and appear
in the result compensated dimensionally by a factor of $1/Q^2$.  Note that
it {\it is \/} possible for $f_1$ to pick up multiplicative factors of
$M^2/Q^2$ during a calculation.  Twist tells us the leading, not the
exclusive, $Q^2$ dependence of an invariant piece of a light-cone bilocal
operator.  As a second example, consider
\be
	\langle P \vert \bar{\psi}(\lambda n) \psi (0) \vert P \rangle
	=M e (\lambda ).
\ee
$e(\lambda)$ has twist-{\it three\/} due to the factor $M$ which must
be introduced to preserve dimensions.  Finally, consider the matrix element
of a gluonic operator
\be
	\langle P \vert G_\mu ^{\phantom{\mu}\alpha} (\lambda n ) 
	G_{\alpha\nu} (0) \vert P
	 \rangle=p_\mu p_\nu f_1 (\lambda ) + (p_\mu n_\nu + p_\nu n_\mu) f_2
	 (\lambda ) M^2	+n_\mu n_\nu f_3 (\lambda ) M^4,
\ee
which has a twist content that can be worked out by the reader.
\subsubsection{Spin and Twist}

 Counting twist in the case of polarized targets (or fragments) has an
added complication.  The Lorentz tensors which describe a hadron's spin can
appear in the Lorentz decomposition of matrix elements --- their role in
determining twist must be explained.   The objects of interest in polarized
scattering (or fragmentation) are forward scattering matrix elements on a
null plane: 
$\langle P,\epsilon \vert \theta (\lambda n,0) \vert P,\epsilon \rangle$.
The matrix elements are bilinear in $\epsilon$ and $\epsilon^*$, 
where $\epsilon$ and $\epsilon^*$ are the generalized spinors describing
the target (Dirac spinors for spin $1/2$, polarization vectors for
spin 1, {\it etc.\/}). The matrix element is a tensor function of
$\epsilon$ and $\epsilon^*$.  For spin-$1/2$ the only (non-trivial) tensors
which can be built from $u\times \bar u$ are $\bar{u} \gamma ^\mu u=2
P^\mu$, a vector, and $\bar{u} \gamma ^\mu \gamma _5 u=2 S^\mu$, an axial
vector.  We have already analyzed $P^\mu$ (it gets decomposed into $p^\mu$
and $n^\mu$). To expose the twist content of terms proportional to $S^\mu$,
express it in terms of $p^\mu$ and $n^\mu$:
\be
	S^\mu = (S\cdot n)p^\mu + (S\cdot p)n^\mu + S_\perp^\mu.
\ee
Since $S\cdot p=-M^2 S\cdot n /2$, it is clear that the second term
contributes at twist-four.  The transverse spin term is more subtle. 
Because we have chosen to normalize $S^2=-M^2$,  $[S_\perp]=1$ and because
there are no transverse momenta in the problem, $S^\mu_\perp$ contains a
hidden factor of the target mass.  At the end of the day this factor will
manifest itself in a suppression by ${M\over Q}$.  So we conclude that
appearances of $S_\perp$ accompany {\it twist-three\/} distributions.   An
example is provided by:
\be
	\langle PS \vert \bar{\psi} (0) \gamma ^\mu \gamma _5
	\psi (\lambda n) \vert PS \rangle =
	S\cdot n\ p^\mu g_1 (\lambda ) + S_\perp ^\mu \ g_T(\lambda )
	+S\cdot p\ n^\mu g_3 (\lambda ).
\ee
According to dimensional analysis, $g_1$ is a twist-two object, $g_T$ has
twist-three and $g_3$ is a twist-four function.  When combined with the
analysis of the following section, one finds that the function we have
labeled $g_1$ is the scaling limit of the ``$g_1$'' defined in \S 1. 
Similarly, $g_T$ turns out to be $g_1+g_2$.  We discard $g_3$ because we
are not concerned with twist-four.

The same method of analysis can be extended to higher spins.  For a
spin-$1$ target, all polarization information is contained in the
spin-density matrix
$\eta_{\mu\nu}\equiv \epsilon_\mu\epsilon^*_\nu$, which contains scalar
($\epsilon^*\cdot\epsilon$), vector ($S_\mu\equiv{i\over
M^2}\epsilon_{\mu\nu\alpha\beta}P^\nu
\epsilon^{*\alpha}\epsilon^\beta$), and tensor
($\hat\eta_{\mu\nu}\equiv\epsilon^*_\mu
\epsilon_\nu+\epsilon^*_\nu\epsilon_\mu
+g_{\mu\nu}{M^2\over 2}$) polarization information.  Note $[S^\mu]=1$ and 
$[\hat\eta_{\mu\nu}]=2$.  To determine the twist of the associated
distributions,
$\eta$ must be projected along
$p^\mu$, $n^\mu$ and transverse directions.\cite{HJM}  For even higher
spins a multipole analysis is more streamlined.\cite{JafMan}  
\subsection{Dominant Diagram in Coordinate Space}
As a final, and physically important example, we take the dominant diagram
identified in \S 2 and use coordinate space methods to compute it.
Since the quark that propagates between currents suffers no interactions
(we are ignoring gluon radiative corrections here), we may use free field
theory.  Working out the commutator of free currents, we get
\bea
	W^{ab}_{\mu \nu}  &=& 
	{1\over 4\pi} \int d^4\xi e^{iq\cdot \xi}
	\langle P,S\vert [ J^a_\mu(\xi) ,
	J^b_{\nu}(0) ] \vert P,S \rangle \nonumber\\
	&=&-\left( {1\over 4\pi} \right)^2 \int d^4\xi 
 e^{iq\cdot \xi} \partial ^\rho
	(\delta (\xi ^2) \epsilon (\xi ^0)) 
	\left\{ S_{\mu \rho \nu \alpha} d^{abc}
	\langle PS \vert A^{c\alpha }(\xi ,0) \vert PS \rangle + \ldots
	\right\},
 \label{eq:gellmann}
\eea
which corresponds to the handbag diagram of  fig.~(\ref{fig:bloem}). The
$\ldots$ represent three more terms, given in eq.~(\ref{eq:puttel}). This
simple free-field picture is modified by:
\begin{itemize}
\item vertex and self energy corrections, which modify the singular
function (fig.~(\ref{fig:boom})). They give rise to logarithmic corrections,
as do the dominant parts of
\item ladder graphs (fig.~(\ref{fig:huis})), and
\item box graphs, which mix in gluons at ${\cal O}(\alpha _s)$
(fig.~(\ref{fig:beestje})).   Finally,
\item in order to preserve color gauge invariance, one has to remember that
the quark propagates in a gluon background (fig.~(\ref{fig:auto})).  
\end{itemize}

On account of the last point, the singular function of free field theory,
$\{\psi(\xi),\bar\psi(0)\}={1\over 2 \pi} \partial \hspace{-2mm}/ \epsilon
(\xi_0) \delta (\xi^2)$,  must be changed to 
\be
	\{\psi(\xi),\bar\psi(0)\}\rightarrow {1\over 2 \pi} 
	\partial \hspace{-2mm}/
	\epsilon (\xi_0) \delta (\xi^2) {\cal P} 
	\left( \exp i\int _0^\xi d\zeta ^\mu
	A_\mu (\zeta ) \right),
\ee
which is the quark propagator in a background gluon field.
[The path ordering (${\cal P}$) is necessary because $A_\mu (\zeta )$ is a
matrix in color space.]  The color field $A^\mu$ 
is that generated by remnants of
the target nucleon and must be viewed as an operator sandwiched between the
target hadron states.  The bilocal operators in eq.~(\ref{eq:gellmann}) 
are therefore
replaced by,
\be
	\bar{\psi}(\xi ) \Gamma\psi (0) \rightarrow
	\bar{\psi} (\xi ){\cal P} \left( \exp i\int _0^\xi d\zeta ^\mu
	A_\mu (\zeta ) \right)\Gamma\psi (0),
\ee
where $\Gamma$ stands for whatever color/flavor/Dirac matrices appear between
$\bar\psi$ and $\psi$.  The $\delta$-function in eq.~(\ref{eq:gellmann}) selects the
light-cone.  If we expand $\xi^\mu$ about the null plane,
$\xi^\mu=\lambda n^\mu+\hat\xi^\mu$, it is easy to see that the terms involving
$\hat\xi^\mu$ are twist-four and higher. One therefore has:
\bea
	\langle PS \vert \bar{\psi} (0) {\cal P} 
	\left( \exp -i\int _0^{\xi^\mu} d\zeta ^\mu
	A_\mu (\zeta ) \right) \psi (\xi ) \vert PS \rangle & = &\nonumber\\
	\langle PS \vert \bar{\psi} (0) {\cal P} 
	\left( \exp -i\int _0^\lambda d\tau
	n\cdot A (\tau n) \right) \psi (\lambda n ) \vert PS \rangle 
	& + &\ldots,
\label{bilocal}
\eea
where the $\ldots$ represent the parts that vanish on the light-cone and
have a twist $\geq 4$. In the light-cone gauge
$n\cdot A=0$, explicit reference to gluons disappears.  However, the
inclusion of the ``Wilson link'', ${\cal P} \left( 
\exp i\int _0^\xi d\zeta ^\mu A_\mu (\zeta ) \right)$, 
is essential in generating higher twist ($t\geq
4$) gluon corrections.

In the unpolarized case, the twist expansion of the bilocal operator matrix
element gives
\be
	\int {d\lambda \over 2\pi} e^{i\lambda x}
	\langle P \vert \bar{\psi} _a (0) \gamma _\mu
	\psi _a (\lambda n) \vert P \rangle \equiv
	2 f_{1a} (x) p_\mu +2 M^2 f_{4a} (x) n_\mu
	\label{eq:gurken}
\ee
and, carrying out the fourier transform in eq.~(\ref{eq:gellmann}), we find
\be
	F_1(x)={1\over 2} \sum _a e_a^2 (f^a_1(x) -f^a_1(-x)),
\ee
where $a=u,d,s,\ldots$ is the flavor index.
The interpretation in terms of the parton model will be given later in this
section. 

A brief summary to this point is:  Up to and including twist-three the
basic objects of analysis in DIS are forward matrix elements of
bilocal products of fields on the light-cone and in light-cone gauge,
\be
	\Gamma (x) = \int {d\lambda \over 2\pi} e^{i\lambda x}
	\langle PS \vert \bar{\psi} (0) \Gamma \psi (\lambda n)
	\vert PS \rangle.
\label{eq:generalbilocal}
\ee
Remember, that important $\ln Q^2$ radiative corrections have been ignored in
pursuit of the twist and spin dependence.
\subsection{Learning from Light-Cone Quantization}
Since the dominant contribution to DIS comes from the light-cone, it is natural
to consider a dynamical formulation in which the light-cone plays a special
role.  At the birth of deep inelastic physics it was recognized that
field theories simplify in some important ways if they are quantized ``on the
light-cone'' rather than at equal times.\cite{Wein,Suss}  Unfortunately some
features which are simple at equal times become difficult on the light-cone. 
Certainly, as we shall see, there is much insight to 
be gained by considering deep
inelastic processes using light-cone quantization.  The larger question --
whether QCD simplifies in essential ways when quantized on the light-cone --
will not be pursued here.

Field theories may be quantized by imposing canonical equal-time
commutation (or anticommutation) relations on the dynamically independent
fields.\cite{Itzub}  Lorentz invariance requires that any other space-like
hyperplane in Minkowski space would serve as well as $\xi^0=0$.  A null-plane,
such as $\xi\cdot n=0$ is the limit of a sequence of space-like surfaces, and 
includes points that are causally connected.  Although a field theory 
quantized on at $\xi\cdot n =0$ could differ from one quantized at $\xi^0=0$,
they coincide for all examples of which I am aware.
Let us study what happens if we attempt to quantize fermions on the surface
$\xi^+=0$.\cite{KS70}  First we must introduce and familiarize 
ourselves with the unusual kinematics of the light-cone.  

\subsubsection{Light-Cone Kinematics}

We have previously introduced light-cone coordinates
$\xi^\pm={1\over\sqrt{2}}(\xi^0\pm \xi^3)$ and $\vec\xi^\perp = (\xi^1,\xi^2)$,
and the metric $g_{\mu\nu}$, with $g_{+-}=g_{-+}=1$, and $g_{ij}=-\delta_{ij}$. 
The partially off-diagonal structure of $g$ makes raising and lowering indices
confusing, {\it viz.\/}, $a^+=a_-$, and so forth.  So we work with upper
(contravariant) indices as much as possible.  

Quantizing at (say) $\xi^+=0$, we are committed to $\xi^+$ as our evolution
variable (just as quantization at $\xi^0=0$ fixes $\xi^0$ as the ``time''). 
$\xi^-$ and $\xi^\perp$ are therefore kinematic, not dynamical variables.  The
conjugate momenta $p^+$ and $\vec p^\perp$ parameterize the 
fourier decomposition of
the independent light-cone fields, just like $\vec p$ in ordinary quantization. 
$p^-$ is the ``Hamiltonian'' for light-cone dynamics.

\subsubsection{Dirac Algebra on the Light-Cone}

The usual selection of $\gamma^0={\rm diag}(1,1,-1-1)$ is prejudiced toward
equal time quantization.  Then a (anti-) particle at rest has only
(``lower'') ``upper'' components in its Dirac spinor.  Much of our analysis is
simplified by choosing a representation for the Dirac matrices tailored to the
light-cone.\cite{KS70}  To represent Dirac matrices compactly, we use the
``bispinor'' notation: let ($\sigma^1,\sigma^2,\sigma^3$) and
($\rho^1,\rho^2,\rho^3$) be two copies of the standard ($2\times 2$) Pauli
matrices.  A $4\times 4$ Dirac matrix can be represented as $\rho^i \otimes
\sigma^j$.  $\rho$ controls the upper-versus-lower two-component space; 
$\sigma$ controls the inner two-component space.  An example will clarify the
notation:  the Dirac-Pauli representation used, for example, by Bjorken and
Drell is,
\bea
	\gamma^0_{BD}&=&\rho^3\otimes 1 = \pmatrix{1&0\cr 0&-1}\nonumber\\
	\gamma^j_{BD}&=&i\rho^2\otimes \sigma^j = 
	\pmatrix{0&\sigma^j\cr -\sigma^j&0}\nonumber\\
	\gamma^5_{BD}&=&\rho^1\otimes 1 = \pmatrix{0&1\cr1&0}\nonumber\\
\eea
The light-cone representation useful for us is instead,
\bea
	\gamma^0&=&\rho^1\otimes \sigma^3 = 
	\pmatrix{0&\sigma^3\cr \sigma^3&0}\nonumber\\
	\vec\gamma^\perp&=&1\otimes i\vec\sigma^\perp = \pmatrix{i
	\vec\sigma^\perp&0\cr
	0&i\vec\sigma^\perp}\\
	\gamma^3&=&-i\rho^2\otimes\sigma^3 = 
	\pmatrix{0&-\sigma^3\cr \sigma^3&0}\nonumber\\
	\gamma^5&=&\rho^3\otimes \sigma^3 = \pmatrix{\sigma^3&0\cr0&-\sigma^3},
	\nonumber\\
\label{eq:mygamma}
\eea
where $\perp\,=1\, {\rm  or }\,2$.
It is easy to check that eq.~(\ref{eq:mygamma}) satisfy the usual algebra,
$\{\gamma^\mu,\gamma^\nu\}=2g^{\mu\nu}$, and
$\gamma^5=i\gamma^0\gamma^1\gamma^2\gamma^3$.

Operators which project on the upper and lower two component subspaces play a
central role in light-cone dynamics.  Define ${\cal P}_\pm$ by,
\bea
	{\cal P}_\pm &=&{1\over 2}\gamma^\mp\gamma^\pm = 
	{1\over 2}(1\pm\alpha_3)\nonumber\\
	\gamma ^\pm &=& {1\over \sqrt{2}} (\gamma^0\pm\gamma^3),\nonumber\\
\eea
with the properties:
\bea
	{\cal P}_-{\cal P}_+&=&{\cal P}_+{\cal P}_-=0\nonumber\\
	{\cal P}_\pm^2&=&{\cal P}_\pm\nonumber\\
	{\cal P}_-+{\cal P}_+&=&1\nonumber\\
	{\cal P}_+=\pmatrix{1&0\cr 0&0}&\quad &
	{\cal P}_-=\pmatrix{0&0\cr 0&1}\nonumber\\
\eea
The ``light-cone projections'' of the Dirac field, $\psi_+\equiv{\cal P}_+\psi$
and $\psi_-\equiv{\cal P}_-\psi$ are known as the 
``good'' and ``bad'' light-cone
components of $\psi$ respectively.  To 
save on subscripts we shall frequently
replace $\psi_\pm$ as follows,
\be
	\psi_+\Rightarrow\phi \quad \psi_-\Rightarrow\chi
\ee
\subsubsection{Independent Degrees of Freedom}

The importance of ${\cal P}_\pm$ becomes clear when they are used to project the
Dirac equation down to two two-component equations,
\bea
	i\gamma ^- D_- \chi&=&-\vec\gamma^\perp\cdot\vec D^\perp\phi 
	+m\phi \nonumber\\
	i\gamma ^+ D_+ \phi&=&-\vec\gamma^\perp\cdot\vec D^\perp\chi 
	+m\chi,
\label{eq:diraclc}
\eea
where $D_\pm = {\partial\over{\partial\xi^\pm}}-igA^\mp$.
In the light-cone gauge $A^+=0$.  $\xi ^+$ is the evolution (``time'')
parameter, but the first of eq.~(\ref{eq:diraclc}) only involves 
$\partial/\partial\xi^-$, so
it appears that $\chi$ is not an independent dynamical field. 
Instead the Dirac equation constrains  $\chi$ in terms of
$\phi$ and $\vec A_\perp$ at fixed $\xi ^+$,
\be
	i\gamma^-{\partial\over{\partial\xi^-}}\chi
	=-\vec\gamma^\perp\cdot\vec D^\perp\phi +m\phi
\label{eq:psiminus}
\ee
The longitudinal component of
the electric field in electrodynamic is similarly constrained ({\it i.e.\/}
determined at any time) by Gauss's Law in Coulomb gauge, $\vec\nabla\cdot\vec
E=\rho$.  Study of the gluon equations of motion indicates that $A^-$ is also a
constrained variable.  The independent fields are therefore $\phi$ and $\vec
A_\perp$.
$\chi$ should be regarded as composite --- as specified by eq.~(\ref{eq:psiminus})
--- $\chi={\cal F}[\phi,\vec A_\perp]$.

By the way, the reduction of the four-component Dirac field to two propagating
degrees of freedom is not unique to light-cone quantization.  In the usual
treatment of the Dirac equation one finds only two solutions for each energy and
momentum, corresponding to the two spin states of a spin-$1/2$ particle.  The
two-degrees of freedom corresponding to the antiparticle are found in the
solution with energy $-E$ and momentum $-\vec p$.  In fact, the Dirac equation
in momentum space is literally written in the form of a projection
operation, $\Lambda_-\psi=0$, where $\Lambda_\pm={1\over 2m}(p
\hspace{-2mm}/\pm m)$ projects out two of the four components of the Dirac
spinor.

Although the complete quantization of QCD requires much more work, the
implication for the Dirac field is already clear:  the {\it good\/} components
should be regarded as independent propagating degrees of freedom; the {\it
bad\/} components are dependent fields -- actually quark-gluon composites.

The classification of quark spin states depends on the Dirac matrices which a)
commute with ${\cal P}_\pm$ and b) commute with one-another.  Returning to
eq.~(\ref{eq:mygamma}) we see that $\gamma^1,\gamma^2$, and $\gamma_5$ commute
with
${\cal P}_\pm$.  Furthermore, the component of the generator of spin-rotations
along the $\hat e_3$-direction,
\be
	\Sigma^3\equiv {i\over 2}[\gamma^1,\gamma^2]
	=\pmatrix{\sigma^3&0\cr 0&\sigma^3},
\ee
also commutes with ${\cal P}_\pm$.  Note that for a Dirac particle with momentum
in the $\hat e_3$-direction, $\Sigma^3$ measures the {\it helicity\/}.  This set
of operators suggests two different maximal sets of commuting observables:
\begin{itemize}
\item Diagonalize $\gamma_5$ and $\Sigma^3$ --- a {\it chirality\/} or {\it
helicity basis\/},
\noindent 
or
\item Diagonalize $\gamma^1$ (or equivalently, $\gamma^2$) -- a {\it
transversity basis\/}.
\end{itemize}
Let us consider these in turn --
\paragraph{Helicity Basis}
In the helicity basis, both the good and bad components of $\psi$ carry 
{\it helicity \/}
labels -- the eigenvalues of $\Sigma^3$,
\be
	\psi=\pmatrix{\phi_+\cr\phi_-\cr\chi_+\cr\chi_-}.
\ee
Note that upper and lower components of $\psi$ correspond to {\it good\/}
and {\it bad\/} light-cone components respectively.  
Referring back to the form of $\gamma_5$, eq.~(\ref{eq:mygamma}), we see that
helicity and chirality are identical for the good components of $\psi$ but
opposite for the bad components,
\be
	\gamma_5\psi=\pmatrix{+\phi_+\cr -\phi_-\cr -\chi_+\cr +\chi_-}.
\ee
This may look strange at first, but it follows immediately from the composite
nature of $\chi$.  A quantum of $\chi$ with positive helicity is actually
a composite of a transverse gluon and a quantum of $\phi$.  Since the gluon
carries helicity-one (but no chirality), angular momentum conservation
requires that the $\phi$-quantum have negative helicity and therefore negative
chirality.  Remembering this association will help sort out the chirality and
helicity selection rules which appear in the following sections.
\paragraph{Transversity Basis}
Alternatively, we can diagonalize one of the transverse $\gamma$-matrices, to
be specific, $\gamma^1$.  We define eigenstates of the transverse
spin-projection operators, ${\cal Q} _\pm = {1\over 2} (1\mp \gamma _5 \gamma
^1)$, (which commute with ${\cal P}_\pm$), 
\bea
	{\cal Q} _+ \phi &\equiv& \phi _\perp \label{eq:transb} \\
	{\cal Q} _- \phi &\equiv& \phi _\top,
\eea
and similarly for $\chi$.  
Of course $\phi_{\top/\perp}$ are linear
combinations of $\phi_\pm$.  Note however that 
$\phi_{\top/\perp}$
are {\it not\/} eigenstates of the transverse spin operator
$\Sigma^1=\pmatrix{0&i\sigma^1\cr i\sigma^1&0}$, which is not diagonal in the
basis of good and bad components of $\psi$.  So we have to be careful that we do
not carelessly confuse {\it transversity\/}, the quantum number associated with 
${\cal Q}_\pm$, which is simple in this picture, with {\it transverse spin\/},
which is not.
\subsection{The Parton Model}
Following the path we are on, the parton model is merely the light-cone Fock
space decomposition of the matrix elements which control hard processes.  Since
we have both the matrix elements and the Fock space in hand, it is
straight-forward to construct the parton model.  We will verify that the parton
interpretation emerges as expected for twist-two and then explore twist-three. 
The reader should beware that twist-four is considerably more complicated.  A
parton model picture of twist-four does exist, however much work is required 
to make it obvious.\cite{JS81,EFP82,Jaf1}

The Fock space in the two bases can be constructed by defining operators
that create the appropriate components of $\phi$.  
In the helicity basis we define
$R^\dagger(k^+,\vec k_\perp)$ to create a right-handed (positive
{\it helicity\/}) component of $\phi$ and
$L^\dagger(k^+,\vec k_\perp)$ to create a left-handed (negative helicity)
component of $\phi$, and $R^\dagger_c$ and $L^\dagger_c$, which do the same for
the antiparticle field $\phi_c$.  In the transversity basis we define the
operators $\alpha^\dagger(k^+,\vec k_\perp)$ and 
$\beta^\dagger(k^+,\vec k_\perp)$
that create the $_\perp$ and $_\top$ components of $\phi$, respectively.

\subsubsection{Twist-Two}

We begin with the simplest case -- the spin average, twist-two deep inelastic
scattering which is controlled by the bilocal operator defined
in eq.~(\ref{eq:gurken}),
\be
	\int {d\lambda \over 2\pi} e^{i\lambda x}
	\langle PS \vert \bar{\psi}(0) \gamma _\mu \psi (\lambda n)
	\vert PS \rangle =
	2 \left\{ f_1(x) p_\mu + f_4 (x) n_\mu \right\}.
\ee
We project out the twist-two part, $f_1$, by contracting with $n^\nu$.
\bea
	f_1(x)&=&
	{1\over 2} \int {d\lambda \over 2\pi} e^{i\lambda x} 
	\langle PS \vert \bar{\psi}(0) n\hspace{-2mm}/ \psi (\lambda n)
	\vert PS \rangle \\ &=&
	{1\over \sqrt{2} p^+} \int {d\lambda \over 2\pi} e^{i\lambda x} 
	\langle PS \vert \phi^\dagger (0) \phi (\lambda n)
	\vert PS \rangle.
\eea
where we have used the Dirac algebra to express the quark field in terms of its
light-cone components.   Notice that only the dynamically independent
``good'' light-cone components occur.  If we make a momentum ($k^+,\vec
k_\perp$) decomposition of $\phi$ and separate helicity states, we find,
\be
	f_1(x)= {1\over x} \int d^2k_\perp
	\langle P\vert R^\dagger (xp,\vec{k}_\perp )
	R (xp,\vec{k}_\perp )+
	L^\dagger (xp,\vec{k}_\perp )
	L (xp,\vec{k}_\perp ) \vert P\rangle,
\label{eq:partonmodel}
\ee

This {\it is \/} the parton model as illustrated in fig.~(\ref{fig:parton}):
\begin{figure}
\centerline{\epsffile{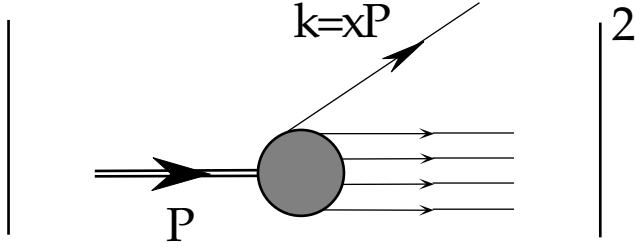}}
	\caption{{\sf The classical parton model. }}
	\label{fig:parton}
\end{figure}        
$f_1$ is
expressed as a sum of probabilities to find a (light-cone quantized) quark with
$p^+=xp$ and any transverse momentum, summed over helicities 
and weighted by the phase space factor
$1/x$.   Perhaps the reader is more familiar 
with the ``infinite momentum frame''
form of the model, where $f_1$ is written as the sum of probabilities to find an
(equal-time quantized) quark with a fraction $x$ of the target's (infinite) 
longitudinal momentum.  The two formulations are equivalent since the boost to
an infinite momentum frame is equivalent to a light-cone formulation.
Since eq.~(\ref{eq:partonmodel}) is valid in any 
frame, it can be used in ({\it e.g.\/})
the lab frame to provide parton distributions which can be associated with quark
models.\cite{Jaf85}  One must, however, be careful to remember that the fields
in eq.~(\ref{eq:partonmodel}) are good light-cone Dirac components quantized at
equal $\xi^+$, not equal $\xi^0$.

An identical calculation for $x<0$ captures antiquark operators and leads to the
standard crossing relation for $f_1$,
\be
	f_1(x)=-\bar f_1(-x)
\ee
where $\bar f_1$ is given by eq.~(\ref{eq:partonmodel}) 
with $R\rightarrow R_c$ and
$L\rightarrow L_c$.  The other (spin-dependent) 
quark distributions are explored in
the following Section.

\subsubsection{Twist-Three}
Now let us apply the same analysis to the simplest 
twist-three distribution function,
\be
	\int {d\lambda \over 2\pi} e^{i\lambda x}
	\langle P \vert \bar{\psi}(0) \psi (\lambda n)
	\vert P \rangle \equiv
	2M e(x).
\label{eq:enaive}
\ee
Decomposing in terms of $\phi$ and $\chi$, we find
\be
	e(x)={1\over 2M}
	\int {d\lambda \over 2\pi} e^{i\lambda x}
	\langle P \vert \phi^\dagger (0) \gamma ^0 \chi (\lambda n)
	+ \chi^\dagger (0) \gamma ^0 \phi (\lambda n)
	\vert P \rangle ,
\ee
which contains the dynamically dependent operator $\chi$.  
If we use the constraint
to eliminate $\chi$ we obtain
\be
	e(x)=-{1\over 4Mx}
	\int {d\lambda \over 2\pi} e^{i\lambda x}
	\langle P \vert \bar\phi(0) 
	n\hspace{-2mm}/D\hspace{-2mm}/ _\perp (\lambda n)
	\phi(\lambda n)
	\vert P \rangle +\mbox{h.c.}
\ee
So $e(x)$ is really a quark-{\it gluon\/} correlation function 
on the light-cone. 
It has no simple Fock-space interpretation in terms of quarks alone, despite the
apparently simple form of eq.~(\ref{eq:enaive}).  

We have happened upon a general (and very 
useful) result:  Every factor of $\chi$ in
the light-cone decomposition of a light-cone correlation function contributes an
additional unit of twist to the associated distribution function,
\bea
	\phi^\dagger \phi \hspace{-.7cm} && 
	\Leftrightarrow \mbox{Twist-2} \nonumber\\
	\phi^\dagger \chi \hspace{-.7cm} && 
	\Leftrightarrow \mbox{Twist-3} \nonumber\\
	\chi^\dagger \chi \hspace{-.7cm} && 
	\Leftrightarrow \mbox{Twist-4}.\nonumber\\
\eea
Likewise each unit of twist introduces an additional independent field in the
null plane correlator:
\bea
	&&      \mbox{Twist-2} \rightarrow \phi^\dagger (0) 
	\phi(\lambda n) \nonumber\\
	&&      \mbox{Twist-3} \rightarrow \phi^\dagger (0) 
	D_\perp (\lambda n) \phi(\mu n) \nonumber\\
	&&      \mbox{Twist-4} \rightarrow \phi^\dagger (0) 
	D_\perp (\lambda n) D_\perp (\mu n)\phi(\nu n) .\nonumber\\
\eea
It is as if $\phi$ had twist-one and $\chi$ had twist-two.

Twist-three is tractable, using the methods that have been developed in
these lectures. Twist-four requires a more extensive analysis based on
operator product expansion methods 
developed during the 1980's.\cite{JS81,EFP82}.  With these general tools in 
hand, we turn in the next section to the analysis of the specific 
distributions which appear in deep inelastic scattering of leptons.

\section{ Deep Inelastic Scattering and Generalized 
Distribution Functions II}
\setcounter{equation}{0}
In this section we use the tools developed in \S 3 to classify and 
interpret the quark distribution functions which appear in the analysis of
DIS.  The topics will include the classification and 
parton interpretation of the
three leading twist quark distribution functions; a discussion of the physics
of the less well known {\it transversity\/} distribution, $h_1$;
a review of transverse spin in hard processes; a short digression on higher
spin targets and gluon distribution functions; and a summary of the physics
associated with the twist-three transverse spin distribution, $g_2$.

\subsection{Helicity Amplitudes}
\begin{figure}
	\centerline{\epsffile{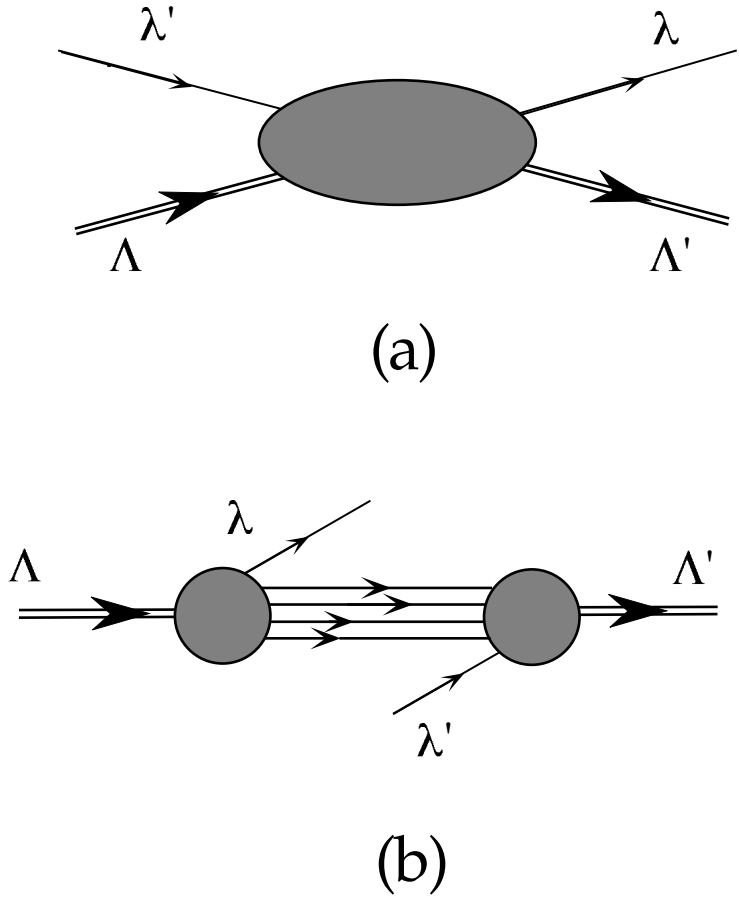}}
	\caption{{\sf Helicity structure a) of the parton-hadron forward
	scattering amplitude; b) of the $u$-channel discontinuity which
	contributes to a parton distribution function.}}
	\label{fig:uchannel}
\end{figure}

Part of the task is simply to enumerate the independent distribution
functions at twist-two and twist-three.  This is simplified by viewing
distribution functions as discontinuities in forward parton-(quark or gluon)
hadron scattering.  Suppressing all momentum indices, each quark distribution
can be labeled by four helicities:  a target of helicity $\Lambda$ emits a
parton of helicity $\lambda$ which then participates in some hard scattering
process.  The resulting parton with helicity $\lambda'$ is reabsorbed by a
hadron of helicity $\Lambda'$.  The process of interest to us is actually a
{\it u-channel\/} discontinuity of the forward parton-hadron scattering
amplitude ${\cal A}_{\Lambda\lambda,\Lambda'\lambda'}$ as shown in 
fig.~(\ref{fig:uchannel}). 
Note the ordering of indices -- although
$\Lambda$ and
$\lambda'$ are the incoming helicities, it is convenient to label the amplitude in
the sequence: initial hadron, struck quark, final hadron, returned quark.

Since the parton-hadron amplitude results from squaring something
like $\left <X\vert\psi\vert PS\right >$, the amplitude must be diagonal in
the target spin.  However spin eigenstates (in particular, {\it transverse\/} spin
eigenstates) are linear superpositions of helicity eigenstates, so the $\{{\cal
A}\}$ {\it do not\/} have to be diagonal in the target helicity.  Only forward
scattering is of interest, so the initial and final helicities must be the same,
\be
	\Lambda+\lambda'=\Lambda'+\lambda.
\ee
Also, the parity and time reversal invariance of the strong interactions place
constraints on the $\{{\cal A}\}$,
\bea
	{\cal A}_{\Lambda \lambda, \Lambda' \lambda' } & = &
	{\cal A}_{-\Lambda -\lambda , -\Lambda' -\lambda'} \\
	{\cal A}_{\Lambda \lambda, \Lambda' \lambda'} & = &
	{\cal A} _{\Lambda'\lambda', \Lambda \lambda } 
\eea
respectively.

Clearly the helicity counting outlined above applies
equally well to good and bad light-cone components of quark or gluon fields. 
Therefore we can use it together with the methods of the previous section
to enumerate quark distribution function through twist-three. 
To work through twist-three we will have to consider the case of one good and
one bad light-cone component.  We will identify any bad light-cone fields in
helicity amplitudes by an asterix on the helicity label.  Thus ${\cal
A}_{0{1\over 2}, 0{1\over 2}^*}$ corresponds to emission of
a good light-cone component and absorption of a bad one.

\subsection {Quark Distributions in Targets with Spin-0, 1/2 and 1}
\subsubsection{Spin-$0$ Target}
Only $\Lambda=0$ is available.  Parity equates ${\cal
A}_{0{1\over 2}, 0{1\over 2}}$ and ${\cal
A}_{0-{1\over 2}, 0-{1\over 2}}$.  Time reversal equates
${\cal A}_{0{1\over 2}, 0{1\over 2}^*}$ and ${\cal
A}_{0{1\over 2}^*, 0{1\over 2}}$.  So there is only one
distribution function at twist-two and one at twist-three.  The twist-two
function is none other than $f_1$ associated with the bilocal operator
$\bar\psi(0)n\hspace{-2mm}/\psi(\lambda n)$ and conserves quark chirality
(``chiral even'').  The twist-three function is $e$, associated with the scalar
bilocal operator $\bar\psi(0)\psi(\lambda n)$ and flips quark chirality
(``chiral odd'').  These properties are summarized by
\newline
\begin{table}[ht]
\center{
\begin{tabular}{|c c c c c c c|}\hline
Twist & $\Lambda$ & $\lambda$ &  & $\Lambda'$ & $\lambda'$ & Chirality\\
\hline   Two & $0$    & $1/2\phantom{^*}$     &              
&  $0$   & $1/2$ & Even\\
Three & $0$    & ${1/2}^*$ &              &  $0$   & $1/2$  & Odd         \\
\hline
\end{tabular}} 
\caption{\sf Quark distributions in a spin-0 hadron through twist-three.}
\label{tbl:spn0}
\end{table}
\subsubsection{Spin-$1/2$ Target}
In the spin-$1/2$ case, for each twist 
assignment there are three independent helicity 
amplitudes.  The reader may wish to verify that parity and time reversal 
invariance relate the many helicity amplitudes to the six listed in the table 
below (through twist-three).  We leave the interpretation of these six 
distribution functions for the next section
where they are discussed in detail.
\newline
\begin{table}[ht]
\center{
\begin{tabular}{|c c c c c c c|}\hline
Twist & $\Lambda$ & $\lambda$ &  & $\Lambda'$ & $\lambda'$ & Chirality\\
\hline  
Two & $\phantom{+}1/2$    & $\phantom{+}1/2\phantom{^*}$     &              & 
$\phantom{+}1/2$   &
$\phantom{+}1/2$ & Even \\
Two & $\phantom{+}1/2$    & $-1/2\phantom{^*}$               &              & 
$\phantom{+}1/2$   & $-1/2$ & Even \\
Two & $\phantom{+}1/2$    & $\phantom{+}1/2\phantom{^*}$               &
& $-1/ 2$              & $ -1/2$ & Odd\\ \hline
Three & $\phantom{+}1/2$    & $\phantom{+}1/2^*$     &              & 
$\phantom{+}1/2$   & $\phantom{+}1/2$ & Odd\\
Three & $\phantom{+}1/2$    & $-1/2^*$               &              & 
$\phantom{+}1/2$   & $-1/2$ & Odd \\
Three & $\phantom{+}1/2$    & $\phantom{+}1/2^*$               &              &
 $-1/2$              & $ -1/2$ & Even\\ \hline
\end{tabular}} 
\caption{\sf Quark distributions in a spin-1/2 hadron through twist-three.}
\label{tbl:spn1/2}
\end{table} 
\vspace{4mm}
\subsubsection{Spin-$1$ Target}
A massive spin-one target has three independent helicity 
states.  A new complication
appears at twist-three: two helicity flip distributions arise which are not
related by any of the symmetries of QCD.  One can easily check that no such
complication occurs for spin ${1\over 2}$.
There is much interesting physics in these spin-one structure 
functions, however time will
not permit us to work through it here.  
Instead we refer the interested reader to the
original literature.\cite{HJM,JafMan,JafMan1,Artru}
\begin{table}[ht]
\center{
\begin{tabular}{|c c c c c c c|}\hline
Twist & $\Lambda$                    & $\lambda$ & & $\Lambda'$    &
$\lambda'$ & Chirality\\ \hline  
Two & $1$ & $\phantom{+}1/2\phantom{^*}$  &  & $1$ & 
$\phantom{+}1/2\phantom{^*}$  &
Even  \\ Two & $1$ & $-1/2\phantom{^*}$  &  & $1$  & 
$-1/2\phantom{^*}$ & Even \\
Two & $0$ & $\phantom{+}1/2\phantom{^*}$  &  & $0$ & 
$\phantom{+}1/2\phantom{^*}$  &
Even \\  Two & $0$ & $-1/2\phantom{^*}$  &  & $1$  & 
$\phantom{+}1/2\phantom{^*}$  &
Odd \\
\hline Three & $1$ & $\phantom{+}1/2^*$  &  & $1$ & 
$\phantom{+}1/2\phantom{^*}$  &
Even  \\ Three & $1$ & $-1/2^*$  &  & $1$  & $-1/2\phantom{^*}$ & Even \\
Three & $0$ & $\phantom{+}1/2^*$  &  & $0$ & 
$\phantom{+}1/2\phantom{^*}$  & Even \\ 
Three & $0$ & $-1/2^*$  &  & $1$  & $\phantom{+}1/2\phantom{^*}$  & Odd \\ 
Three & $0$ & $-1/2\phantom{^*}$  &  & $1$  &
 $\phantom{+}1/2^*$  & Odd \\ \hline
\end{tabular}} 
\caption{\sf Quark distributions in a spin-1 hadron through twist-three.}
\label{tbl:spn1}
\end{table}
\vspace{4mm}
\subsection{Quark Distribution Functions for the Nucleon}
The distribution functions for a spin-${1\over 2}$ 
target deserve special attention
because protons and neutrons are the principal targets of interest.   In
Table~\ref{tbl:nucleon} the quark distribution functions  for a nucleon target
are listed through twist-three. They are classified 
according to their twist (or light-cone 
projection) and their helicity.   
\renewcommand{\arraystretch}{2}
\begin{table}[ht]
\center{
\begin{tabular}{|c|c|c|c|c|}\hline
twist $O(1/Q^{t-2})$      & Name & Helicity Amplitude   
& Measurement & Chirality \\
\hline
Two &  $f_1$ & ${\cal A}_{1{1\over 2},1{1\over 2}}+
{\cal A}_{1-{1\over 2},1-{1\over 2}}$ &
Spin average & Even\\ \hline
Two &  $g_1$ & ${\cal A}_{1{1\over 2},1{1\over 2}}-
{\cal A}_{1-{1\over 2},1-{1\over 2}}$ &
Helicity difference & Even\\ \hline
Two &  $h_1$ & ${\cal A}_{0{1\over 2},1 -{1\over 2}}$ &
Helicity flip & Odd\\ \hline
Three &  $e$ & ${\cal A}_{1{1\over 2}^*,1{1\over 2}}+
{\cal A}_{1-{1\over 2},1-{1\over 2}}$ &
Spin average & Odd\\ \hline
Three &  $h_L$ & ${\cal A}_{1{1\over 2}^*,1{1\over 2}}-
{\cal A}_{1-{1\over 2},1-{1\over 2}}$ &
Helicity difference & Odd\\ \hline
Three &  $g_T$ & ${\cal A}_{0{1\over 2}^*,1 -{1\over 2}}$ &
Helicity flip & Even\\ \hline
\end{tabular} 
}
\caption{\sf Nucleon structure functions classified 
according to their twist and target helicities.}
\label{tbl:nucleon}
\end{table}

The distribution functions $f_1$, $g_1$ 
and $g_T$ are familiar because they can be
measured in lepton scattering.  The others 
are less well known, but are essential to
understand the nucleon spin substructure
in deep inelastic processes.  All of them are
defined by the matrix elements of quark
bilocal operators,
\bea
	\int {d\lambda\over{2\pi}}e^{i\lambda x}
	\left<PS\vert\bar\psi(0)\gamma_\mu\psi
	(\lambda n)\vert PS\right> & = &  
	2\left\{ f_1(x) p_{\mu} + M^2 f_4(x) n_{\mu} \right\},\nonumber\\
	\int {d\lambda\over{2\pi}}e^{i\lambda x}
	\left<PS\vert\bar\psi(0)\gamma_\mu\gamma_5\psi
	(\lambda n)\vert PS\right> & = &
	 2\left\{ g_1(x) p_{\mu} S\cdot n + g_T(x)S_{\perp\mu}
	 + M^2 g_3(x) n_\mu S\cdot n \right\},\nonumber\\
	\int {d\lambda\over{2\pi}}e^{i\lambda x}\left<PS\vert\bar\psi(0)\psi
	(\lambda n)\vert PS\right> & = &
	 2 e(x),\nonumber\\
	\int {d\lambda\over{2\pi}}e^{i\lambda
	x}\left<PS\vert\bar\psi(0)i\sigma_{\mu\nu}\gamma_5\psi 
	(\lambda n)\vert PS\right> & = &
	 2\{h_1(x)(S_{\perp\mu} p_{\nu} - S_{\perp\nu}p_{\mu})/M \nonumber\\
	 & + & h_L(x) M (p_{\mu}n_{\nu} - p_{\nu}n_{\mu}) S\cdot n \nonumber \\
	& + &h_3(x) M (S_{\perp\mu} n_{\nu} - S_{\perp\nu}n_{\mu})\} , 
\label{eq:nucstructure}
\eea            
Some twist-four distributions ($f_4$, $g_3$, and $h_3$) appear in these matrix
elements.  However, they are joined by many other quark-quark and quark-gluon
distributions from which they cannot 
be separated, so there is no point in keeping track
of them in this analysis.
\subsubsection{Nucleon Spin Structure at Twist-Two}
$f_1$, $g_1$ and $h_1$ are twist-two, {\it i.e.\/} they 
scale modulo logarithms.  They
can be projected out of the general decompositions, eq.~(\ref{eq:nucstructure}),
\bea
	f_1(x)&=& \int {d\lambda\over{4\pi}}e^{i\lambda
	x}\langle P\vert\bar\psi(0)n\hspace{-2mm}/ 
	\psi(\lambda n)\vert P\rangle \nonumber \\
	g_1(x)&=& \int {d\lambda\over{4\pi}}e^{i\lambda
	x}\langle PS_\parallel\vert\bar\psi(0)n\hspace{-2mm}/\gamma_5
	\psi(\lambda n)\vert PS_\parallel\rangle \nonumber \\
	h_1(x)&=& \int {d\lambda\over{4\pi}}e^{i\lambda x}\langle
	PS_\perp\vert\bar\psi(0)[S\hspace{-2mm}/_\perp,n\hspace{-2mm}/]
	\gamma_5\psi(\lambda n)\vert PS_\perp\rangle
\label{eq:nuctwist2}
\eea

To understand their physical
significance --- in particular, to see why a 
third quark distribution in addition to
$f_1$ and $g_1$ is necessary to describe the
 nucleon's quark spin substructure at leading twist in the
parton model --- it suffices to decompose them
 with respect to a light-cone Fock space
basis.  If we use the {\it helicity\/} basis, then
\bea
	f_1(x) &=& \frac {1}{x} \left <P|R^\dagger(xp)R(xp) 
	+ L^\dagger(xp) L(xp)|P\right>,\nonumber\\
	g_1(x) &=& \frac {1}{x} \left <P\hat e_3|R^\dagger(xp)R(xp) -
	 L^\dagger(xp)
	L(xp)|P\hat e_3\right>,\label{eq:g1} \nonumber\\
	h_1(x) &=& \frac {2}{x} Re\left <P\hat e_1|L^\dagger(xp)R(xp)|P\hat
	 e_1\right>,
	\label{eq:bloemkool}
\eea
in analogy with eq.~(\ref{eq:partonmodel}), 
where we have integrated out the dependence on transverse
momentum.  Here $\hat e_3$ and $\hat e_1$ are unit 
vectors parallel and transverse,
respectively, to the target nucleon's three-momentum.  Clearly  
$f_1$ and $ g_1$ can be interpreted in 
a probabilistic way:  $f_1$ measures quarks independent of their 
helicity and $g_1$
measures the helicity asymmetry.  But $h_1$ does not appear to 
have a probabilistic
interpretation, instead it mixes right and left handed    
quarks. 

If instead we use a {\it transversity\/} basis, 
diagonalizing $\gamma^1$, we find,
\bea
	f_1(x) & = & \frac {1}{x} \langle P|\alpha^\dagger(xp)\alpha(xp)+
	\beta^\dagger(xp)\beta(xp)|P\rangle, \nonumber\\
	g_1(x) & = & \frac {2}{x} {\rm Re} 
	\langle P\hat e_3|\alpha^\dagger(xp)\beta(xp)|P\hat
	e_3\rangle, \nonumber\\
	h_1(x) & = & \frac {1}{x} 
	\langle P\hat e_1|\alpha^\dagger(xp)\alpha(xp) - 
	\beta^\dagger(xp)\beta(xp)|P\hat e_1\rangle.
\label{eq:h1}
\eea
Clearly $h_1$ can be interpreted as the probability to find a quark with
spin polarized along the transverse spin of a polarized nucleon minus the
probability to find it polarized oppositely.  $f_1$ still has the same
interpretation, while now
$g_1$ lacks a clear probabilistic interpretation.  
Of course the structure here is
merely that of a $2\otimes 2$ -- spin density matrix, with the assignments
$f_1\leftrightarrow 1$, $g_1\leftrightarrow\sigma_3$, and 
$h_1\leftrightarrow\sigma_1$ in the basis of helicity eigenstates.  
The remaining element, $\sigma_2$, is related to $h_1$ by rotation about the 
$\hat e_3$-axis.  In non-relativistic situations, spin and space operations 
(Euclidean boosts, {\it etc.\/}) commute and it is easy to show that 
$g_1=h_1$, so $h_1$ is a measure of the relativistic nature of the quarks 
inside the nucleon.

\begin{figure}
	\centerline{\epsffile{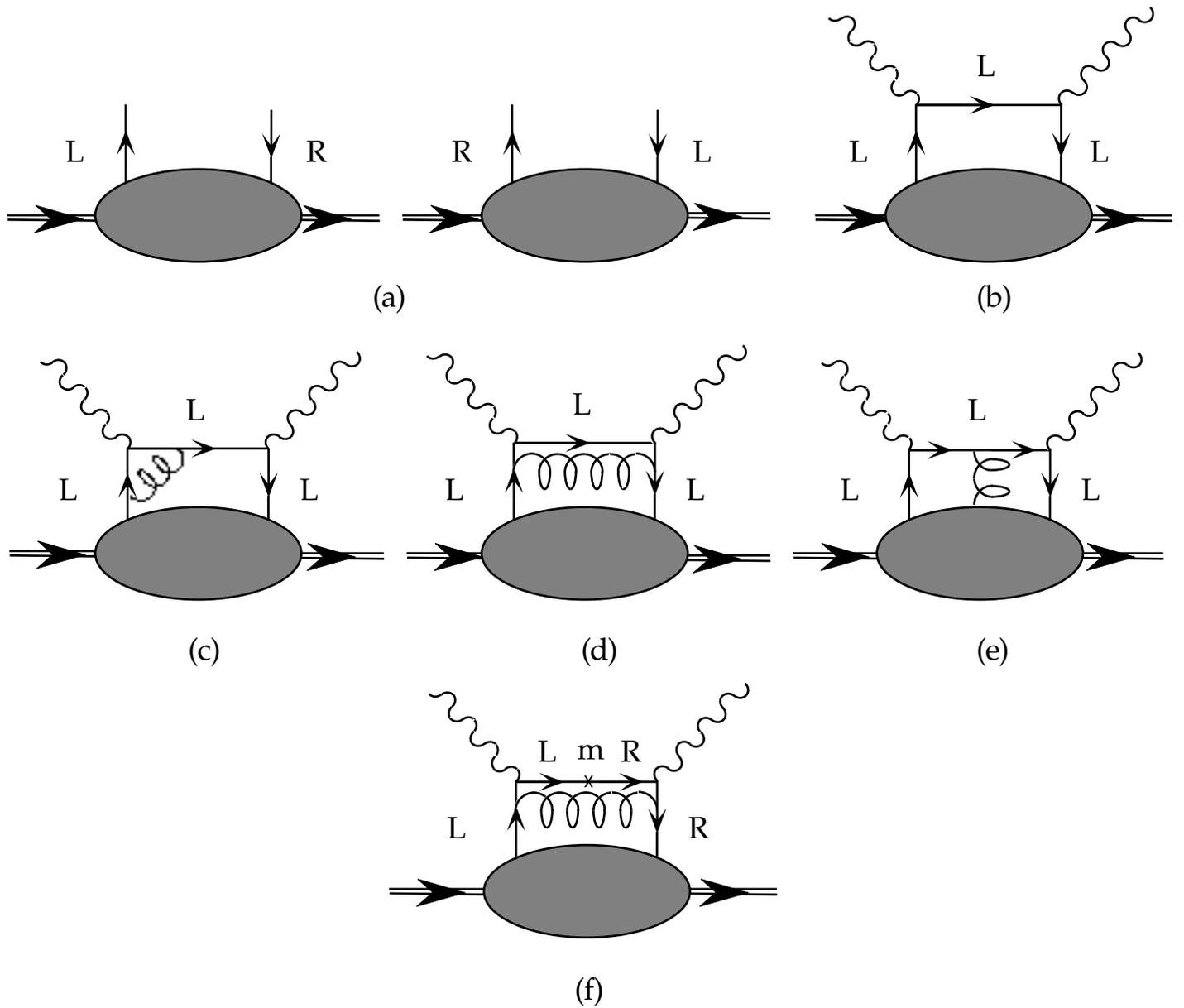}}
	\caption{{\sf Chirality in deep inelastic scattering:
	a) Chirally odd contributions to $h_1(x)$; b)-e) Chirally
	even contributions to deep inelastic 
        scattering (plus $L \leftrightarrow R$
	for electromagnetic currents); f) Chirality flip by mass insertion. }}
	\label{fig:DCH}
\end{figure}

The chirally odd structure functions like $h_1$ fig.~(fig. \ref{fig:DCH}a)
are suppressed in DIS. The dominant handbag diagram fig.~(fig. \ref{fig:DCH}b)
as well as the various decorations which generate $\log Q^2$ dependences,
$\alpha _{QCD}(Q^2)$ corrections and higher twist corrections, examples 
of which are shown in figs.~\ref{fig:DCH}c-e involve only chirally-even
quark distributions because the quark couplings to the photon and 
gluon preserve chirality. Only the quark mass insertion, 
fig.~(\ref{fig:DCH}f), flips chirality.  So up to corrections of order $m_q/Q$,
$h_1(x,Q^2)$ decouples from electron scattering.

\begin{figure}
\centerline{\epsffile{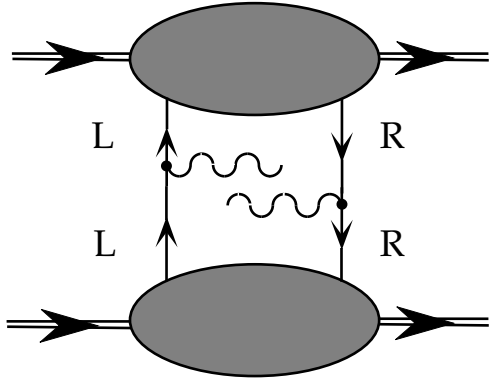}}
	\caption{{\sf Chirality in Drell-Yan 
	(plus $L \leftrightarrow R$) production of lepton pairs
	.}}
	\label{fig:YCH}
\end{figure}

There is no analogous suppression of $h_1(x,Q^2)$ in deep inelastic processes
with hadronic initial states such as Drell-Yan. The argument can be read from 
the standard parton diagram for Drell-Yan, fig.~(\ref{fig:YCH}). Although
chirality is conserved on each quark line separately, the two quarks'
chiralities are unrelated. It is not surprising, then, that
Ralston and Soper \cite{RS79} found that $h_1(x,Q^2)$ determines the
transverse-target, transverse-beam asymmetry in Drell-Yan.

\subsection{Transverse Spin in QCD}
The simple structure of eqs.~(\ref{eq:bloemkool}) and (\ref{eq:h1} )
shows that transverse spin effects and longitudinal spin effects are on a
completely equivalent footing in perturbative QCD.  On the other hand, $h_1$ 
was unknown in the early days of QCD when only deep inelastic lepton 
scattering was studied in detail.

Not knowing about $h_1$, many authors, beginning with Feynman\cite{Fey70}, 
have attempted
to interpret $g_T$ as the natural transverse spin 
distribution function. Since $g_T$ is twist-three and interaction 
dependent, this attempt led to the erroneous impression that transverse 
spin effects were inextricably associated with off-shellness, transverse
momentum and/or quark-gluon interactions
The resolution contained in the present analysis is summarized in
Table~\ref{tbl:transverse} where the symmetry between transverse and longitudinal
spin effects is  apparent. Only ignorance of $h_1$ and $h_L$ prevented the 
appreciation of this symmetry at a much earlier date.
\begin{table}[ht]
	\center{
	\begin{tabular}{|c|c|c|}\hline
			 & Longitudinal & Transverse \\   
			 &    Spin      &    Spin    \\ \hline
	Twist-2          & $g_1(x,Q^2)$ & $h_1(x,Q^2)$\\ \hline
	Twist-3          & $h_L(x,Q^2)$ & $g_T(x,Q^2)$\\ \hline  
	\end{tabular} 
	       }
	\caption{{\sf The transverse and longitudinal
	spin distribution functions through twist-three.}}
	\label{tbl:transverse}
\end{table}

Since experiments to measure $h_1$ are being planned, now is the time for 
theorists to make predictions.  At this time, however, not much is known about 
either the general behavior of $h_1$ or its form in models.  Here is a 
summary, presented in parallel with $g_1$ for the purpose of comparison.

\begin{figure}
\centerline{\epsffile{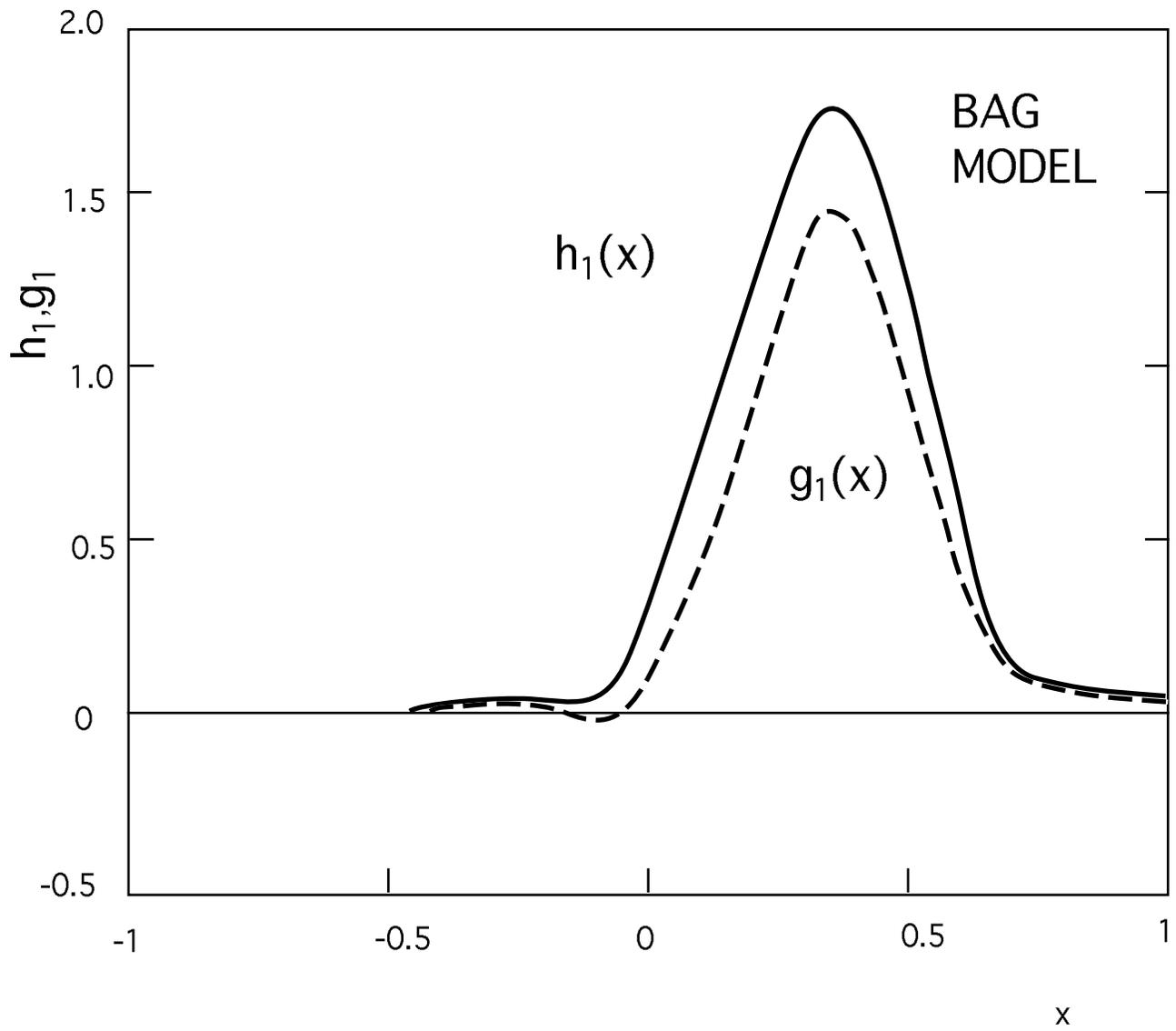}}
	\caption{{\sf Bag model calculation for $h_1$ and $g_1$ 
	from \protect\cite{JafJi1952,JafJi2005}. }}
	\label{fig:bag}
\end{figure}

\begin{itemize}
\item Inequalities:
\bea
	\vert g_1(x,Q^2) \vert & \le f_1(x,Q^2) \nonumber\\
	\vert h_1(x,Q^2) \vert & \le f_1(x,Q^2) 
\eea
for each flavor of quark and antiquark.  These follow from the positivity of 
parton probability distributions (see eqs.~(\ref{eq:bloemkool}) and 
(\ref{eq:h1})).  Another inequality, proposed by Soffer\cite{Sof94} has 
attracted attention recently,
\be
	f_1^a(x,Q^2)+g_1^a(x,Q^2)\ge 2\vert h_1^a(x,Q^2)\vert.
	\label{eq:soffer}
\ee
valid for each flavor ($a$) of quark and antiquark.
Soffer's inequality is invalidated by QCD radiative 
corrections,\cite{JafJi2402}
in much the same way as the Callan-Gross relation, $F_2=2xF_2$.  Despite this 
problem, the inequality may prove to be a useful qualitative guide to the 
magnitude of $h_1$.  A recent discussion of QCD 
radiative corrections to Soffer's 
inequality may be found in \cite{Cont}.

\item Physical interpretation: 
The structure function $h_1(x,Q^2)$ measures transversity. It is
chirally odd and related to a bilocal generalization of the tensor operator,
$\bar{q} \sigma _{\mu \nu} i\gamma _5 q$.  
On the other hand $g_1(x,Q^2)$ measures helicity.
It is chirally even and related to a bilocal generalization of the axial 
charge operator, $\bar{q} \gamma _\mu \gamma _5 q$. 
Although $h_1$ is spin-dependent, it is not directly related to the quark or 
nucleon spin.  It would be very useful 
to have a better idea of the dynamical and 
relativistic effects which generate differences between $g_1$ and $h_1$.

\item Sum rules: If we define a ``tensor charge'' in analogy to the axial 
vector charge measured in $\beta$--decay,
\be
	2S^i \delta q^a (Q^2) \equiv \langle PS \vert
	\bar{q} \sigma ^{0i} i\gamma _5 {\lambda ^a\over 2} q \Big| _{Q^2}
	\vert PS\rangle,
\ee
where $\lambda ^a$ is a flavor matrix and $Q^2$ is a 
renormalization scale, then $\delta q^a(Q^2)$ is related to an integral over
$h_1^a(x,Q^2)$,
\be
	\delta q^a(Q^2)=\int _0^1 dx ( h_1^a(x,Q^2) - h_1^{\bar{a}} (x,Q^2) )
\ee
where $h_1^a$ and $h_1^{\bar{a}}$ receive contributions from quarks and
antiquarks, respectively.
The analogous expressions for $g_1(x,Q^2)$ involve axial charges,
\bea
	2s^i \Delta q^a (Q^2) & \equiv & \langle PS \vert
	\bar{q} \gamma ^{i} \gamma _5 {\lambda ^a\over 2} q \Big| _{Q^2}
	\vert PS\rangle \\
	\Delta q^a(Q^2) & = & \int _0^1 dx 
	( g_1^a(x,Q^2) + g_1^{\bar{a}} (x,Q^2) ).
\eea
Note the contrast: $h_1(x,Q^2)$ is not normalized to a piece of the angular
momentum tensor, so $h_1$, unlike $g_1$, cannot be interpreted 
as the fraction of the nucleons' spin found on the quarks' spin. Note the
sign of the antiquark contributions: $\delta q^a$ is charge-conjugation
odd, whereas $\Delta q^a$ is charge-conjugation even.  $\delta q^a$ gets no 
contribution from quark-antiquark pairs in the {\it sea\/}.
All tensor charges
$\delta q^a$ have non-vanishing anomalous dimensions \cite{Kod79},
but none mix with gluonic operators under renormalization because they are 
chirally odd and gluon operators are even.  In contrast,
the flavor non-singlet axial charges, $\Delta q^a, a\neq0$, have vanishing
anomalous dimensions, whereas the singlet axial charge 
$\Delta q^0$ has an anomalous dimension arising from the
triangle anomaly.\cite{Kod279}

\item Evolution:  It is worth re-emphasizing that $h_1$ has unusual QCD 
evolution properties.  All of the local operators associated with $h_1$ have 
non-vanishing leading order anomalous dimensions.  On the other hand, no gluon 
operators contribute to $h_1$ in any order, because $h_1$ is chiral odd.  So 
$h_1$ is a {\it non-singlet\/} structure function 
-- it evolves homogeneously with $Q^2$, but none of its moments are $Q^2$ 
independent.

\item Models: $h_1$ and $g_1$ are identical in non-relativistic 
quark models, but
differ in relativistic models like the bag model - see fig.~(\ref{fig:bag}).
\end{itemize}
\subsection{Twist-Three: Physics with $g_2(x,Q^2)$}
There are several reasons to be particularly interested in the transverse spin 
dependent structure function, $g_2(x,Q^2)$ 
\begin{enumerate}
\item it can be measured in deep inelastic scattering; 
\item it is unique among higher twist distributions in that 
it dominates the cross section in a specific kinematic domain --- at $90^\circ$
all twist-two effects decouple, see eq.~(\ref{eq:spindepxsection}); 
\item it is related to interesting quark gluon matrix elements; 
\item it should obey an interesting sum rules.
\end{enumerate}
For a review of the properties of $g_2$, see ref.~\cite{Jafg2}
\subsubsection{Operator Product Expansion}

The spin structure functions $g_1$ and $g_2$ parameterize the antisymmetric 
part of the hadronic tensor $W_{\mu\nu}$ as shown in eq.~(\ref{eq:WA}). 
Applying the methods of \S 2 we can relate the moments of the 
antisymmetric part of $W_{\mu\nu}$ to the matrix elements of quark operators 
accurate through twist-three.  Consider the antisymmetric
part of the forward Compton amplitude $T_{\mu\nu}$: 
\bea
	T_{[\mu\nu]} &=& \frac 
	{1}{2}\left[T_{\mu\nu}-T_{\nu\mu}\right],\nonumber\\
	&=& \left<PS|{\cal T}_{[\mu\nu]}(q)|PS\right>,\quad{\rm 
	where}\nonumber \\
	{\cal T}_{[\mu\nu]}(q)&=&\int d^4\xi e^{iq\cdot\xi} 
	T(J_{[\mu}(\xi)J_{\nu]}(0)).
\eea
The leading twist (twists-two and -three) contributions come 
from the series of operators 
\be
	{\cal T}_{[\mu\nu]} = i \epsilon_{\mu\nu\lambda\sigma}q^{\lambda}
	    \sum\limits_{n=0,2,4\dots} q^{\mu_1}\dots q^{\mu_n} 
	    \Theta^\sigma_{\{\mu_1\dots\mu_n\}} 
	    \left(\frac {-2}{q^2}\right)^{n+1}C_n(q^2). 
\ee
with
\be
	\Theta_{\sigma \{\mu_1\dots\mu_n\}}\equiv  i^n \bar{\psi}
       \gamma_{\sigma}\gamma_5 D_{\{\mu_1}\dots D_{\mu_n\}} \psi - {\rm 
	traces},
	\label{eq:mixedsym}
	\footnotemark          
\ee
	\footnotetext{The terms denoted ``traces'' are whatever is necessary 
	to subtract from the displayed term in order to render the resulting 
	operator traceless and will be suppressed in the following.} 
Here $\{ \hspace{2mm}, \hspace{2mm} \}$ denotes symmetrization and 
$[ \hspace{2mm},\hspace{2mm} ]$ denotes antisymmetrization of 
the enclosed indices.  The string of symmetric indices originates in the 
expansion of the bilocal operator about $\xi=0$, and contraction of each index 
with $q^\mu$ selects the totally symmetric part.  On the other hand, the 
index on the $\gamma$--matrix is not symmetrized.
$C_n(q^2)$ is the 
coefficient function and is unity to lowest order.
Flavor factors such as squared charges are suppressed. 

Eq.~(\ref{eq:mixedsym}) can be split into a totally symmetric part and a part 
of lower symmetry,
\be
	\Theta_{\sigma \{\mu_1\dots\mu_n\}}\equiv  
	\Theta_{\{\sigma \mu_1\dots\mu_n\}}  +  \Theta_{[\sigma 
	\{\mu_1]\mu_2\dots\mu_n\}},
\ee
where the first term has twist-two (dimension $n+3$, spin $n+1$):   
\be
	\Theta_{\{\sigma \mu_1\dots\mu_n\}} = 
	\frac{1}{n+1}\left\{
	\Theta_{\sigma \{\mu_1\dots\mu_n\}}
       +\Theta_{\mu_1\{\sigma\mu_2\dots\mu_n\}}       
       +\Theta_{\mu_2\{\mu_1\sigma\dots\mu_n\}}+\dots
	\right\},
\ee
and the second term, with no totally symmetric part, has twist-three
(dimension $n+3$, spin $n$):      
\bea
	\Theta_{[\sigma \{\mu_1]\mu_2\dots\mu_n\}} &=& 
	\frac{1}{n+1}\Big\{
	\Theta_{\sigma \{\mu_1\dots\mu_n\}}
       -\Theta_{\mu_1\{\sigma\mu_2\dots\mu_n\}}\nonumber\\       
       &+&\Theta_{\sigma\{\mu_1\dots\mu_n\}}
       -\Theta_{\mu_2\{\mu_1\sigma\dots\mu_n\}}
       +\dots\Big\}.
\eea
The proton matrix elements of these operators are 
\bea
	\left<PS|\Theta_{\{\sigma \mu_1\dots\mu_n\}}|PS\right> &=& 
	\left\{S_{\sigma} P_{\mu_1}\dots P_{\mu_n} 
	+ S_{\mu_1}P_{\sigma}\dots P_{\mu_n} + 
	\dots\right\}\frac{a_n}{n+1},\nonumber\\
	\left<PS|\Theta_{[\sigma \{\mu_1]\mu_2\dots\mu_n\}}|PS\right> &=&
	\left\{(S_{\sigma}P_{\mu_1}-S_{\mu_1}P_{\sigma})P_{\mu_2}\dots P_{\mu_n}
	\right.\nonumber\\  
	&+& \left.(S_{\sigma}P_{\mu_2}-
	S_{\mu_2}P_{\sigma})P_{\mu_1}\dots P_{\mu_n} 
	+\dots \right\}\frac{d_n}{n+1}.
\eea
\subsubsection{Wandzura -- Wilczek Decomposition of $g_2$}

Extracting the relation between these matrix elements and the moments of $g_1$ 
and $g_2$ is an exercise in the methods of \S 2.
Substituting these matrix elements into the definition of $T_{[\mu\nu]}$, 
writing a dispersion relation in terms of $g_1$ and $g_2$ and equating terms 
in the Taylor expansion in $1/x$ in the BJL--limit, we find
\bea
	\int\limits_0^1 dx x^n g_1(x,Q^2) &=& \frac 14 a_n \quad n 
	= 0,2,4,\dots,\nonumber\\
	\int\limits_0^1 dx x^n g_2(x,Q^2) 
	&=& \frac 14 \frac{n}{n+1}\left(d_n - a_n)
	\right),\quad n=2,4\dots, 
	\label{eq:koe}
\eea
Notice that the same operators which determine $g_1$ make an appearance in the 
moments of $g_2$.  It follows that $g_2$ can be decomposed into two parts, one 
which is fixed by $g_1$, and another --- the ``true twist-three'' part --- 
associated with the operator of mixed symmetry,
\be
	g_2(xQ^2) = -g_1(x,Q^2) + \int\limits_x^1\frac {dy}{y} g_1(y,Q^2) + 
	\bar g_2(x,Q^2). 
\ee
Wandzura and Wilczek proposed this decomposition in 1977.\cite{WW77}  They 
went further and suggested that $\bar g_2$ might be zero.  From another 
perspective, $\bar g_2$ is the {\it interesting} part of $g_2$.  It's moments,
\be
	\int\limits_0^1 dx x^n \bar g_2(x,Q^2) = \frac{n}{4(n+1)}d_n(Q^2),
\ee
are twist-three and 
measure quark-gluon correlations, as was pointed out 
by Shuryak and Vainshteyn\cite{SV82}. 
They showed that the $QCD$ equations of motion  
can be used to trade the antisymmetry of 
$\Theta_{[\sigma \{\mu_1]\mu_2\dots\mu_n\}}$ for factors of the 
gluon field strength $G_{\mu\nu}$ and the QCD coupling constant 
$g$. The result is:
\bea
	\Theta_{[\sigma \{\mu_1]\mu_2\dots\mu_n\}} &=&
	\frac{g}{8} {\cal S}_n\left\{ \sum\limits_{l=0}^{n-2} 
	i^{n-2}\bar{ \psi} D_{\mu_1}..D_{\mu_l} \tilde G_{\sigma\mu_{l+1}} 
	D_{\mu_{l+2}}..D_{\mu_{n-1}}\gamma_{\mu_n}\psi \right.\nonumber \\
	&+&
	\left.\frac{1}{2} \sum\limits_{l=0}^{n-3}i^{n-3} 
	\bar {\psi} D_{\mu_1}..D_{\mu_l} (D_{\mu_{l+1}} G_{\sigma\mu_{l+2}}) 
	D_{\mu_{l+3}}..D_{\mu_{n-1}}\gamma_{\mu_n}\gamma_5\psi \right\}, 
\label{eq:sv}
\eea
where $\tilde G_{\alpha\beta}=\frac 12 \epsilon_{\alpha\beta\lambda\sigma}
G^{\lambda\sigma}$ and ${\cal S}_n$ symmetrizes the indices $\mu_1\dots\mu_n$. 

Eq.~(\ref{eq:sv}) is quite formidable.  A simpler example might help explain how 
manipulation of the equations of motion exposes the interaction dependence of 
higher twist operators.  Consider the twist-three operator,
\be
	X_{\mu\nu}= \bar {\psi} D_{\mu}D_{\nu} \psi
\ee
as an example.  $X$ is clearly twist-three (dimension five, 
spin no greater than two).  One might be tempted to make a ``parton model'' 
for the matrix element of $X$, by replacing $D_\mu\rightarrow\partial_\mu$ and 
evaluating $X$ in a beam of collinear quarks.  That, however, would be a 
mistake, since application of the identity $D_{\mu} = \frac 12\{\gamma_{\mu},
D \hspace{-2.5mm} / \}$, and the QCD equations of motion, $D \hspace{-2.5mm} / 
\psi =\bar{\psi} D \hspace{-2.5mm} /  = 0$ 
and $[D_{\mu},D_{\nu}]=gG_{\mu\nu}$, yields
\be
	X_{\mu\nu} = \frac{g}{2} \bar{\psi}G_{\mu\lambda}\gamma^{\lambda} 
	\gamma_{\nu}\psi. 
\ee
So it is clear that $X$ and its matrix elements are interaction dependent and 
measure a quark-gluon correlation in the target hadron.
This is a special case of a general result that operators 
of twist $\ge 3$ can always be written in a form in which they are manifestly 
interaction dependent. In particular one has:  
\be
	\int _0^1 x^2 \bar{g}_2(x,Q^2) \propto
	\langle PS \vert {1\over 8} g {\cal S}_{\mu_1 \mu_2} \bar{\psi}
	\tilde{G}_{\sigma \mu_1} \gamma _{\mu_2} \psi \vert PS\rangle.
\ee
Model builders or lattice enthusiasts who want to predict $g_2$ must confront 
such matrix elements.

\subsubsection{The Burkhardt Cottingham Sum Rule} 

A striking consequence of the light-cone analysis of $g_2$ is the apparent sum 
rule,\cite{BC70} 
\be
\int\limits_0^1 dx g_2(x,Q^2) = 0,
\ee
which follows directly from eq.~(\ref{eq:koe}) with $n=0$.  
If true, the sum rule 
requires $g_2$ to have a node (other than at $0$ and $1$).

The Burkhardt-Cottingham (BC) Sum Rule looks to be a 
consequence of rotation invariance.  To see this, return to 
eq.~(\ref{eq:nucstructure}) and consider the $\gamma_\mu\gamma_5$ case in the 
laboratory frame (where $S^0=0$).  First, set $\hat S = M\hat e_3$, 
contract the 
free Lorentz index with $n^\mu$ and integrate overall $x$, leaving
\be
	\int\limits_{-1}^1 dx g_1(x,Q^2)=\langle P \hat e_3\vert\bar 
	q(0)\gamma^3\gamma_5 q(0)\vert_{Q^2}\vert P \hat e_3\rangle.
\ee
Next repeat the process with $\hat S = M\hat e_1$ and $\mu=1$, with the 
result,
\be
	\int\limits_{-1}^1 dx g_T(x,Q^2)=\langle P \hat e_1\vert\bar
	q(0)\gamma^1\gamma_5 q(0)\vert_{Q^2}\vert P \hat e_1\rangle.
\ee
The right hand sides of these two equations are equal in the rest frame 
by rotation invariance, so
\be
	\int_{-1}^1 dx g_T(x,Q^2) = \int_{-1}^1 dx g_1(x,Q^2),
\ee
whence
\be
	\int_{-1}^1 dx g_2(x,Q^2) = 0,
	\label{eq:BC}
\ee
apparently a consequence of rotation invariance.  

The subtlety in this derivation is that the integral in eq.~(\ref{eq:BC})
goes from $-1$ to $1$ including $x=0$.  As we have 
defined it, $g_2(x,Q^2)$ is the limit of a function of $Q^2$ and $\nu$ and 
therefore might contain a distribution ($\delta$--functions, {\it etc.\/}) at
$x=0$.  Suppose $g_2$ has a $\delta$--function contribution at $x=0$,
\be
	g_2(x,Q^2)=g_2^{\rm observable}(x,Q^2)+ c\delta(x).
\ee
Then since experimenters cannot reach $x=0$, the BC sum rule reads
\be
	\int_0^1 dx g_2^{\rm observable}(x,Q^2) = -{1\over 2}c,
\ee
which is useless.

This pathology --- a $\delta$--function at $x=0$ --- is not as arbitrary as it 
looks.  Instead it is an example of a disease known as a {\it``$J=0$ fixed pole 
with non-polynomial residue''.\/} First studied in Regge 
theory,\cite{FoxFr70,BGJ}
a $\delta (x)$ 
in $g_2(x,Q^2)$ corresponds to a {\it real constant term in a spin flip 
Compton amplitude which persists to high energy\/}.  
There is no fundamental reason to exclude such a constant. 
On the other hand the sum rule is known to be satisfied in QCD perturbation 
theory through order $O(g^2)$. The sum rule has been studied recently by 
several groups who find no evidence for a
$\delta (x)$ in perturbative QCD.\cite{AnKoun,Burk,Alta}
So at least provisionally, we must regard this as a reliable sum rule.
At least one other sum rule of interest experimentally, the Gerasimov, Drell, 
Hearn Sum Rule for spin dependent Compton scattering has the same potential
pathology. For further discussion of the BC sum rule see ref.~\cite{Jafg2}.

\subsubsection{The Evolution of $g_2$}

The following discussion can be regarded as a ``theoretical interlude'' and 
may be omitted by the casual reader.

The $Q^2$ evolution of quark distribution functions is an unavoidable
complication in perturbative QCD.  Data are inevitably taken at 
different $Q^2$, making it
difficult to evaluate sum rules (which require data at some definite $Q^2$)
without some information about $Q^2$--evolution.
At leading twist the subject is well
understood and we have ignored it.  The evolution of $g_2$ is more
complicated.  
Eq.~(\ref{eq:nucstructure}) 
provides as deceptively simple operator representation of $g_T$,
\bea
	g_T(x,q^2) &=& g_1(x,q^2) + g_2(x,q^2) \nonumber \\
           &=& \int {d\lambda\over 4\pi}e^{i\lambda x}
	\langle PS_\perp\vert\bar\psi(0) S\hspace{-2mm}/_\perp\gamma_5\psi
	(\lambda n)\vert_{Q^2}\vert PS_\perp\rangle
\label{eq:gt}
\eea
It looks as though $g_T$ is determined by a single operator which evolves
homogeneously.  However, the operator in eq.~(\ref{eq:gt}) is equivalent to
the series of quark gluon operators given in eq.~(\ref{eq:sv}).  
Worse still, all
these operators mix under renormalization.  The number of operators that mix
grows linearly with $n$.  Back in the 1970's, Ahmed and Ross calculated the
anomalous dimension matrix element for the evolution of the transverse
quark axial current operator in eq.~(\ref{eq:gt}).\cite{AhmRo} 
In 1982 Shuryak and
Vainshteyn pointed out that this was but one element of an $n\times n$
anomalous dimension {\it matrix\/}\cite{SV82}.  Later Ratcliffe, Lipatov
{\it et al.\/} and others 
calculated the full anomalous dimension matrix.\cite{Rat,Lip,JiChou,Kod3}

\begin{figure}
\centerline{\epsffile{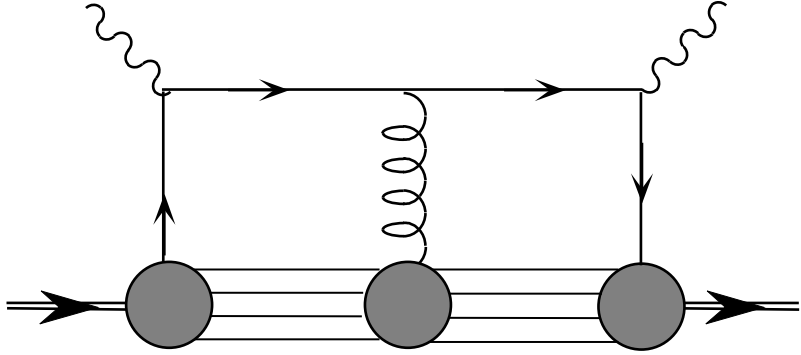}}
	\caption{{\sf Quark/gluon diagram contributing to the two-variable
	parton distribution functions $G$ and $\tilde G$.}}
	\label{fig:g2partons}
\end{figure}

Underlying
this matrix renormalization group evolution is a somewhat simpler parton
picture.  The fundamental objects for studying the evolution of $g_2$ are
{\it two-variable\/} parton distribution functions defined by double
light-cone fourier transforms\cite{JafJi91}.
\bea
	\int {d\lambda\over 2\pi}{d\mu\over 2\pi}
	e^{i\lambda x + i\mu(y-x)}\langle
	PS\vert\bar q(0)iD^\alpha(\mu n)n\hspace{-2mm}/q(\lambda n)
	\vert_{Q^2}\vert
	PS\rangle &=&2i\varepsilon^{\alpha\beta\sigma\tau}n_\beta 
	S_\sigma p_\tau G(x,y,Q^2) 
	\nonumber\\ 
	&+&\ldots \nonumber \\
	\int {d\lambda\over 2\pi}{d\mu\over 2\pi}e^{i\lambda x 
	+ i\mu(y-x)}\langle
	PS\vert\bar q(0)iD^\alpha(\mu n)n\hspace{-2mm}/\gamma_5q(\lambda
	n)\vert_{Q^2}\vert PS\rangle 
	&=&2i\varepsilon^{\alpha\beta\sigma\tau}n_\beta
	S_\sigma p_\tau\tilde G(x,y,Q^2)\nonumber\\
	&+&\ldots \nonumber \\
	\label{eq:openg2}
\eea
where $\ldots$ denote other tensor structures of twist greater than three. 
The functions $G(x,y,Q^2)$ and $\tilde G(x,y,Q^2)$ are generalizations of
parton distributions describing the amplitude to find quarks and
gluons with momentum fractions $x$, $y$ and $x-y$ in the target nucleon. 
They can be represented diagrammatically as in
fig.~(\ref{fig:g2partons}).
These ``higher twist 
distribution functions'' share many properties with leading twist.  Only good 
light-cone components and 
collinear momenta appear -- transverse momentum has been eliminated in favor 
of the interactions which generate it.  The variables $x$, $y$, and $x-y$ 
take on only physical values, $-1\le (x, y, x-y)\le 1$, corresponding to 
emission or absorption of quarks, antiquarks or gluons.\cite{Jaf1}

At any value of $Q^2$, $g_T(x,Q^2)$ can be projected out by integrating out 
the variable $y$,
\be
	g_T(x,Q^2)={1\over 2x}\int_{-1}^1 dy\{G(x,y,Q^2)-G(y,x,Q^2)+\tilde G(x,y,Q^2)
	-\tilde G(y,x,Q^2)\}
\ee
The reason for introducing $G$ and $\tilde G$ is that they obey natural 
generalizations of the Gribov-Lipatov-Altarelli-Parisi (GLAP)
evolution equations.  Schematically,
\be
	{d\over d\ln Q^2}G(x,y,Q^2)=
	{\alpha_{QCD}\over 2\pi}\int_x^1\int_y^1 dx'dy'{\cal 
	P}(x,x',y,y')G(x',y',Q^2)
\ee
for some ``splitting function'', ${\cal P}$.  In order to evolve $g_T$ from 
one $Q^2$ to another it is necessary to know $G$ and $\tilde G$ as a function 
of both $x$ and $y$.  However, measurement of $g_T$ does not supply that 
information.  $g_T$ is, in a sense, a ``compressed'' structure function.  
Information essential to evolution has been integrated out of it.
The associated ``open'' distributions, $G$ and $\tilde G$ evolve simply.  $g_T$ 
does not.

The situation is not entirely hopeless, however.  It appears that the 
evolution of $g_2$ may simplify in the $N_c\rightarrow\infty$ limit, 
especially at large $x$.\cite{ABH}  Ali, Braun and Hiller showed that in this 
limit $\bar g_2$ obeys a standard GLAP equation, albeit with non-standard 
anomalous dimensions.  As measurements of $g_2$ improve, theorists will be 
forced to expend more effort to clarify the nature of $Q^2$ evolution of 
$g_T$.

\subsubsection{The Shape of $g_2$}

With the first measurements of $g_2$ now available,\cite{g2expt} and more 
accurate measurements expected soon, it seems timely to review model 
predictions.  First, here is an agenda of progressively more detailed 
questions for experimenters as they relate to theory
\begin{itemize}
\item Is $g_2$ zero?  

The answer is in on this one:  the new SLAC data show $g_2\ne 0$.

\item Is $\int_0^1 dx g_2(x,Q^2)$ zero as required by the BC sum rule?

The SLAC data are consistent with the BC sum rule, albeit with large errors.

\item If $\int_0^1 dx g_2(x,Q^2)=0$, then $g_2$ must have at least one 
non-trivial zero (besides $x=1$ and perhaps $x=0$).  Is there just one such 
node?  

The SLAC data are consistent with one node, again within large errors.

\item If there is only one non-trivial node, what is the sign of 
\be
	M_2[g_2]\equiv\int_0^1 dx g_2(x,Q^2)?
\ee

With only one non-trivial node, the sign of $M_2$ determines the gross 
structure of $g_2$.  If $M_2$ is positive $g_2$ is negative at small $x$ and 
positive at large $x$, and {\it vica versa\/} if $M_2$ is negative.
The Wandzura-Wilczek 
(WW) contribution to $g_2$ gives $M_2^{WW}<0$.

The SLAC data favor $M_2<0$.

\item Is there a signal for $\bar g_2$ or does the WW
contribution account for all of $g_2$?

The SLAC data agree rather well with the WW prediction, however 
the accuracy is such that a fairly significant $\bar g_2$ term would not yet 
be detectable.  Some observers claim to be able to see a deviation from 
WW with $\overline M_2>0$.

\item As more accurate data become available it will be possible to subtract 
away the WW contribution to reveal $\bar g_2$.
\end{itemize}

\begin{figure}
\centerline{\epsffile{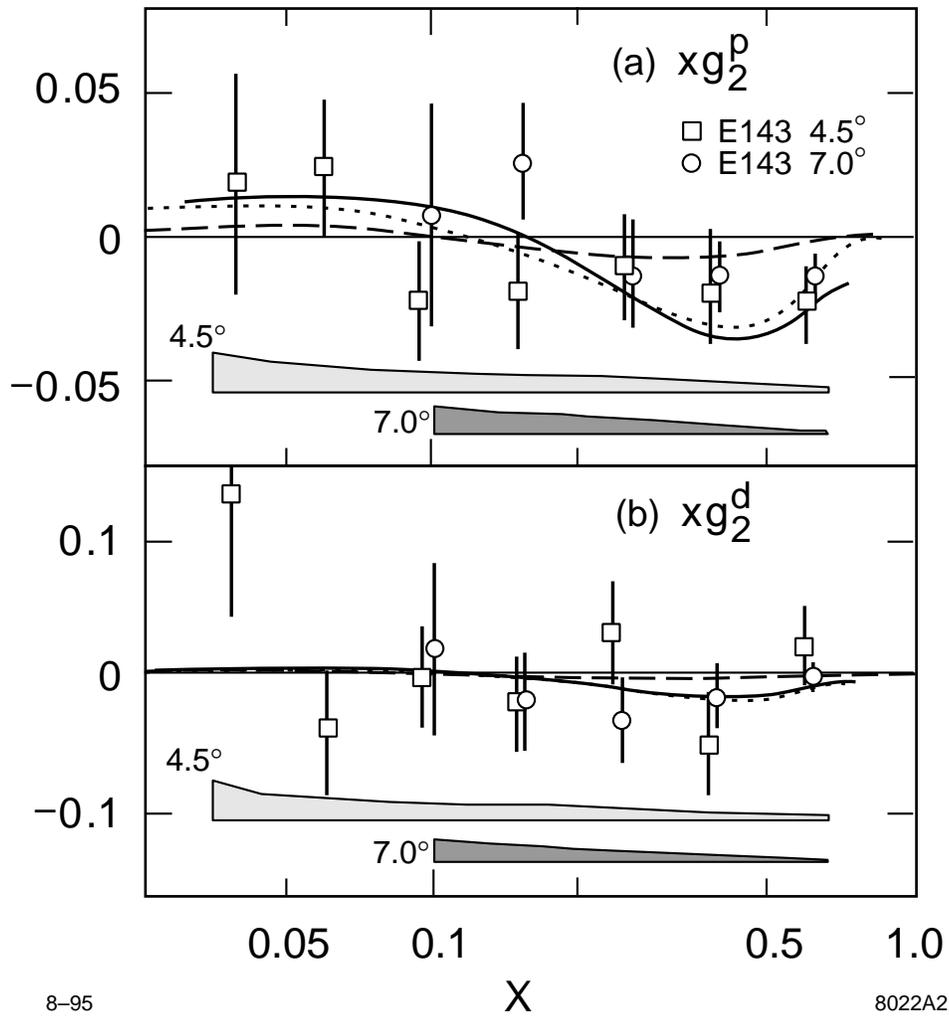}}
	\caption{{\sf Data on $g_2$ from the SLAC E143 collaboration compared
	with the WW contribution and the bag model estimates of Strattmann.}}
	\label{fig:E143g2}
\end{figure}

$g_2$ has been studied in quark models and using QCD Sum Rules.  
Perhaps the most thorough analysis, 
including QCD evolution, has been performed by Strattman.\cite{Str}  He has 
taken several versions of the MIT bag model, calculated $g_2$ and then evolved 
the resulting distribution using the method of Ali, Braun and Hiller 
to experimentally interesting $Q^2$.  His 
estimates for the proton are shown in comparison with the SLAC data in
fig.~(\ref{fig:E143g2}).  

His estimates of 
$\bar g_2$ are small compared to $g_2^{WW}$ and cannot be excluded by the
existing data.  The neutron $g_2$ is very small 
in quark and bag models for the same reason that the neutron's
$g_1$ is so small:  the correlations of charge and spin in the $SU(6)$ symmetric
neutron tend to  cancel for the neutron.  The second moment of $\bar g_2$ has
also been estimated using QCD sum  rules.\cite{QCDsrg2}  Surprisingly
ref.~\cite{QCDsrg2} finds $\overline  M_2[g_2^{\rm proton}]$ consistent with zero
and $\overline M_2[g_2^{\rm  neutron}]$ negative and significantly different
from zero.  Existing data cannot rule out this behavior --- we shall have to 
wait for the HERMES facility at HERA to provide more accurate data.
$\bar g_2$  depends on quark gluon correlations within the nucleon,
which are not likely to be perfectly described in such simple models, so these 
predictions should be more as a guide than a prediction of expected behavior.

\section{The Drell-Yan Process}
\setcounter{equation}{0}

Although deep inelastic scattering has been the source of much insight into
nucleon structure, it has many limitations:  polarized targets can only be
probed with electromagnetic currents (neutrino scattering from polarized
targets being impractical); gluon distributions do not couple directly,
but instead must be inferred from careful study of the evolution of quark
distributions; chiral odd distributions like $h_1$ decouple.  None of these
limitations afflict deep inelastic processes with hadron initial states. 
These include not only the original Drell-Yan process, $pp\rightarrow
\ell\bar\ell+X$ via one photon annihilation, but also generalizations to
annihilation via $W^{\pm}$ and $Z^0$, and parton-parton scattering
resulting in jet production or production of hadrons or photons at high
transverse momentum.  QCD predictions for all of these processes are
obtained by combining quark/gluon distribution and fragmentation functions
with hard scattering amplitudes calculated perturbatively.  This formalism
is treated in standard texts --- our object is to explain how the spin,
twist and chirality classification developed in previous sections can be
applied to Drell-Yan processes.  In this section we will treat only
the ``classic'' Drell-Yan process, $pp\rightarrow \gamma^* + X\rightarrow
\ell\bar\ell +X$.  The generalizations of our spin/twist/chirality
analysis to other processes is fairly straightforward and yields
interesting predictions for a variety of processes.  The original treatment of 
polarized Drell-Yan at this level was made by Ralston and Soper.\cite{RS79}

When last we considered Drell-Yan in \S 2.4, we noted that the
dominance of the leading light-cone singularity could be overwhelmed by
rapid phase oscillations in the matrix element.
Fig.~(\ref{fig:aap}) shows a contribution to Drell-Yan which is obviously
proportional to the product of quark distribution functions each of which
has the form of eq.~(\ref{eq:generalbilocal}).  Each oscillates rapidly
along the light-cone $\propto e^{i\alpha P\cdot\xi}$, where $P$ is either
of the two external hadron momenta.  In this section we will use the
formalism of the previous two sections to compute fig.~(\ref{fig:aap}), to
show that it generates large (scaling) contributions to Drell-Yan, and to
classify them with respect to spin and twist.
\subsection{Operator Analysis}
We begin with the Drell-Yan tensor, $W_{\mu\nu}$, from
eq.~(\ref{eq:braak}), with momenta and spin more carefully labeled (and
{\small IN} labels suppressed):
\be
	W_{\mu \nu}={s\over 2} \int d^4 \xi e^{iq\cdot \xi }
	\langle P_A S_A P_B S_B \vert J_\mu (0) J_\nu (\xi) \vert
	P_A S_A P_B S_B \rangle.
\label{eq:wmunufull}
\ee
\begin{figure}
\centerline{\epsffile{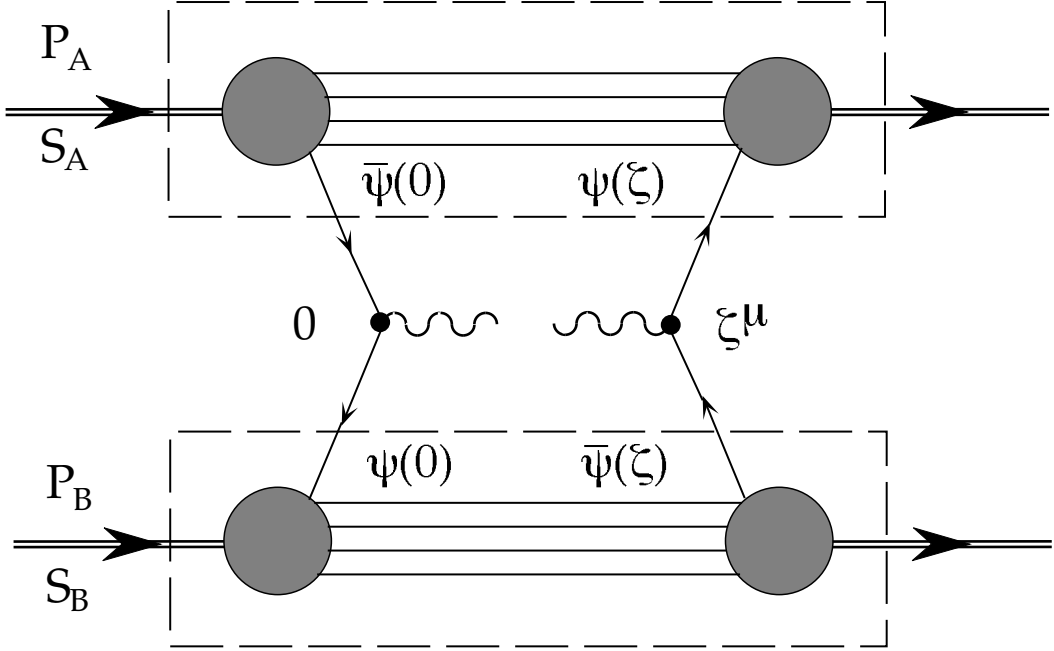}}
	\caption{{\sf Dominant contribution to the Drell-Yan process.}}
	\label{fig-DY}
\end{figure}        
The dominant contribution for the Drell-Yan process is shown in
fig.~(\ref{fig-DY}).  From the diagram it seems that $W_{\mu\nu}$ reduces to a
product of light-cone quark correlation functions, as indicated by the markings
on the figure.  The essential features are 1) that nothing propagates between
the two currents, so there is no singularity as
$\xi^\mu\rightarrow 0$; and 2) each current has a quark line landing on
each hadron.  The latter suggests that the diagram can be factored into
products of quark-hadron amplitudes by making a Fierz transformation in
order to couple the spin, color and flavor indices on the quarks in a more
appropriate order.  First we write out the currents in terms of quark
fields, limiting ourselves to terms symmetric in $\mu\leftrightarrow\nu$
(which survive contraction with the lepton tensor, $L_{\mu\nu}$),
\be
	J^{\{ \mu} (0) J^{\nu \}} (\xi )=
	-\bar{\psi} _k(0) \psi _l (\xi )
	\bar{\psi}_i(\xi ) \psi _j (0) 
	({\bf 1} \gamma ^{\{ \mu} )_{kj}
	({\bf 1} \gamma ^{\nu \}} )_{il}.
\ee
Note the $-$ sign from anticommuting quark fields.
We wish to recouple indices so quarks acting in the same hadron are
coupled to one another --- $(kj)(il)\rightarrow (kl)(ij)$.  The color
Fierz transformation is simple,
\be
	{\bf 1}_{kj}{\bf 1}_{il}=
	{1\over 3} ({\bf 1}_{kl}{\bf 1}_{ij} + 2  
	{\lambda ^a_{kl} \over 2}
	{\lambda ^a_{ij} \over 2}),
\label{eq:colorfierz}
\ee
where ${\bf 1}$ is the $3\times 3$ unit matrix in color space.  [This
result is easily derived by multiplying both sides by the matrices 
${\bf 1}_{kj}$ and $\lambda_{kj}$, in turn, and using multiplication
properties of the $\lambda$'s.]  
The second term in eq.~(\ref{eq:colorfierz}) vanishes when matrix elements
are taken in color singlet hadron states.
The Fierz transformation for the Dirac matrices is more complicated but
better known,
\bea
	\left( \gamma ^{\{ \mu} \right) _{kj}
	\left( \gamma ^{ \nu\}} \right) _{il} 
	&=& {1\over 4} \Bigg[
	\left( \gamma ^{\{ \mu} \right) _{kl}
	\left( \gamma ^{ \nu\}} \right) _{ij} + 
	\left( \gamma ^{\{ \mu} \gamma _5 \right) _{kl}
	\left( \gamma ^{ \nu\}} \gamma _5 \right) _{ij}  \nonumber\\
	&+& \left( \sigma ^{\alpha \{ \mu } \right) _{kl}
	\left( \sigma ^{ \nu\}}_{\ \alpha} \right) _{ij} \Bigg]
	+{1\over 4} g^{\mu \nu} \left[ \cdots \right],
	\label{eq:diracfierz}
\eea
where the omitted terms are traces of the terms shown explicitly.
One is left with bilocal light-cone correlation functions of the form:
vector$\times$vector, axial vector$\times$axial vector and
tensor$\times$tensor.  The flavor Fierz transformation is straightforward
and is left to the reader.

After reorganizing the product of currents to group fields that act in the
same hadron together, the matrix element in eq.~(\ref{eq:wmunufull})
factors into the product of quark distribution functions.  
However, the coordinate interval $\xi^\mu$ is not
constrained to be lightlike.  In ref.~\cite{JafJi2005} each bilocal matrix
element is expanded about the tangent to the light-cone defined by the
large momentum $P_A$ or $P_B$, and it is shown that the light-cone
contributions dominate.  The matrix element in hadron $A$ is a non-trivial
function of $P_A\cdot\xi$ along the surface $\xi^2=0$ and $P_B\cdot\xi=0$,
while the matrix element in hadron $B$ is a non-trivial function of
$P_B\cdot\xi$ along the corresponding tangent to the light-cone.  We refer
to ref.~\cite{JafJi2005} for further details.  Accepting this, $W_{\mu\nu}$
reduces to the product of quark distribution functions, one for hadron
$A$, another for hadron $B$.  The $V\times V$, $A\times A$, $T\times T$
structure of eq.~(\ref{eq:diracfierz}) is reflected in the structure of
the resulting product of distribution functions.
The $V\times V$ part is described by:
\be
	W_V^{\mu \nu} = {1\over 3} (2\pi)^4 \delta ^2 (\vec{Q}_\perp)
	\sum _a e_a^2 f_1^a(x) f_1^{\bar{a}} (y) 
	\left( p_A^\mu p_B ^\nu + p_A^\nu p_B ^\mu - g^{\mu \nu}
	p_A \cdot p_B \right),
\label{eq:vv}
\ee
where
\bea
	P_A^\mu &=& p_A^\mu +{M^2\over s} p_B^\mu \\
	P_B^\mu &=& p_B^\mu +{M^2\over s} p_A^\mu,
\eea
with $p_A^2=p_b^2=0$ and $2p_A\cdot p_B=s$.
This result is valid up to corrections of order $1/Q^2$ or $1/s$.  There
are no order $1/Q$ or $1/\sqrt{s}$ (twist-three) corrections ---  another
example of the general result that only even twists appear in spin average
deep inelastic phenomena.  The $\delta(Q_\perp^2)$ in eq.~(\ref{eq:vv})
reflects the fact that to leading twist and leading order in
$QCD$ radiative corrections, it is as though all partons move parallel to
the parent hadron.  Our calculation can only be used to study observables
that integrate over $\vec Q_\perp$.  Eq.~(\ref{eq:vv}) is the original
result of Drell and Yan:  quarks that annihilate to form a virtual photon
of squared-mass $Q^2$ and three-momentum $Q^3$ must have $xy=Q^2/s$ and
$x-y=2Q^3/\sqrt{s}$.

In the same spirit the $A\times A$ contribution is
\bea
	W_A^{\mu \nu} &=& -{1\over 3} (2\pi)^4 \delta ^2 (\vec{Q}_\perp)
	\Bigg\{ \sum _a e_a^2 g_1^a(x) g_1^{\bar{a}} (y)
	{p_A\cdot S_B p_B\cdot S_A \over (p_A\cdot p_B)^2} 
	\left( p_A^\mu p_B ^\nu + p_A^\nu p_B ^\mu - g^{\mu \nu}
	p_A\cdot p_B \right),\nonumber\\ 
	&+& \sum _a e_a^2 g_1^a(x) g_T^{\bar{a}} (y)
	{p_B\cdot S_A \over p_A\cdot p_B} 
	\left( p_A^\mu S_{B\perp} ^\nu + p_A^\nu 
	S_{B\perp} ^\mu \right) \\ \nonumber
	&+& \sum _a e_a^2 g_T^a(x) g_1^{\bar{a}} (y)
	{p_A\cdot S_B \over p_A\cdot p_B} 
	\left( p_B^\mu S_{A\perp} ^\nu + p_B^\nu 
	S_{A\perp} ^\mu \right) \Bigg\} , \nonumber
\label{eq:aa}
\eea
The first line is twist-two and contributes only when both initial hadrons are
longitudinally polarized.  Not surprisingly, this contribution measures
$g_1\otimes g_1$.  The latter two lines are twist-three, suppressed by
the factor $S_\perp$, and contribute only when one hadron has longitudinal
polarization and the other transverse.  It is worthwhile relating the spin,
twist and chirality structure of this result to the classification scheme
developed in \S 's 3 and 4.  Products of the form $g_1\otimes g_1$ and
$g_1\otimes g_T$ conserve quark chirality and contribute at orders of $1/Q^2$
which reflect the twist assignments we made in \S 4.  Axial vector bilocals
can only generate $g_1$ and $g_T$, so we can be confident that we have not
missed other contributions.  In fact, this result could have been written down
{\it a priori\/}, up to coefficients of order unity simply by carefully
considering the selections rules and twist assignments developed earlier in
these notes.  What is not at all obvious, however, is that other twist-three 
quark/gluon operators, of the form discussed in the previous section, and not 
directly related to $g_T$, do not arise when gluonic corrections to the diagram 
of fig.~\ref{fig-DY} are computed.  We take up this question below.

Finally, the $T\times T$ contribution takes the form,
\bea
	W_T^{\mu \nu} &=& -{1\over 3} (2\pi)^4 \delta ^2 (\vec{Q}_\perp) \Bigg\{
	\Big[S_{A\perp}\cdot S_{B\perp}
	\left( p_A^\mu p_B ^\nu + p_A^\nu p_B ^\mu - g^{\mu \nu}
	p_A\cdot p_B \right) \nonumber\\
	&+& (p_A\cdot p_B)
	\left( S_{A\perp}^\mu S_{B\perp} ^\nu +
	 S_{A\perp}^\nu S_{B\perp} ^\mu \right) \Big]{1\over M^2}
	 \sum _a e_a^2 h_1^a(x) h_1^{\bar{a}} (y) \nonumber\\
	&-&\sum _a e_a^2 h_1^a(x) h_L^{\bar{a}} (y)
	{p_A\cdot S_B \over p_B\cdot p_A} 
	\left( p_A^\mu S_{A\perp} ^\nu + p_A^\nu 
	S_{A\perp} ^\mu \right) \\ \nonumber
	&-& \sum _a e_a^2 h_L^a(x) h_1^{\bar{a}} (y)
	{S_A\cdot p_B \over p_A\cdot p_B} 
	\left( p_B^\mu S_{B\perp} ^\nu + p_B^\nu 
	S_{B\perp} ^\mu \right) \Bigg\} . \nonumber
	\label{eq:tt}
\eea
Here the first two lines are twist-two --- they scale (modulo logarithms) in the
deep inelastic limit.  They contribute only when both initial hadrons are
transversely polarized and provide a leading twist probe of transversity
distributions.  The second two lines are twist-three and contribute when one
hadron is transversely polarized and the other longitudinally.  Once again the
classification scheme of \S's 3 and 4 is illustrated.  Note,
for example, that 
transverse-longitudinal polarization receives contributions from
both $g_1\otimes g_T$ and $h_1\otimes h_L$.

\begin{figure}
	\centerline{\epsffile{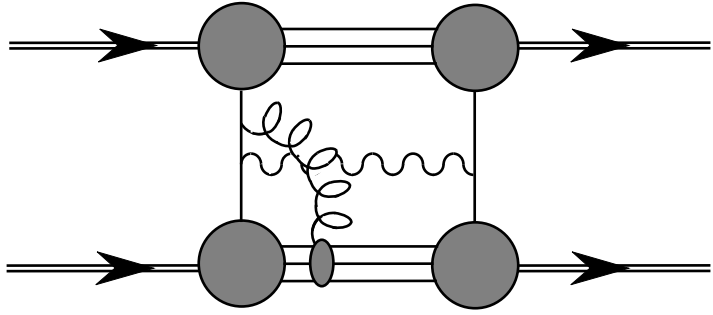}}
	\caption{{\sf New class of diagrams, needed to restore color
	and electromagnetic gauge invariance.}}
	\label{fig-newdia}
\end{figure}

Unfortunately things are not quite as simple as they have been presented so
far.  Two questions of gauge invariance arise --- one straightforward and the
other rather subtle:
\begin{itemize}
\item  The factorization of $W_{\mu\nu}$ into products of bilocal operators
acting in different hadrons is not color gauge invariant.  Evidence of this is
the {\it absence\/} of the Wilson link,
\begin{equation}
	{\cal P}\left(\exp i\int_0^\xi
	d\zeta^\mu A_\mu(\zeta)\right)
\end{equation}
between the quark fields at $0$ and $\xi$.  The
problem is that we have not included those diagrams that represent each quark
propagating in the color field of the remnants of the other hadron.  The same
problem was discussed in \S 3 for the case of deep inelastic scattering,
where it is easier to handle 
because the operator product expansion will preserve
gauge invariance if treated with sufficient care.  In this case we must find and
analyze the diagrams which restore color gauge invariance through twist-three
{\it and\/} be certain that they do not generate any contributions to
$W_{\mu\nu}$ beyond rendering the identification of Drell-Yan matrix elements
with quark distributions functions gauge invariant.  An example of the type of
diagram that does the trick is shown in fig.~(\ref{fig-newdia}).  The interested
reader is referred to ref.~\cite{JafJi2005} for details.
\item  The results we have just quoted (eqs.~(\ref{eq:aa}), (\ref{eq:vv}),
and (\ref{eq:tt})) violate
electromagnetic gauge invariance at twist-three.  It is easy to see that the
twist-three terms do not satisfy $W_{\mu\nu}q^\nu= 0$.  Once again the answer
lies in diagrams like fig.~(\ref{fig-newdia}).  In this case, when the Wilson
link which appears in the bilocal operator is expanded to first order away from
the light-cone, a set of contributions with explicit transverse gluon fields
arise.  Using the equations of motion these can be related back to the product
of {\it good\/} and {\it bad\/} quark fields which define the twist-three
distributions
$g_T$ and $h_L$.  With sufficient care one finds exactly the terms necessary to
restore electromagnetic gauge invariance.  Once again a fuller discussion can be
found in ref.~\cite{JafJi2005}.
\end{itemize}
\subsection{Polarized Drell-Yan: a Brief Summary}

Since the equations and the analysis in this section have become rather
complicated, it is useful to extract the simple predictions for spin asymmetries
to provide a summary and to reference the workers who originally derived these
results.

\begin{figure}
	\centerline{\epsffile{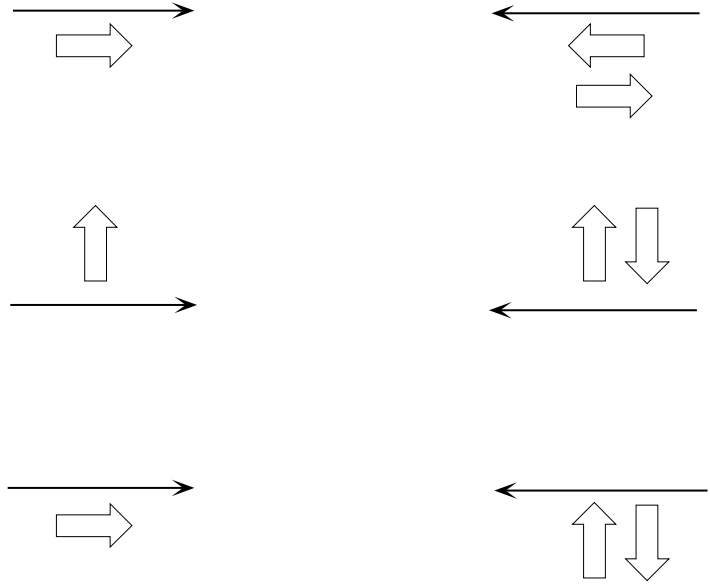}}
	\caption{{\sf Various polarization configurations, studied
	in the Drell-Yan process.
	The small arrows denote the spin projections of the
	two nucleons relative to the direction of the nucleon
	beams (long arrows). For one of the nucleons the difference
	of the up and down projections is taken. One distinguishes
	longitudinal (top), transverse (middle) and longitudinal-transverse
	(bottom) scattering.}}
	\label{fig:PDY}
\end{figure}

In the polarized Drell-Yan process, three cases appear:
longitudinal-longitudinal ($LL$), transverse-transverse ($TT$) and
longitudinal-transverse ($LT$), as illustrated in fig.~(\ref{fig:PDY}).  If
instead of virtual photon production, we had considered Drell-Yan production of
$Z^0$ or $W^\pm$, then a {\it longitudinal\/}, leading twist, parity violating
single spin asymmetry proportional to $g_1\otimes f_1$ appears.  There is no
analogous {\it transverse\/} asymmetry because the product $h_1\otimes f_1$
cannot conserve quark chirality.
The longitudinal asymmetry, which was first studied by Close and Sivers
\cite{CS77}, is given by:
\be
	A_{LL}={ \sum _a e_a^2 g_1^a(x) g_1^{\bar{a}}(y)
	\over \sum _a e_a^2 f_1^a(x) f_1^{\bar{a}}(y) }.
\ee
Ralston and Soper first discovered transversity, defined a twist-two
distribution and expressed the transverse asymmetry as:\cite{RS79}
\be
	A_{TT}={\sin^2 \theta \ \cos 2 \phi \over 1+\cos^2\theta}
	{\sum _a e_a^2 h_1^a(x) h_1^{\bar{a}}(y)
	\over \sum _a e_a^2 f_1^a(x) f_1^{\bar{a}}(y) }.
\ee
The angles are defined in the lepton center of mass frame.
The longitudinal-transverse asymmetry has been investigated by Jaffe and
Ji \cite{JafJi2005} and can be written as:
\be
	A_{LT}={2\sin 2\theta \cos \phi \over 1+\cos^2\theta}
	{M \over \sqrt{Q^2}}
	{ \sum _a e_a^2 ( g_1^a(x) y g_T^{\bar{a}}(y)
	-xh_L^a(x) h_1^{\bar{a}} (y) )
	\over \sum _a e_a^2 f_1^a(x) f_1^{\bar{a}}(y) }.
\ee
Clearly, it is a twist-three observable, which in principle allows for
a measurement of $h_L$.

\section{Annihilation and Quark Fragmentation \protect\\Functions}
\setcounter{equation}{0}

As a final application we give a brief introduction to the 
classification and uses of the spin dependent fragmentation functions which
determine the distribution of final state hadrons in deep inelastic
processes.  There are strong reasons to want to develop a better
understanding of hadron fragmentation processes.  In \S 1 we
mentioned the possibility of studying the spin structure of unstable
hadrons like the $\Lambda, \rho$ and $D^*$.  Another reason is that
parity violating processes like $W^\pm\rightarrow q\bar q\rightarrow$ hadrons
provide probes of spin structure unavailable in deep
inelastic scattering, where the analogous experiment would be neutrino
scattering from a polarized target.  A final reason is that the
selection of a particular hadronic fragment can serve as a filter for
an interesting quark or gluon distribution function.
The material in this section is based primarily on
refs.~\cite{JafJi2158,Ji} and \cite{BurkJaf}
where more details can be found.

\subsection{Single Particle Inclusive Annihilation} 

\begin{figure}
 \centerline{\epsffile{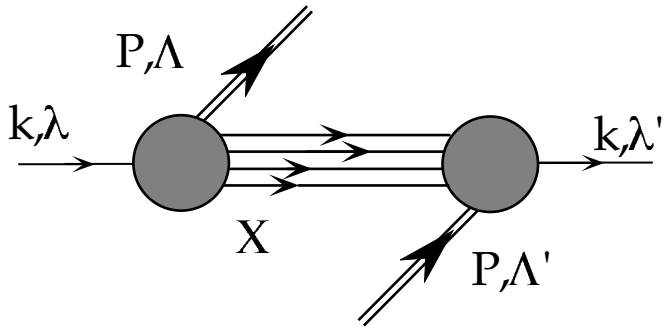}}
	\caption{{\sf Quark fragmentation function in a helicity basis.}}
	\label{fig:xx}
\end{figure}

The simplest quark fragmentation function
is represented diagrammatically in fig.~(\ref{fig:xx}).  
More complicated
fragmentation processes, such as coherent fragmentation of several quarks
and gluons, do not contribute until order $1/Q^2$, beyond our interest.  
First we consider the helicity classification in analogy to \S 4.1--4.3.
In the figure a quark of momentum
$k$ and helicity $\lambda$ fragments into a hadron of momentum $P$ and helicity
$\Lambda$ plus an unobserved final state $X$.  The process then repeats in
reverse as the unobserved system, $X$, plus the hadron of momentum $P$ and
helicity $\Lambda'$ reconstitute the quark 
of momentum $k$ and helicity $\lambda'$.  The scattering
$k+P\to k+P$ is forward, {\it i.e.\/} collinear.  For definiteness, we take
the momentum of the quark--hadron system to be aligned along the $\hat
e_3$--axis.  Then helicity is conserved as a consequence of angular
momentum conservation about this axis:  
$\lambda-\Lambda=\lambda'-\Lambda'$.  As in deep
inelastic scattering, the initial and final 
hadron helicities $\Lambda$ and $\Lambda'$
need not be equal because the hadron need not have been in a helicity
eigenstate; likewise for the quark.  As in 
scattering, the quark lines may correspond
either to good or bad light-cone components.

Some of the results of refs.~\cite{JafJi2365} 
and \cite{Ji} are as follows.
Quark fragmentation functions of the form shown in
fig.~(\ref{fig:xx}) and the equivalent gluon fragmentation
functions (without further active parton lines) are sufficient to
characterize hadron production in hard processes, 
provided:  i) one studies leading
twist (${\cal O}(1/Q^0)$) in any hard process, or 
ii) one studies an effect in deep
inelastic lepton scattering at the lowest twist at 
which it arises, and one ignores QCD
radiative corrections.\cite{Ji}   Each appearance of a bad
component of the quark field costs one power of $\sqrt{Q^2}$ in the deep
inelastic limit ({\it i.e.\/} it increases the twist by unity).
As in scattering, for produced hadrons of spin-1/2, helicity differences
are observed in longitudinal spin asymmetries; helicity flip is observed
in transverse spin asymmetries.  Since perturbative 
QCD cannot flip quark chirality
(except through quark mass insertions which we assume 
to be negligible for light
quarks), chirally--odd quark distribution and fragmentation functions
must occur in pairs.

Fragmentation functions can be labeled uniquely by specifying the
helicity of quarks and hadrons and the light-cone projection of the quarks
in direct analogy to the classification of distribution functions.  We denote 
fragmentation functions in a helicity basis by $\hat{\cal A}$.
Parity invariance of QCD requires:
$\hat {\cal A}_{\lambda\Lambda,\lambda'\Lambda'}= \hat {\cal
A}_{-\lambda-\Lambda,-\lambda'-\Lambda'}$.  Time reversal
invariance, which further reduces the number of independent quark
distribution functions {\it does not\/} generate relationships among the
$\{\hat {\cal A}\}$ because it changes the {\it out\/}--state $(PX)_{\rm out}$ 
in fig.~(\ref{fig:xx}) to an {\it in\/}--state.  As in the 
scattering case, we denote the
appearance of bad light-cone components by an asterix 
on the appropriate helicity index.

As a simple example, consider production of a 
scalar meson like the pion.  Through order $1/\sqrt{Q^2}$
there are three independent fragmentation functions:  $\hat {\cal A}_{{1\over
2}0,{1\over 2}0}$, $\hat {\cal A}_{{1\over 2}0,{1\over 2}^*0}$, and
$\hat {\cal A}_{{1\over 2}^*0,{1\over 2}0}$.  The first is twist-two and scales
in the $Q^2\rightarrow\infty$ limit, the latter two are twist-three and are
suppressed by $1/\sqrt{Q^2}$ in the $Q^2\rightarrow\infty$ limit.
The first function, $\hat {\cal A}_{{1\over 2}0,{1\over 2}0}$, is
proportional to the traditional fragmentation function $D(z,Q^2)$.  It has
the same twist, light-cone, helicity 
and chirality structure as $f_1(x,Q^2)$, so to
avoid an explosion of notation we denote it by $\hat f_1(z,Q^2)$
[We will follow the same convention for other fragmentation functions.]:
\begin{equation}
	\hat f_1(z,Q^2) \propto \hat{\cal A}_{{1\over 2}0,{1\over 2}0}
	\label{eq:fhat}
\end{equation}
If we were studying quark {\it distribution}
functions, the other two helicity amplitudes 
would be equal by time-reversal invariance.
Here, there are two independent fragmentation functions.
\begin{eqnarray}
	\hat e_1(z,Q^2) &\propto& \hat {\cal A}_{{1\over2}0,{1\over 2}^*0}
	+ \hat {\cal A}_{{1\over 2}^*0,{1\over 2}0}\nonumber\\
	\hat e_{\bar 1}(z,Q^2) &\propto& 	
	\hat {\cal A}_{{1\over 2}0,{1\over 2}^*0}
	- \hat {\cal A}_{{1\over 2}^*0,{1\over 2}0}\nonumber\\
	\label{eq:ehat}
\end{eqnarray}
We have found that the helicity classification of fragmentation functions is
identical to that of distribution 
functions at leading twist.  At twist-three, however, there are more 
fragmentation functions due to the absence of time reversal constraints.

The application to spin--$1/2$ is analogous to 
the classification of spin--$1/2$
distribution functions given in 
Table~\ref{tbl:spn1/2} of \S 4.2.2 except that each 
twist-three fragmentation function 
comes in two forms, one even and the other odd under
time reversal.  We suspect that the 
T-even functions are more important than the T-odd
since the latter vanish in the absence of final state interactions.
So fragmentation at twist-two requires 
$\hat f_1(z)$, $\hat g_1(z)$
and $\hat h_1(z)$ with helicity, 
transversity and chirality properties identical to
the analogous distribution function.  
The only twist-three fragmentation function
likely to be of much interest is 
the spin-averaged, chiral-odd, time reversal even
function $\hat e_1(z)$. 

In order to relate particular deep inelastic processes to quark
distribution and fragmentation functions and to study them in models of
non-perturbative QCD, it is necessary to have operator representations for
them.  This formalism is developed in refs.~\cite{JafJi2365,Ji}.  
Here we display the
results for the three leading twist fragmentation functions 
for a spin-$1/2$ hadron, and
for the twist-three spin average function, $\hat 
e_1$.  Generalizations to spin-$1$ can
be found in ref.~\cite{Ji}.  The generic 
expression for a fragmentation function takes
the form,
\bea
	\hat{\Gamma}(z)={\rm Tr}\left\{ \Gamma_{\alpha\beta} 
			  \sum_X \int\frac{d\lambda}
			  {2\pi} e^{-i\lambda/z} 
			  \left<0|\psi_{\beta}(0)|PX\right>
			  \left< PX|\bar{\psi}_{\alpha}(\lambda n)|0\right>
			  \right\}.
\eea
where
$\Gamma_{\alpha\beta}$ stands for an 
arbitrary Dirac matrix.  This result holds under
the condition that the diagram of 
fig.~(\ref{fig:xx}) dominates.  To obtain $\hat f_1$,
$\hat g_1$ and $\hat h_1$ one chooses 
$\Gamma=n \hspace{-2mm} /,  n \hspace{-2mm} /
\gamma_5$ and $\sigma^j_{\nu} n^{\nu}i \gamma_5$ respectively,
\begin{eqnarray} 
	\hat f_1(z)&=& 
		  \frac 12 \sum_X \int\frac{d\lambda}
		  {2\pi} e^{-i\lambda/z} 
		  \left<0|n \hspace{-2mm} / \psi (0)|PX\right>
		  \left< PX|\bar{\psi}(\lambda n)|0\right>,\nonumber\\ 
	\hat g_1(z)&=& 
		  \frac 12 \sum_X \int\frac{d\lambda}
		  {2\pi} e^{-i\lambda/z} 
		  \left<0|n \hspace{-2mm} / \gamma_5\psi (0)|PSX\right>
		  \left< PSX|\bar{\psi}(\lambda n)|0\right> ,\nonumber\\
	\frac {S_{\perp}{^j}}{M}  \hat h_1(z)&=&
		  \frac 12 \sum_X \int\frac{d\lambda}
		  {2\pi} e^{-i\lambda/z} 
		  \left<0|\sigma^j_{\nu} n^{\nu}i 
		  \gamma_5\psi (0)|PS_{\perp}X\right>
		  \left< PS_{\perp}X|\bar{\psi}(\lambda n)|0\right>\nonumber\\ 
	\label{eq:fghhat}
\end{eqnarray}
At twist-three the equations of motions can be used to express the structure 
functions in terms of independent degrees of freedom, quantized at $\xi^+=0$. 
For example one can obtain two expressions for the chiral odd, T-even
spin independent
fragmentation function $\hat e_1(z)$.  One involving
$\psi\bar\psi$ and another where bad light-cone components have been traded for
transverse derivatives and gluon degrees of freedom,
\begin{eqnarray}
	M\hat e_1(z)&=&\frac 12 \sum_X \int\frac{d\lambda}
	{2\pi} e^{-i\lambda/z} 
	\langle 0|\psi (0)|PX\rangle
	\langle PX|\bar{\psi}(\lambda n)|0\rangle\nonumber\\
	\hat e_1(z)&=&\frac{z}{4M} \sum_X \int\frac{d\lambda}{2\pi}
	e^{-i\lambda/z} 
	\{\langle 0|i n \hspace{-2mm} / 
	D\hspace{-3mm}/ _\perp (0)\psi_+(0)|PX\rangle
	\langle PX|\bar{\psi}_+(\lambda n)|0\rangle\nonumber\\
	&-&\langle 0|\psi_+(0)|PX\rangle
	\langle PX|\bar{\psi}_+(\lambda n)iD\hspace{-3mm}/ _\perp ( \lambda n) 
	n \hspace{-2mm} / |0\rangle\}.\nonumber\\
\end{eqnarray}
With these ingredients we are now prepared to explore a few applications of spin
dependent fragmentation functions.

\subsection{Polarized $q\rightarrow\Lambda$ Fragmentation Functions from  $e^+
e^-\to\Lambda+X$}
This section is based on work done with M.~Burkardt.\cite{BurkJaf}
In the symmetric quark model, the $\Lambda$-baryon has a rather simple
spin-flavor wavefunction. All its spin is carried by the $s$-quark, while the
$ud$-pair is coupled to $S=0$, $I=0$:  $\Delta u^\Lambda=\Delta d^\Lambda=0$
and $\Delta s^\Lambda=1$. While the quark model identifies  the
$\Lambda$-spin with the spin of the
$s$-quark, the data on the quark spin structure of the nucleon suggests
that the actual situation is more complex.  If we take the latest SMC/SLAC
data on the $\int_0^1 dx g_1^{ep}(x)$, combine it with $\beta$-decay data
and assume exact $SU(3)_{flavor}$ symmetry for baryon axial charges, we find
\begin{eqnarray}
\Delta u^\Lambda = \Delta d^\Lambda &=&{1\over 3}(\Sigma - D)
= -0.23 \pm 0.06\nonumber\\
\Delta s^\Lambda &=&{1\over 3}(\Sigma+2D) = +0.58 \pm
0.07\nonumber\\
\label{lambdaspin}
\end{eqnarray}
It would be exciting to test the $SU(3)_{flavor}$ assumption by observing
deep inelastic scattering from a $\Lambda$ target.  Unfortunately we
have to settle for observing the $\Lambda$ as a fragment in annihilation
processes.  The parity violating, self analyzing
decay of the final state $\Lambda$ makes it particularly easy to study its
polarization in fragmentation processes.  Measurement of the helicity
asymmetries for semi-inclusive production of $\Lambda$'s in $e^+e^-$
annihilation near the $Z^0$ resonance allows a complete determination
of the spin-dependent fragmentation functions for the different quark
flavors into the $\Lambda$.  In the event that these could be measured it
would be very interesting to compare the spin fractions measured in
fragmentation, $\Delta\hat s^\Lambda$, $\Delta\hat d^\Lambda$, and
$\Delta\hat u^\Lambda$, with the predictions, eq.~(\ref{lambdaspin}) in
order to get a better understanding of the role of spin in the fragmentation
process.

We are concerned here with twist-two helicity asymmetries, described by the
fragmentation function $\hat g_1(z,Q^2)$.  For simplicity we adopt a more
conventional parton-model notation where we define $d^{\Lambda(L)}_{q(L)}$
to be the probability that a left handed $\Lambda$ fragments into a left
handed quark, {\it etc.\/}  Then the unpolarized differential
cross section for
$e^-e^+\rightarrow \Lambda+X$ is obtained by summing over the
cross sections for $e^+e^-\rightarrow q\bar{q}$, weighted with
the probability $d^\Lambda_q(z,Q^2)$ that a quark with momentum
${1\over z}P$ fragments into a $\Lambda$ with momentum $P$.  As usual, we
suppress the $Q^2$ dependence generated by radiative corrections in QCD,
\begin{equation}
	{d^2\sigma^\Lambda\over d\Omega\, dz} = \sum_q {d\sigma^q\over d\Omega}
	d^\Lambda_q(z).
\label{eq:unpol}
\end{equation}

There is a single polarized fragmentation function for each flavor of quark
or antiquark,
\begin{eqnarray}
	\Delta \hat{ q}(z) &=&
	d^{\Lambda(L)}_{q(L)}(z) - d^{\Lambda(R)}_{q(L)}(z)\nonumber\\
	&=& d^{\Lambda(L)}_{q(L)}(z) - d^{\Lambda(L)}_{q(R)}(z)\nonumber\\
	 \Delta \hat{ \bar{q}}(z) &=&
	d^{\Lambda(L)}_{\bar{q}(L)}(z) - 
	d^{\Lambda(R)}_{\bar{q}(L)}(z) \nonumber\\
	 &=&d^{\Lambda(L)}_{\bar{q}(L)}(z) -
	d^{\Lambda(L)}_{\bar{q}(R)}(z),
	\label{eq:fragfractions}
\end{eqnarray}
and furthermore isospin invariance requires that $\Delta\hat u(z)=\Delta\hat
d(z)$ and $\Delta\hat{\bar u}(z)=\Delta\hat{\bar d}(z)$, so the number of
independent fragmentation functions is reduces to four --- {\it e.g.\/}
$\Delta{\hat u}, \Delta{\hat{\bar u}}, 
\Delta{\hat s}$ and $\Delta{\hat{\bar s}}$.
 
As an example we quote the prediction for $\Lambda$ production in
$e^-e^+$ annihilation via photons.  In this case
one has to start from polarized $e^-$ (or $e^+$) in order to fix the
polarization of the quarks. One finds for the helicity asymmetric
cross-section,
\begin{equation}
	{d^2\sigma^{e^-(L)e^+\rightarrow \Lambda(L)X}\over d\Omega \,dz}
	-{d^2\sigma^{e^-(L)e^+\rightarrow \Lambda(R)X}\over d\Omega \,dz}
	={\alpha^2\over 2s}\cos \theta
	\left[ {5\over 9}\left( \Delta \hat{ u}(z) + \Delta \hat{
	\bar{u}}(z)\right)
	+ {1\over 9}\left( \Delta \hat{ s}(z) + \Delta \hat{
	\bar{s}}(z)\right)\right]
\end{equation}
where $L,R$ denotes the helicity of the
$e^-$ and the $\Lambda$.  So annihilation into a single photon allows one
combination of the four independent fragmentation functions to be measured. 
As described in ref.~\cite{BurkJaf}, annihilation at the $Z^0$ and just off
the $Z^0$ peak where $\gamma-Z$ interference is largest will allow
independent detection of all four quark and antiquark fragmentation
functions.  For further discussion of this process, backgrounds and
experimental possibilities see ref.~\cite{BurkJaf} and recent papers by the
$Aleph$ and $Delphi$ collaborations.

\subsection{Observing $h_1(x,Q^2)$ in Electroproduction}
This section is based on work done with X.~Ji.\cite{JafJi2158,Ji}
Chirally odd quark distributions are difficult to
measure because they are suppressed in totally--inclusive deep inelastic
scattering.  So far, the only practical way to determine $h_1(x,Q^2)$
we have discussed has been Drell-Yan production with
transversely polarized target and beam.

As an application of the fragmentation function formalism
-- one of many -- we show how a chirally odd {\it fragmentation 
function\/} can be exploited to enable a measurement of $h_1(x,Q^2)$ to be
obtained in polarized electroproduction of pions from a transversely 
polarized nucleon. This is an experiment which could be performed at several
existing facilities. Related suggestions involving semi--inclusive
production of $\Lambda$--hyperons and of two pions have been discussed
previously.\cite{Collins,Artru}  The proposal outlined here is simpler since
it involves only one particle in the final state and does not require
measurement of that particle's spin.   The price we pay for this
simplicity is suppression by a power of $\sqrt{Q^2}$.

\begin{figure}
	\centerline{\epsffile{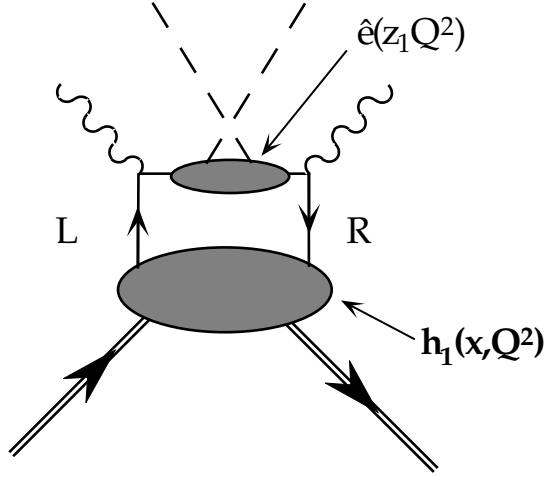}}
	\caption{{\sf Single particle inclusive scattering $ep\rightarrow ehX$. 
	The labels $L$ and $R$ reflect the chiral odd nature of $h_1$.}}
	\label{fig:muis}
\end{figure}

We consider pion production in the current fragmentation region
of deep-inelastic scattering with longitudinally polarized leptons on a
polarized nucleon target.  The simplest diagram for the process is
shown in fig.~(\ref{fig:muis}), where a quark struck by
the virtual photon fragments into an observed pion
plus other unobserved hadrons. The cross section is proportional to a trace
and integral over the quark loop which contains the quark distribution
function and fragmentation function.  Due to chirality conservation at the
hard (photon) vertex, the trace picks up only
the products of the terms in which the distribution and
fragmentation functions have the same chirality.
When the nucleon is longitudinally polarized
(with respect to the virtual-photon momentum),
the twist-two, chirally even distribution $g_1(x)$ can couple
with the twist-two chirally even fragmentation function
$\hat f_1(z)$, producing a leading contribution ${\cal O}(1/Q^0)$
to the cross section. On the other hand, in the case of a transversely
polarized nucleon, there is no leading-order contribution.
At the next order, the nucleon's transversity
distribution $h_1(x)$ can combine with the twist-three
chirally odd fragmentation function $\hat e_1(z)$,
and similarly $g_T(x)$ can combine with the chirally even
transverse-spin distribution $\hat f_1(z)$. Both couplings produce
$1/\sqrt{Q^2}$ contributions to the cross section.
It is simple to see, however, that fig.~(\ref{fig:muis}) alone
does not produce an electromagnetically-gauge-invariant
result.  This is a typical example of the need to consider
multi-quark/gluon processes beyond twist-two.
In the present case (twist-three), however, the contributions from coherent
scattering can be expressed, with novel use of QCD equations of motion, in
terms of the distributions and fragmentation functions defined from quark
bilinears.  This is a specific example of another rule quoted in \S 6.2:
``Parton diagrams (without further active parton lines) are sufficient to
characterize hadron production in hard processes, provided:  one studies an
effect in deep inelastic lepton scattering at the lowest twist at which it
arises, and one ignores QCD radiative corrections.''
The combined result is gauge invariant, as can be seen from the resulting
nucleon tensor,
\begin{equation}
	\hat W^{\mu\nu} = -i\epsilon^{\mu\nu\alpha\beta}
	{q_{\alpha}\over \nu}
	[(S\cdot n)p_\beta \hat G_1(x,z) +
	{S_{\perp \beta}} \hat G_T(x,z)]
	\label{eq:tensor}
\end{equation}
The two structure functions in $\hat W^{\mu\nu}$ are related to
parton distributions and fragmentation functions,
\begin{eqnarray}
	\hat G_1(x,z) &=&{1\over 2} \sum_a 
	e_a^2 g_1^a(x) \hat f_1^a(z)\nonumber\\
	\hat G_T(x,z) &=&{1\over 2} \sum_a e_a^2 \Big [g^a_T(x) \hat f^a_1(z)
        + {h^a_1(x) \over x} {\hat e^a(z)\over z} \Big ]\nonumber\\
\end{eqnarray}
where the summation over $a$ includes quarks and antiquarks
of all flavors.

To isolate the spin-dependent part of the deep-inelastic
cross section we take the difference of cross sections
with left-handed and right-handed leptons, we use
\begin{equation}
{d^2\Delta \sigma \over dE'd\Omega} = 
{\alpha_{\rm em}^2 \over Q^4} {E'\over EM_N}
\Delta \ell^{\mu\nu}\hat W_{\mu\nu}
\end{equation}
It is convenient to express the cross section in terms of
scaling variables in a frame where lepton beam defines the
$\hat e_3$-axis
and the $\hat e_1-\hat e_3$ plane contains the nucleon polarization vector,
which has a polar angle $\alpha$.
In this system, the scattered lepton has polar angles
$(\theta, \phi)$ and therefore the momentum transfer ${\bf q}$
has angles $(\theta, \pi-\phi)$. Then we obtain an expression for the
semi-inclusive process quite similar to that for the total inclusive
scattering defined by eq.~(\ref{eq:spindepxsection}).
\begin{eqnarray}
	{d^4 \Delta \sigma \over dx\,dy\,dz\,d\phi}
	&=& {8\alpha^2_{\rm em} \over Q^2}
	\Big[ \cos\alpha (1-{y\over 2}) G_1(x,z)\nonumber\\
	&+& \cos\phi\sin\alpha\sqrt{(\kappa -1)(1-y)}
	\left(G_T(x,z) - G_1(x,z)(1-{y\over 2})\right)\Big]\nonumber\\
	\label{eq:final}
\end{eqnarray}
where $\kappa = 1 + 4x^2M^2/Q^2$ in the second term
signals the suppression by a factor of $1/\sqrt{Q^2}$ associated with
the structure function $G_T$. The existence
of $G_1$ in the same term is due to a small
longitudinal polarization of the nucleon relative to ${\bf q}$
when its spin is perpendicular to the lepton beam.

Eq.~(\ref{eq:final}) is the main result of this section. As a check, we
multiply by $z$, integrate over it and sum over all hadron species.
Using the well-known momentum sum rule,
\be
	\sum_{\rm hadrons} \int dz z \hat f_1^a(z) = 1,
\ee
and the sum rule,
\be
\sum_{\rm hadrons}\int dz \hat e^a_1(z) = 0
\ee
which is related to the fact that the chiral condensate
vanishes in the perturbative QCD vacuum,
we get the result for {\it total\/} inclusive scattering, given in
eq.~(\ref{eq:spindepxsection})  The
similarity between the inclusive and semi-inclusive cross sections
suggests that they can be extracted conveniently
from the same experiment.

The aim of this example was to show that an unfamiliar fragmentation
function ($\hat e_1$) could be employed to obtain a measurement of an
interesting, if unfamiliar, distribution function ($h_1$).  It is apparent
from eq.~(\ref{eq:final}) that we have been only partially successful:
although the $h_1^a$ distribution for each quark flavor appears in
eq.~(\ref{eq:final}), the sum
over flavors couples it to the unknown flavor dependence of $\hat e_1^a$.
Perhaps flavor tagging can be used at large--$z$ to identify the
contributions of individual quark flavors.  For $x$ in the valence region
(where one can ignore antiquarks in the nucleon), and $z\rightarrow 1$, the
dominant fragmentation, $u\rightarrow\pi^+$, $d\rightarrow\pi^-$,
$s\rightarrow K^-$,  may allow one to trigger on the contributions
of $u$, $d$ and $s$ quarks separately.\cite{Mank}  
One might be concerned that the
unknown fragmentation function, $\hat e_1$, might not respect the dominant
fragmentation selection rules, which have only been tested for the
spin-averaged, twist-two fragmentation function,
$\hat f_1$.  However, the coherent gluon interactions which distinguish the
twist-three $\hat e_1$ from $\hat f_1$ are flavor independent and should
not alter the selection rules.  Although this may be a difficult path to
measuring $h_1$, so are all the others.  This one owes its existence to our
improved understanding of the spin and twist dependence of quark
fragmentation functions, including the spin-average twist-three
fragmentation function $\hat e_1$.

\newpage

\bibliography{newjafref}
\end{document}